\documentclass[aps,prx,twocolumn,superscriptaddress,longbibliography]{revtex4-2}

\usepackage{mathrsfs}
\usepackage{amsmath}      
\usepackage{amssymb}      
\usepackage{bm}           
\usepackage{graphicx}  
    \graphicspath{{Figures/}}
\usepackage{hyperref}     
\usepackage{physics}      
\usepackage{braket}
\usepackage[dvipsnames]{xcolor}
\usepackage{soul}      
\usepackage{siunitx}     
\usepackage[caption=false]{subfig} 
\usepackage{dsfont}

\newcommand{\lp}{\left(}
\newcommand{\rp}{\right)}
\newcommand{\lb}{\left[}
\newcommand{\rb}{\right]}

\definecolor{darkgreen}{rgb}{0.0, 0.5, 0.0}

\hypersetup{
    colorlinks=true,
    linkcolor=blue,
    citecolor=blue,
    urlcolor=blue
}

\begin{document}

\title{Dynamical metastability and transient topological magnons \\ 
in interacting driven-dissipative magnetic systems}

\author{Vincent P. Flynn}
\affiliation{Department of Physics, Boston College, Chestnut Hill, Massachusetts 02467, USA}
\affiliation{Department of Physics and Astronomy, Dartmouth College, Hanover, New Hampshire 03755, USA}

\author{Lorenza Viola}
\affiliation{Department of Physics and Astronomy, Dartmouth College, Hanover, New Hampshire 03755, USA}

\author{Benedetta Flebus}
\affiliation{Department of Physics, Boston College, Chestnut Hill, Massachusetts 02467, USA}

\date{\today}

\begin{abstract} 
Metastability, i.e., partial relaxation to long-lived, quasi-stationary states before true asymptotic equilibrium sets in, emerges ubiquitously in classical and quantum dynamical systems as a result of timescales separation. In open quantum systems, an intrinsically nonequilibrium analogue, \textit{dynamical metastability}, can additionally originate from the spectral geometry of a non-Hermitian evolution operator. In noninteracting bosonic models, this mechanism produces boundary-sensitive anomalous relaxation, transient amplification, and topologically mandated long-lived edge modes, all of which are enhanced with increasing system size. Here we extend dynamical metastability into the nonlinear, interacting regime and identify magnetic heterostructures as a natural experimental platform for its exploration. We introduce an interacting spin Lindbladian whose linearized magnon dynamics map onto a dynamically metastable Hatano-Nelson chain, and show that dynamical metastability in the noninteracting limit seeds genuinely nonlinear phenomena, including size-dependent spin dipping and anomalous attraction to unstable equilibria. Exponentially localized long-lived edge states associated to topologically mandated Dirac bosons persist under nonlinearities and disorder.  We further analyze the magnetization dynamics in magnetic multilayers within the classical Landau-Lifshitz-Gilbert-Slonczewski (LLGS) framework, identifying Dzyaloshinskii-Moriya interaction, nonlocal damping, and spin-transfer torque as control parameters governing bulk-boundary stability mismatch and band topology. While all the distinctive dynamical phenomena identified for the above spin Lindbladian reappear in this experimentally relevant setting, the LLGS framework also supports multistability and limit cycles that are absent in the quantum model. Our results constitute the first systematic study of dynamical metastability in nonlinear magnetization dynamics, directly relevant to spin-torque oscillator arrays, magnonic devices, and beyond.
\end{abstract}

\maketitle

\section{Introduction}
\label{sec: intro}

\subsection{Context and motivation}

The discovery and classification of phases in both traditional condensed-matter systems and synthetic many-body platforms is a central theme across quantum sciences, with implications ranging from fundamental physics advances to new functionalized materials and devices \cite{Sachdev2023,Goyal2025}. In closed systems at equilibrium, phases are organized by order parameters and symmetry breaking in the Landau--Ginzburg framework \cite{LandauPT1937}, or by global invariants protecting symmetry-protected topological phases and topological order \cite{RyuSPT2016}. In such settings, conservation laws and detailed balance constrain fluctuations, and phase boundaries are associated with singular changes in the equilibrium free energy.

Driven-dissipative quantum systems fall outside this paradigm \cite{DiehlUniversal2025}. With openness and non-Hermiticity fundamentally altering concepts such as symmetry and energy gaps, the identification of phases and their transitions is no longer captured by a free-energy minimization and is effectively encoded in the spectral properties of non-unitary dynamical generators. As a result, extensive efforts are underway to generalize topological classification and phase structure to this novel class of nonequilibrium systems \cite{DiehlTopDiss2011, BardynTopDiss2013, GongNHTopoPhase2018, KawaNHSymTop2019, Mori2020, FlynnBosoranasPRL2021, UedaLSkin2021, BergholtzEPTop2021, KalthoffNEPT2022, SelfOsccQED_Dreon2022, KPZPolariton_Fontaine2022, FlynnBosoranasPRB2023, Rako2024, UghrelidzeDM2024,Wanjura2025}
and to further characterize the interplay between non-Hermiticity and nonlinearities or interactions in critical and transport phenomena \cite{SpagnoloMSCMP2017,NunnenNLHN2024,BeggNRCrit2024,MandaNLHN2024}. These developments are revealing a rich landscape of nonequilibrium physics where hallmark features of non-Hermitian (more generally, non-normal) dynamical systems -- such as exceptional points, extreme sensitivity to boundary conditions, and the non-Hermitian skin effect (NHSE) -- play a crucial role \cite{YaoNHSE2019,SlagerNHSE2020,OkumaNHSE2020,OkumaPS2020,OkumaAnom2021,NoriEPs2020,NoriLEPs2020,MarkoPhantom2021,MarkoNHSE2022}.
Beyond their fundamental interest, such effects underpin emerging applications in quantum technologies, notably, topological amplification \cite{PorrasTopoAmp2019,WanjuraTopAmp2020,WanjuraTopAmp2021,PorrasIO2021,BrunelliTopAmp2023} 
and quantum sensing \cite{BudichSensor2020,DeCarloNHSense2022,XiaoNHSense2024,ClerkSensor2018}.

In the search for a conceptual bridge between equilibrium and nonequilibrium physics, \emph{metastability} -- loosely speaking, the emergence of long-lived, quasi-stationary behavior preceding eventual relaxation to a true equilibrium -- plays a central role. In classical stochastic dynamics and thermodynamic systems, metastability typically reflects a separation of timescales induced by free-energy barriers \cite{BinderPT1987}. Ubiquitously encountered in modeling brain activity \cite{RossiMS2024}, metastability is a common occurrence in classical soft matter, with glasses providing a paradigmatic example \cite{Binder2011}. In open quantum systems evolving according to continuous-time Markovian dynamics, metastability has been systematically linked to spectral gaps and slow manifolds of the corresponding Liouvillian generator \cite{GarrahanMS2016,GarrahanMS2021,MoriMS2021,Macieszczak20UnravMS24,XiangSwitchMS2025}, with a similar picture holding for discrete-time Markovian quantum systems as well \cite{DiscreteTime}. Notably, recent results are also pointing to a rigorous characterization of metastable quantum states in terms of area laws of mutual information \cite{bergamaschi2025}.

Alongside this traditional notion, a distinct and intrinsically nonequilibrium mechanism -- dubbed \emph{dynamical metastability} \cite{FlynnBosoranasPRL2021} -- has been identified in driven-dissipative Markovian systems with strong boundary sensitivity. Specifically, in a class of quadratic (i.e., mean-field, or noninteracting) bosonic systems in one spatial dimension, long-lived or transiently amplifying dynamics has been found to originate {\em not} from thermodynamic barriers or spectral gaps, but rather from the geometry and topology of {\em non-Hermitian rapidity spectra} under open versus periodic boundary conditions (BCs) -- which ultimately dictate the interplay between dynamical stability properties and system size \cite{FlynnBosoranasPRB2023,UghrelidzeDM2024}. This interplay can in fact give rise to a series of anomalous transient phenomena: two-step relaxation due to bulk-boundary dissipative gap mismatch, dynamical metastability with long-lived amplification, and topological dynamical metastability featuring (so-called Majorana or Dirac) boundary-localized modes protected by rapidity-band winding. These effects have been explored entirely within the noninteracting regime, leaving open the crucial question of how they evolve once nonlinearities and interactions are present -- as inevitable to a lesser or greater extent in realistic settings.

Magnetic and spintronic systems provide a natural and timely setting in which to address this question. Although long viewed primarily as application-driven platforms for spin transport and microwave functionalities, their experimental maturity is precisely their strength: decades of development have yielded exceptional control over damping mechanisms, spin-torque injection, interlayer exchange, and chiral interactions, along with probes that directly resolve both spectral properties and real-time dynamics. This level of control has already enabled the experimental realization of far-from-equilibrium, dissipation-driven collective dynamics in extended magnetic arrays, including synchronized networks of spintronic nano-oscillators \cite{kiselev2003microwave, lebrun2017mutual, kumar2023robust, kumar2025} and spintronic neuromorphic architectures \cite{zahedinejad2022memristive, grollier2020neuromorphic}. These same features have also recently sparked a surge of interest in engineering non-Hermitian phenomena and topological phases in magnetic systems \cite{hurst2022non,flebus2023recent,GunninkMagZM2023}: however, the potential for leveraging the interplay of coherent and dissipative interactions with the intrinsic nonlinearities of magnetization dynamics to explore novel nonequilibrium phenomena remains largely untapped.

These considerations motivate our central question, at the intersection of non-Hermitian topology, metastability, and nonlinear dynamics: \textit{Which anomalous transient phases and boundary-sensitive dynamics survive -- or qualitatively change -- once nonlinear magnetization dynamics is taken fully into account?}

\subsection{Structure and summary of main results}

In this work, we address the above question by investigating driven-dissipative magnetic systems whose linearized magnon dynamics closely resemble those of a paradigmatic Hatano-Nelson (HN) chain \cite{Hatano1996}, yet whose full evolution is strongly nonlinear and far from equilibrium.  Specifically, after recalling in Sec.\,\ref{sec: DM} the general framework for dynamical metastability in quadratic bosonic dynamics, and introducing the necessary concepts and tools, we study two complementary models that allow us to disentangle universal features from platform-specific nonlinear physics, while also providing a route towards experimental implementation:

$\bullet$ {\em Magnetization dynamics in an interacting quantum spin-chain model}  (Sec.\,\ref{sec: Lind}). Here, the driven-dissipative dynamics are governed by a Markovian (Lindblad) master equation which, in the dilute-magnon limit of linearized dynamics, is designed to exactly match the setting previously analyzed for the noninteracting HN Lindbladian. In this way, a well-defined connection between interacting spin systems and Markovian free-boson models is established. Restoring nonlinear spin interactions, we derive the corresponding semiclassical equations of motion (EOM), construct the nonlinear spin-wave steady state phase diagram, and investigate how the anomalous transient phenomena identified in the linear regime persist, deform, or terminate in the presence of nonlinear interactions (see also Fig.\,\ref{fig: EOMHierarchy} for a pictorial summary of the relevant dynamical regimes we consider, and their interconnections). Notably, we demonstrate that {\em finite-frequency} boundary-localized spin configurations that serve as magnonic analogues of Dirac-type topological edge modes still exist with a lifetime that increases with system size, as long as interactions remain sufficiently small.  We also uncover genuinely nonlinear transient phenomena, including pronounced {\em spin dipping} and {\em transient attraction} to unstable equilibria, that have no counterpart in the linear theory considered thus far. 

$\bullet$ {\em Classical magnetization dynamics in a ferromagnetic heterostructure model} (Sec.\,\ref{sec: LLG}). In this case, we describe the dynamics through the phenomenological Landau--Lifshitz--Gilbert--Slonczewski (LLGS) formalism \cite{Gilbert2004,SlonSTT1996,BergerGilb2001} widely employed across spintronics and magnonics \cite{Lakshmanan2011,FlebusMagRoad2024}. We show that the linearized LLGS equations inherit tunable nonreciprocity and a bulk-boundary stability gap mismatch, albeit with a distinctive dependence upon the underlying equilibria. Likewise, we demonstrate how the fully nonlinear dynamics exhibit the same set of anomalous transients of the Lindbladian model -- including a long-lived edge state which is the classical-magnetization counterpart of the Dirac boson mentioned above for the Lindbladian setting. Crucially, we identify the experimentally accessible control knobs -- namely, the interfacial Dzyaloshinskii-Moriya interaction, nonlocal damping and spin-transfer torque -- that can tune these effects within parameter regimes accessible to existing or near-future spintronic devices \cite{HsuSkym2018,ZhangTunableDMI2021,NembachDMI2022,ZhangObsDMI2025,HuangRKKYDMIPRM2025}. 

While our results position magnetic systems as a versatile platform for exploring the interplay of non-Hermitian topology, metastability, and nonlinear dynamics, our comparison between the Lindbladian and LLGS descriptions also unveils fundamental differences between the two approaches. As remarked, both models reproduce the same non-Hermitian spectral phenomenology at the linearized level, and the same metastable phenomena emerge. The intrinsically amplitude-dependent dissipation in the LLGS framework causes the nonequilibrium steady-state phase diagram to be qualitatively different, however: the LLGS dynamics support limit cycles and multistable phases that are absent in the interacting Lindbladian model. This intersects with an open theoretical frontier, namely, the development of many-body open-system theories capable of bridging semiclassical nonlinear magnetization dynamics and fully quantum, interacting Lindbladian evolution on equal footing \cite{Wieser2013,BranislavLLGLind2024,BranislavKeldyshLLG2024,WatanabeLLGLind2025,UhrigLLGLind2025}.

We conclude in Sec.\,\ref{sec: end} by reiterating our key findings and their broad significance, along with mentioning some important open questions.  Relevant technical details are included in three separate appendices. In particular, Appendix \ref{app: DBlife} provides an estimate of both the characteristic localization length and lifetime of the predicted Dirac boson modes, which may be of independent interest in the context of topological metastability.

\section{Dynamical metastability in driven-dissipative bosonic systems}
\label{sec: DM}

In this section, we provide a self-contained account of dynamical metastability and related concepts through the lens of a prototypical example: a driven-dissipative, noninteracting bosonic HN chain. For a more extended discussion of the general framework, we refer the reader to Refs.\,\cite{FlynnBosoranasPRL2021,FlynnBosoranasPRB2023,UghrelidzeDM2024}. 

\subsection{Anomalous relaxation and \\ dynamical metastability}
\label{sub: ARDM}

We consider an open-quantum system model that describes a one-dimensional lattice of $N$ bosonic degrees of freedom, coupled to nearest neighbors both coherently and incoherently, and evolving under Markovian dynamics. If we denote by $a_j$ the annihilation operator of the $j$'th boson, the state of the system obeys a master equation $\dot{\rho} = \mathcal{L}(\rho)$ generated by a quadratic bosonic Lindbladian (QBL). That is, in units where $\hbar =1$, $\mathcal{L}(\rho) = -i[H,\rho] + \mathcal{D}(\rho)$, with $H$ being the quadratic Hamiltonian and $\mathcal{D}$ the Markovian dissipator given by
\begin{subequations}
\label{eq: HNQBL}
\begin{align}
    H &= \sum_j \omega a_j^\dag a_j -\sum_{j}\lb (J+iD) a_j^\dag a_{j+1} + \text{H.c.}\rb \label{eq: HNHam}, \\
\mathcal{D} &= 2\sum_{j}\lp\kappa_-\mathcal{D}[a_j] + \Gamma\mathcal{D}[a_j + a_{j+1}]+\kappa_+\mathcal{D}[a_j^\dag]\rp
\label{eq: HNdiss},
\end{align}
\end{subequations}
respectively. Here, $\mathcal{D}[A](\rho) \equiv A\rho A^\dag - \{A^\dag A,\rho\}/2$, $\omega$ is the spatially uniform onsite frequency, and  $J$ and $D$ are the real and imaginary parts of the complex hopping amplitude, respectively. The parameters $\kappa_-$, $\Gamma$, and $\kappa_+$ correspond to the nonnegative rates of single-particle loss, correlated decay, and single-particle gain, respectively.

An essential step toward defining dynamical metastability is to characterize the {\em dynamical stability} properties of the system described by the QBL defined by Eqs.\,\eqref{eq: HNHam}-\eqref{eq: HNdiss}. More precisely, we wish to determine which choices of parameters ensure bounded evolution for all observable expectation values, independent of initial conditions. This is relevant, for instance, if a given QBL describes the linearized magnon evolution of an interacting spin system about some equilibrium configuration. In such a case, dynamical stability indicates local attractivity of the latter (that is, any sufficiently nearby initial condition will converge to said equilibrium), while instability characterizes its local repulsivity. 
For a QBL, dynamical stability can be diagnosed by studying the evolution of the annihilation operators $a_j$ in the Heisenberg picture \cite{ProsenThirdQ2010,YikangThirdQ2022,FlynnBosoranasPRB2023}. 
In our case, we find the EOM
\begin{align}
\label{eq: HNEOM} 
    \dot{a}_j
    = \mathcal{L}^\star(a_j) = -\lp \kappa_\text{eff} +i\omega\rp a_j+J_R a_{j-1} + J_L a_{j+1}, \;
\end{align}
where $\mathcal{L}^\star$ denotes the Heisenberg picture generator, and we have introduced the effective local loss rate $\kappa_\text{eff}\equiv \kappa_- + 2\Gamma -\kappa_+$, the effective right-hopping amplitude $J_R \equiv iJ+D -\Gamma$, and the effective left-hopping amplitude $J_{L} \equiv iJ-D-\Gamma $. We recognize the EOM in Eq.\,\eqref{eq: HNEOM} as a  generalization of those obtained from the well-known non-Hermitian HN Hamiltonian \cite{Hatano1996}. In particular, since 
\begin{align}
\label{eq: HNnonrec}
|J_L|^2 - |J_R|^2 = 4\Gamma D,
\end{align}
 the system undergoes  asymmetric hopping characteristic of the HN chain whenever $\Gamma$ and $D$ are both nonzero. If $\Gamma\neq 0$ and $D>0$ $(D<0$), then wavepackets will propagate towards the left (right), since $|J_L|>|J_R|$ ($|J_L|<|J_R|$) in this case. Despite the manifest similarities, it is essential, however, to emphasize that Eq.\,\eqref{eq: HNEOM} follows from an exact, {\em fully quantum} Markovian evolution - including the effects of quantum jumps -  rather than via post-selection or other schemes for achieving dynamics under an effective non-Hermitian Hamiltonian. 

For our analysis, it is convenient to recast Eq.\,\eqref{eq: HNEOM} as a matrix equation
\begin{align}
\label{eq: phiEOM}
    \dot{\phi} = \mathbf{A} \,\phi, \quad \phi \equiv [a_1,\ldots,a_N]^T,
\end{align}
where $\mathbf{A}$ is a {\em dynamical matrix} whose form depends, in addition to the model parameters, on the BCs that are being assumed. The form of Eq.\,\eqref{eq: phiEOM} is a consequence of the ``weak" global U(1) symmetry of the Lindbladian. Namely, $\mathcal{L}$ is invariant under the transformation $a_j\mapsto e^{i\theta}a_j$, for arbitrary $\theta \in \mathbb{R}$. In open quantum systems, a weak symmetry is any unitary or antiunitary transformation $S$ that commutes with the Lindbladian, i.e., $\mathcal{L}(S\rho S^{-1}) = S\mathcal{L}(\rho)S^{-1}$, for any $\rho$ \cite{BucaSymm2012,AlbertSymm2014}. Weak U(1) symmetry guarantees, in particular, that annihilation operators will not be coupled to creation operators (a situation commonly encountered in Bogoliubov-de Gennes systems), thus reducing the dimension of the system of equations from $2N$ to $N$, as in Eq.\,\eqref{eq: phiEOM}. Throughout this work, we will focus on two types of BCs: open BCs (OBCs) and periodic BCs (PBCs). In the case of OBCs, we set $a_0 = a_{N+1} = 0$ in Eq.\,\eqref{eq: HNEOM}, while for PBCs we take $a_0 = a_N$ and $a_{N+1} = a_1$ \footnote{Note that imposing these conditions at the level of the generator $\mathcal{L}$ would produce different EOM. This is because the term $2\Gamma$ appearing in the effective loss rate $\kappa_\text{eff}$ originates from having two dissipatively coupled neighbors at rate $\Gamma$. Hence, truncating boundaries at the generator level would result in a effective loss rate of $\kappa_-+\Gamma-\kappa_+$ at the boundaries. Physically, we may imagine adding an additional boundary loss term with strength $\Gamma$ to account for this discrepancy and produce the desired EOM. In general, however, this difference does not meaningfully change the results presented.}. These two BCs result in $\mathbf{A}$ belonging to different matrix classes: namely, for OBCs, $\mathbf{A}$ is known as a Toeplitz matrix, while for PBCs it is a circulant matrix \cite{TrefethenPS2005}. 

As shown in Ref.\,\cite{ProsenThirdQ2010,YikangThirdQ2022}, the dynamical stability properties of the many-body system are almost completely determined, in a sense we clarify below, by the stability of Eq.\,\eqref{eq: phiEOM}, viewed as a linear time-invariant dynamical system. Let the {\em stability gap} $\Delta$ be defined as the largest real part of the eigenvalues of $\mathbf{A}$. If $\Delta<0$, the system is dynamically stable and possesses a unique steady state \cite{YikangThirdQ2022}.  If $\Delta>0$, then it is dynamically unstable and does not possess a steady state. If $\Delta=0$, the system is either stable or unstable, depending on whether there are infinitely many or zero steady states, respectively. 

\begin{figure}[t!]
    \centering
    \includegraphics[width=\linewidth]{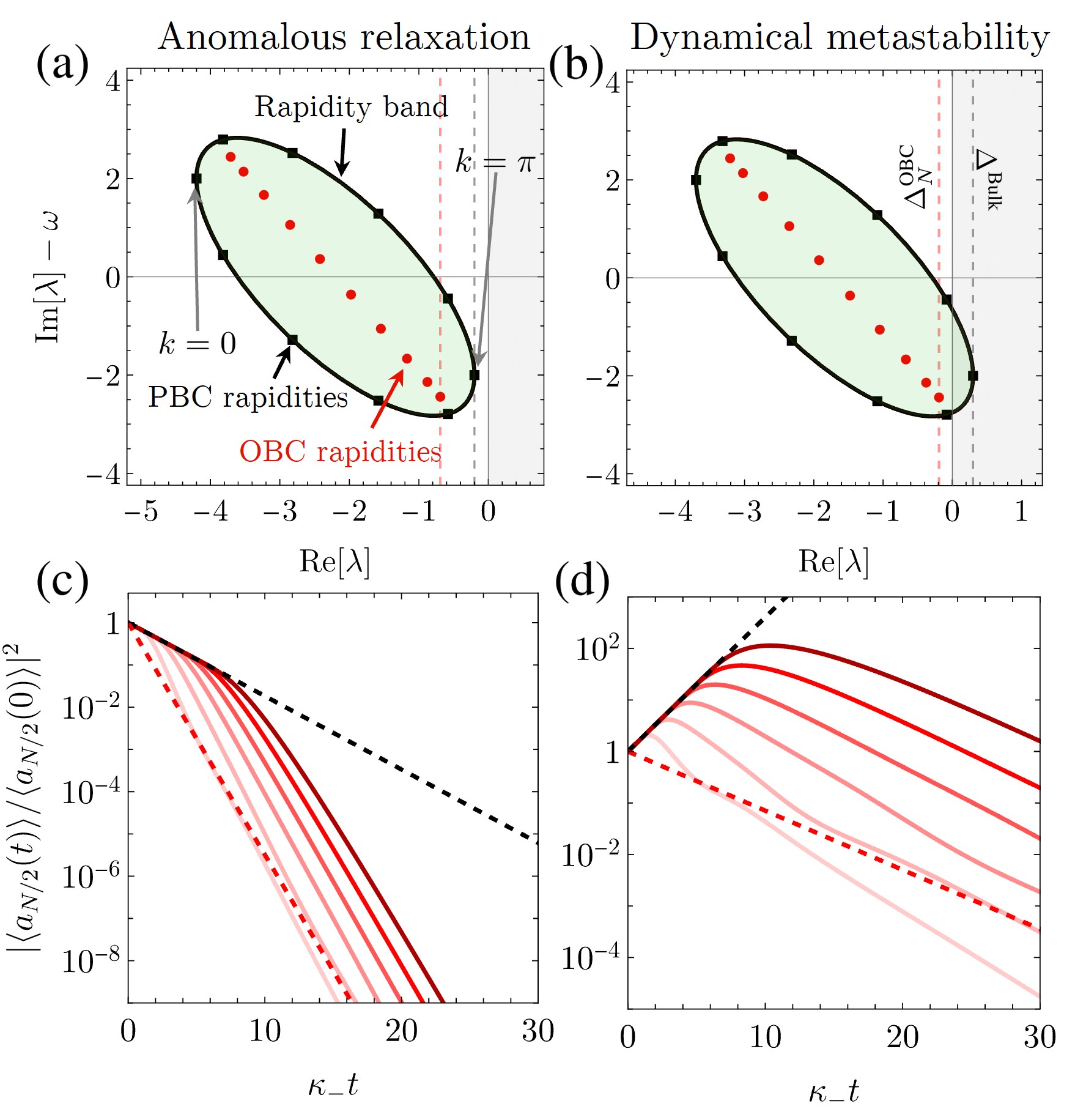}
    \vspace*{-3mm}
\caption{\textbf{(a)-(b)} Example rapidity spectra for the bosonic HN chain, described by Eq.\,\eqref{eq: bkEOM}, in different dynamical stability regimes: $\Delta^\text{OBC}_N < \Delta^\text{Bulk} < 0 $ for (a) and $\Delta^\text{OBC}_N<0< \Delta^\text{Bulk}$ for (b). In both cases, the rapidity band $A(k)$ is depicted in black, with black squares marking the discrete PBC rapidities $A\,(2m\pi/N)$, $m=0,\ldots,N-1$. The red disks correspond to the OBC rapidities in Eq.\,\eqref{eq: HNOBCRaps}. The red (gray) dashed lines indicate the OBC (PBC) stability gaps $\Delta^\text{OBC}_N$ ($\Delta^\text{Bulk}$) with the gray region corresponding to the unstable half-plane $\text{Re}[\lambda]>0$. In (a), both gaps are negative, while in (b), the OBC gap is negative and the PBC gap is positive. In all cases, $J=D=\kappa_-=\Gamma =1$ and $N=10$. The frequency $\omega$ is arbitrary (see the discussion in Sec.\,\ref{sec: frameTDM} on rotating frames). In (a), $\kappa_+=0.8$, while for (b) $\kappa_+=1.3$. 
\textbf{(c)-(d)} Anomalous transient dynamics of the bulk site $|\langle a_{N/2}(t)\rangle|^2$ in each dynamical regime. The red curves correspond to the OBC dynamics, with darker curves featuring larger $N$, with $N$ increasing from 10 to 40 in steps of 6. The slope of the red dashed line is the OBC asymptotic relaxation rate $|\Delta^\text{OBC}_N|$, while the slope of the black dashed line is the PBC relaxation (amplification) rate in the case of (c) ((d)) $|\Delta^\text{Bulk}|$.  Interestingly, the observed amplification timescale turns out to be much longer than the one predicted by Eq.\,\eqref{eq: amptime}, which is only $\kappa_- \tau_\pi \sim 6.9$ for $N=100$. In both (c-d), the system is initialized in a state satisfying $\braket{b_k(0)}  \propto \delta_{k\pi}$.}
\label{fig: HNSpecARDM}
\end{figure}

Spectra of Toeplitz and circulant matrices have been extensively studied \cite{TrefethenPS2005,GBTJPA2017}. In the PBC case, the spectrum can be found by moving to a  plane-wave (eigen)basis, $b_k = N^{-1/2}\sum_j e^{-ikj}a_j$, with $k=2\pi m/N$ and $m$ an integer. These momentum-space modes evolve according to
\begin{align}
\label{eq: bkEOM}
    \dot{b}_k &= \mathcal{L}^\star(b_k) =  A(k)\, b_k, \\
A(k) &= -(\kappa_\text{eff}+i\omega)+J_Re^{-ik} + J_L e^{ik}.
\label{eq: Ak}
\end{align}
Equivalently, we may write $A(k) \equiv -(r_k + i\omega_k)$, with
\begin{align}
    r_k &= \kappa_\text{eff} +2\Gamma \cos(k), 
    \label{eq: rk}
    \\
    \omega_k &= \omega - 2 \,\big[ J\cos(k) + D\sin(k)\big], 
    \label{eq: omk}
\end{align}
representing the decay rate and frequency of mode $k$, respectively. The function $A(k)$ is known as the {\em rapidity band} of the system, while one more generally refers to the eigenvalues $\lambda_n$ of $\mathbf{A}$, with $n=1, \ldots, N$, as {\em rapidities}. For the model described by Eqs.\,\eqref{eq: HNHam} and \eqref{eq: HNdiss}, the rapidity band traces out a skewed ellipse in the complex plane, which is shown along with the discrete finite-size rapidities in Fig.\,\ref{fig: HNSpecARDM}(a-b). The area of this ellipse is precisely $4\pi D\Gamma$ which, as we have noted, is proportional to the degree of nonreciprocity $|J_L|^2-|J_R|^2$. Thus, the rapidity band enclosing a nonzero area is equivalent to the existence of nonreciprocity in this model.     

Commonly associated with nonreciprocity is the {\em non-Hermitian skin-effect} (NHSE), whereby an extensive number of eigenstates populate the boundaries of the system. The NHSE arises in this model as evidenced by the OBC rapidities:
\begin{align}
    \lambda_m = -\kappa_\text{eff} + 2\sqrt{J_LJ_R} \, \cos\lp\frac{m\pi}{N+1}\rp, \quad m=1,\ldots, N.
    \label{eq: HNOBCRaps}
\end{align}
As can be seen in Fig.\,\ref{fig: HNSpecARDM}(a-b), these OBC rapidities do not lie on the bulk rapidity band and thus correspond to exponentially localized edge states. It is also worth noting that the rapidities of the model described by Eqs.\,\eqref{eq: Ak} and \eqref{eq: HNOBCRaps} closely resemble the complex energies of the non-Hermitian HN Hamiltonian, once an appropriate factor of $i$ is  accounted for. This correspondence is not accidental, and can be made precise by observing that the single-particle wavefunction amplitudes evolving under the non-Hermitian HN Hamiltonian obey an equation analogous to that of the bosonic annihilation operators in Eq.\,\eqref{eq: HNEOM}. Specifically, the dynamical matrix in $\mathbf{A}$ in Eq.\,\eqref{eq: bkEOM} is equal to the HN single-particle Hamiltonian, modulo a factor of $i$. This correspondence then explains why the NHSE, a well-known feature of the HN Hamiltonian, arises in the Lindbladian model as well. In fact, since the eigenvectors of the dynamical matrix correspond to eigenoperators of the Heisenberg EOM in Eq.\,\eqref{eq: HNEOM}, the NHSE at the level of the dynamical matrix manifests as a Liouvillian skin-effect at the many-body level \cite{UedaLSkin2021}. 

From the perspective of dynamics, the NHSE presents an interesting scenario. Let $\Delta^\text{OBC}_N$ denote the stability gap of the size-$N$ system under OBCs, and let $\Delta^\text{Bulk} \equiv \sup_k \text{Re}[A(k)]$ denote the bulk stability gap, computed under PBCs. It can be shown from Eq.\,\eqref{eq: HNOBCRaps} that, even in the limit of infinite size $N\to\infty$, the OBC gaps remain {\em strictly} smaller than the bulk gap. Mathematically,
\begin{align}
\label{eq: stabbound}
    \lim_{N\to\infty} \Delta^\text{OBC}_N < \Delta^\text{Bulk}.
\end{align}
In the case where {\em both} $\Delta^\text{OBC}_N$ and $\Delta^\text{Bulk}$ are strictly negative, as shown by the dashed lines in Fig.\,\ref{fig: HNSpecARDM}(a), we say the system is {\em anomalously relaxing}. This regime is characterized by a two-step relaxation process of observable expectation values. In the transient, before a bosonic excitation can ``detect" the boundaries, the finite system decays at a rate set by $|\Delta^\text{Bulk}|$, until eventually relaxing at its ``true'' asymptotic (in time) relaxation rate, set by $|\Delta_N^\text{OBC}|$. While this behavior is reflected in generic initial conditions \cite{FlynnBosoranasPRB2023,UghrelidzeDM2024}, it is most extreme when the system is initialized in a bulk eigenstate (i.e., a plane-wave $b_k$) with decay rate $r_k = \Delta^\text{Bulk}$. For our model, the slowest-relaxing bulk mode is $k=\pi$, whose trajectories are plotted in Fig.\,\ref{fig: HNSpecARDM}(c). The duration of the anomalous relaxation increases as system size increases, thus restoring the physically intuitive principle that a large system behaves as though it were infinite, despite the spectral discontinuity implied by Eq.\,\eqref{eq: stabbound}. 

To further understand this behavior, let $\mathcal{L}_N^\star$ denote the Heisenberg-picture Lindbladian of the $N$-site system under OBCs. Then, we have 
\begin{align}
\label{eq: pwps}
    \dot{b}_k = \mathcal{L}_N^\star(b_k) = A(k)b_k -\frac{1}{\sqrt{N}}\lp J_R a_1 + e^{-ik} J_L a_N\rp.
\end{align}
Equation\,\eqref{eq: pwps} shows that, under OBCs, the momentum-space modes behave like eigenmodes, up to a correction due to boundary terms, that decays to zero as $N^{-1/2}$. This property is not unique to nonreciprocal systems, nor particularly surprising. The anomalous transient behavior arises from its interplay with the spectral disagreement between OBCs and the bulk band.  Indeed, as $N\to\infty$, the transient regime of anomalously slow relaxation increases in duration, as demonstrated in Fig.\,\ref{fig: HNSpecARDM}(c). This observation can be aptly reformulated in the mathematical language of {\em pseudospectra}. The $\epsilon$-pseudospectrum of an operator $\mathbf{X}$ is the set of approximate eigenvalues of $\mathbf{X}$, i.e., the set of $\lambda$ such that there is a unit norm vector $\vec{v}$ (called the pseudoeigenvector) satisfying $\norm{(\mathbf{X}-\lambda)\,\vec{v}}< \epsilon$ for some norm $\norm{\cdot}$ \cite{TrefethenPS2005}. In our case, we see that $\epsilon\sim N^{-1/2}$ and so the anomalously relaxing plane-waves become more and more like exact normal modes as $N\to\infty$. 

Beyond anomalous relaxation, Eq.\,\eqref{eq: stabbound} allows for another interesting dynamical scenario, corresponding to a dynamically unstable  bulk, i.e., $\Delta^\text{Bulk}>0$, while the OBC system remains stable for all system sizes $N$, i.e., $\Delta^\text{OBC}_N <0$. We refer to such systems as  {\em dynamically metastable} \footnote{This is actually the definition of ``Type I'' dynamical metastability. See Ref.\,\cite{UghrelidzeDM2024} for a definition of Type II.}. The word ``metastable" reflects the fact that the system behaves as if it were in a different dynamical stability phase (namely, an unstable one) than it actually is (namely, a stable one) for a transient duration that is long (compared to, e.g., the time scale set by the local loss rate $1/\kappa_-$). The explanation for this behavior is exactly the same as that for anomalous relaxation, except that now there are plane-wave modes with $\text{Re}[A(k)]>0$. In this case, Eq.\,(14.13)   in Ref.\,\cite{TrefethenPS2005} provides a method for estimating the timescale of amplification. Specifically, if $b_k$ is a mode with $\text{Re}[A(k)] = - r_k>0$ (and $\epsilon\sim N^{-1/2}$), then it will amplify on a timescale at least as long as
\begin{align}
\label{eq: amptime}
    \tau_k \sim \frac{1}{|r_k|}\ln\lp N^{1/2}\left|\frac{r_k}{\omega_0}\right|\rp ,
\end{align}
where $\omega_0$ is a system-dependent, but system-size \textit{independent}, frequency scale (e.g., $\omega_0 = \sqrt{J_L^2+J_R^2}$ for the HN QBL), which diverges as $N\to\infty$. We again plot the trajectory of $b_{k=\pi}$ in Fig.\,\ref{fig: HNSpecARDM}(d). Initially, the mode amplifies at a rate set by $\Delta^\text{Bulk}>0$, until ``detecting" the boundaries and ultimately relaxing at a rate equal to or faster than $|\Delta^\text{OBC}_N|$. As expected from Eq.\,\eqref{eq: pwps}, the duration of this transient  amplification regime increases with system size.

\subsection{Topological metastability and edge modes}
\label{sec: TDMbg}

\subsubsection{Definition and properties}

\begin{figure*}[t!]
    \centering
    \includegraphics[width=0.85\linewidth]{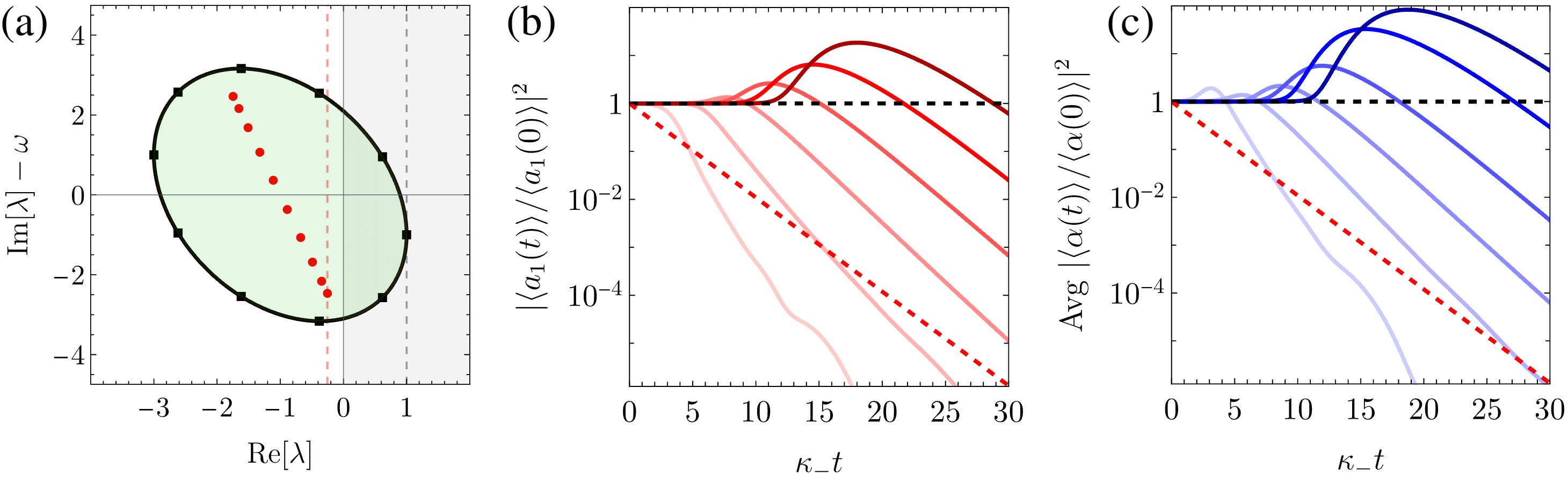}
    \vspace*{-2mm}
    \caption{\textbf{(a)} Example rapidity spectra for the HN QBL in the topologically metastable regime. 
    \textbf{(b)} The evolution of the site-1 intensity $|\braket{a_1}|^2$ for the system described by Eq.\,\eqref{eq: HNQBL} when prepared in the quasi-steady state $\rho_\text{ss}(z=1)$ defined in the main text. \textbf{(c)} The average value of the (normalized) DB expectation value $\braket{\alpha(t)}/\braket{\alpha(0)}$ averaged over 250 random initial conditions. In both (b) and (c), the black dashed line corresponds to the $N\to\infty$ exact trajectory whereby both $\rho_\text{ss}(z=1)$ is steady and $\alpha$ is an exact zero mode. The red dashed line indicates the characteristic OBC decay rate. In both plots, system size increases from $N=10$ to $40$ in steps of $6$ with the curve becoming darker for larger $N$. The model parameters chosen are $J=0.5$, $D=1.5$, $\kappa_-=1$, $\kappa_+=2$, and $\Gamma=1$. The frequency $\omega$ is arbitrary.}
    \label{fig: HNSpecTDM}
\end{figure*}

There is one final scenario to consider. As shown in Fig.\,\ref{fig: HNSpecTDM}(b), it is possible for the rapidity band of a dynamically metastable system to wind around \emph{zero}. When this condition is met (the parameter range under which this happens for the HN Lindbladian is explicitly determined in Eq.\,\eqref{eq: windcond}), two  modes localized on opposite edges from one another emerge: one approximate bosonic zero mode, and one generator of an approximate displacement symmetry. We call this regime {\em topologically metastable} and refer to  the associated edge modes as {\em Dirac bosons} (DBs) due to their tight similarities with the Dirac-fermion edge zero modes familiar from topological insulators \cite{BernevigBook2013}. The definition and basic properties of these bosonic modes may be described as follows.

Consider first the approximate zero mode, described by a bosonic operator $\alpha$ which, without loss of generality, we assume to be localized on the right side of the chain. This scenario implies that i)
\begin{align}
\label{eq: DBalpha}
    \alpha = \sum_{j=1}^N v_j^* a_j,\quad |v_j| \leq e^{-(N-j)/\xi},
\end{align}
for some localization length $\xi$; and ii) under OBCs, the corresponding EOM is
\begin{align}
\label{eq: DBZM}
    \dot{\alpha} = \mathcal{L}_N^\star(\alpha) \sim  e^{-N/\xi}, 
\end{align} 
for all sufficiently large $N$. The localization length $\xi$ roughly characterizes how ``deep" the origin is within the interior of the rapidity band, as expanded upon in Appendix \ref{app: DBlife}. Note that the bosonic requirement, i.e., $[\alpha,\alpha^\dag]=1$, means that the vector of coefficients, $\vec{v} = [v_1,\ldots,v_N]^T$, has unit 2-norm, i.e., 
$\norm{\vec{v}}^2 = \sum_{j}|v_j|^2= 1$. If the right hand-side of Eq.\,\eqref{eq: DBZM} were exactly zero, then $\alpha$ would be a conserved quantity. However, it is exponentially small, but nonzero, and so $\alpha$ is only approximately conserved for finite $N$.

In a similar venue, let $\beta$ be the bosonic operator corresponding to an approximate displacement symmetry generator. This means that i$'$) 
\begin{align}
\label{eq: DBbeta}
    \beta = \sum_{j=1}^N w_j^* a_j,\quad |w_j| \leq e^{-j/\xi},
\end{align}
for the same localization length $\xi$ given in Eq.\,\eqref{eq: DBZM} for $\alpha$; and, ii$'$) if $\mathscr{D}(z) \equiv\exp(z\beta^\dag - z^*\beta)$, $z\in {\mathbb C}$, defines the phase-space displacement operator, the following approximate commutation holds:
\begin{align}
\label{eq: DBSG}
\mathcal{L}_N\big(\mathscr{D}(z)\rho\mathscr{D}^\dag(z)\big) - \mathscr{D}(z)\mathcal{L}_N(\rho)\mathscr{D}^\dag(z) \sim \!|z|e^{-N/\xi},
\end{align}
for all system sizes $N$ and states $\rho$. If the right hand-side were exactly zero, then $\mathscr{D}(z)$ would be a continuous family of exact unitary (weak) symmetries. The approximate case corresponds to an approximate preservation of dynamics under the displacement. 

In our bosonic setting, the mode $\beta$ defines a direction in classical phase space along which the system has an approximate translational symmetry. If, for instance, the bosonic degrees of freedom describe magnons, then $\mathscr{D}(z)$ would map onto a product of rotations $R_{\hat{n}_j}(\vartheta_j)$ for each spin $j$,  with $\vartheta_j\equiv |zw_j|$ and an axis $\hat{n}_j$ in the $xy$-plane, set by $\text{arg}(zw_j)$. An important implication emerges from Eq.\,\eqref{eq: DBSG} when $\rho=\rho_\text{ss}$, the (unique, Gaussian) steady state. Then, $\mathcal{L}(\rho_\text{ss})=0$ and we learn that the family of states $\rho_\text{ss}(z) = \mathscr{D}(z)\rho_\text{ss}\mathscr{D}^\dag(z)$ are {\em quasi-steady states} \footnote{Note that, in general, $\beta$ is not an approximate zero mode, and $\alpha$ does not generate an approximate displacement symmetry. This separation of conserved quantities from symmetry generators arises from a breakdown of Noether's theorem in dissipative quantum systems \cite{FlynnBosoranasPRB2023}.}. In Appendix \ref{app: DBlife}, we derive the following estimate for the lifetime of both $\rho_\text{ss}(z)$ and $\braket{\alpha}$: 
\begin{align}
\label{eq: tauDB}
    \tau_\text{DB} \sim \frac{1}{\Delta^\text{Bulk}}\ln\lp e^{N/\xi}\frac{\Delta^\text{Bulk}}{\omega_0} +1\rp \sim  N/\xi,
\end{align}
where, as in Eq.\,\eqref{eq: amptime}, $\omega_0$ is some system-dependent, but system size-\textit{independent} frequency scale. 
We see then that the lifetime scales linearly with $N$ and that highly localized states live longer. In Fig.\,\ref{fig: HNSpecTDM}, we prepare the HN QBL in a topologically metastable regime and investigate the dynamics of the quasi-steady state $\rho_\text{ss}(z=1)$ in (b) and the expectation value of $\alpha$ over randomized initial conditions in (c). In (b), we see that the boson occupation at site $1$ becomes increasingly more stationary as $N\to\infty$, while in (c), we see that, generically, the expectation value of $\alpha$ remains increasingly more constant as $N\to\infty$. The dual nature of $\alpha$ and $\beta$ manifests clearly in the similarities between (b) and (c). 

As in the case of anomalous relaxation and dynamical metastability, the properties of DBs can be largely explained via the theory of pseudospectra. As elaborated in Appendix \ref{app: DBlife}, the coefficient vectors $\vec{v}$ and $\vec{w}$ which define $\alpha$ and $\beta$ in Eq.\,\eqref{eq: DBalpha} and \eqref{eq: DBbeta}, respectively, satisfy the pseudospectral relationships
  \begin{align*}
      \norm{\mathbf{A}^\dag \vec{v}}<\epsilon_N,\quad \norm{\mathbf{A}\vec{w}}<\epsilon_N,\quad \epsilon_N \sim e^{-N/\xi}. 
  \end{align*}
That is, they are $\epsilon_N$-pseudoeigenvectors of $\mathbf{A}^\dag$ and $\mathbf{A}$, respectively, corresponding to pseudoeigenvalue $0$. Approximate conservation of $\alpha$ follows from this property of $\vec{v}$ in addition to the more general fact that 
  \begin{align*}
      \frac{d}{dt}\lp \vec{u}^{\,\dag} \phi\rp = \mathcal{L}^\star(\vec{u}^\dag\phi) = \lp\mathbf{A}\vec{u}\rp^\dag\phi,
  \end{align*}
for any coefficient vector $\vec{u}$.
Meanwhile, the pseudoeigenvector nature of $\vec{w}$, in conjunction with Eq.\,\eqref{eq: phiEOM}, implies that \textit{any} state with $\braket{\phi(0)} = \vec{w}$ will have a long-lived vector of first moments $\braket{\phi(t)}$. One may verify that, in fact, $\tr[\phi \, \rho_\text{ss}(z)] = z\,\vec{w}$, $\forall z$. However, $\rho_\text{ss}(z)$ additionally has all higher-order moments nearly conserved, resulting in an overall quasi-stationarity of these (Gaussian) states. 

The pseudospectral and topological origin of DBs additionally provides them with a notable degree of robustness. It is a general fact of pseudospectra that small perturbations can only result in a proportionally small change in the pseudospectra \cite{TrefethenPS2005}. This is to be contrasted with the spectra of, e.g., strongly non-reciprocal non-Hermitian matrices. For such systems, small perturbations can result in dramatic nonlinear changes in the spectrum. This ensures that $\epsilon$-pseudoeigenvalues generically survive the introduction of weak disorder, with, at worst, a proportionally weak modification made to the accuracy $\epsilon$. Moreover, circulant matrices, which describe models like the HN QBL under PBCs, are normal matrices. Consequentially, their spectra respond linearly to weak perturbations. Thus, the winding topology of the rapidity bands, which ensure the existence of edge-localized DBs, is  generically robust to perturbations.

Finally, we note that, throughout this paper, the numerical determination of the DB pseudoeigenvectors $\vec{v}$ and $\vec{w}$ in Eqs.\,\eqref{eq: DBalpha} and \eqref{eq: DBbeta}, respectively, was performed by computing the right singular vector (in the sense of the singular value decomposition) corresponding to the minimal singular value of the appropriate matrix  (see also Appendix B1 of Ref.\,\cite{FlynnBosoranasPRB2023}).

\subsubsection{U(1) symmetry and frame-dependent topological metastability: Generalized Dirac bosons}
\label{sec: frameTDM}

As noted in Sec.\,\ref{sec: DM}, a U(1)-symmetric QBL $\mathcal{L}$ is invariant under a global redefinition of the bosonic modes, via a transformation $a_j\mapsto e^{i\theta}a_j$.  This invariance also makes it possible to describe the consequences of moving to a rotating frame in an especially simple way.
Specifically, given any frequency $\Omega\in\mathbb{R}$, we can define rotating modes $a'_j(t) \equiv e^{i\Omega t}a_j(t)$ or, equivalently, the rotating array $\phi'(t) \equiv  e^{i\Omega t} \phi(t)$.  In this frame, the modes obey a slightly modified EOM:
\begin{align}
    \dot{\phi}'(t) = \lp  \mathbf{A} - i \Omega\rp\phi'(t) = \mathbf{A}'\,\phi'(t).
\end{align}
That is, we may think of them as evolving under the influence of the rotating-frame dynamical matrix $\mathbf{A}' \equiv \mathbf{A}-i\Omega \mathds{1}_{N}$, with $\mathds{1}_N$ the $N\times N$ identity matrix. Given only a constant shift to the dynamical matrix, the rapidities transform trivially,   $\lambda_n\mapsto \lambda'_n = \lambda_n-i\Omega $, which simply translates them vertically in the complex plane by $\Omega$. In contrast, if U(1) were to be broken, the annihilation operators would couple to both creation and annihilation operators in a more complicated time-dependent fashion. 

Importantly, moving to a rotating frame does not change the dynamical stability properties of the system, hence it does not change whether the system is anomalously relaxing or dynamically metastable. On the other hand, a vertical translation of the rapidity band can change whether or not it winds about the origin. In light of this, it becomes natural to define a frame-dependent notion of topological metastability. If the rapidity band $A(k)$ of a stable (under OBCs) system winds around some point $i\Omega_0$ on the imaginary axis, then the corresponding rapidity band in a frame that rotates at frequency $\Omega=\Omega_0$, i.e., $\tilde{A}(k) = A(k) - i\Omega_0$, winds around zero. Thus, in this frame, the system is topologically metastable, and the aforementioned DBs emerge. For the purposes of this work, we will still consider these {\em generalized, rotating-frame DBs} of interest. Since our primary application of this formalism will be to ferromagnetic heterostructures (which conventionally maintain a large, nonzero energy gap), the realization of long-lived, nonzero frequency edge modes (as opposed to their zero frequency counterparts) is both physically motivated and, in fact, more natural.

The properties of these rotating-frame DBs become modified upon returning to the lab frame. For instance, the mode $\alpha$, which is long-lived and stationary in the rotating frame, will rotate at a {\em finite} frequency $\Omega$ in the lab frame. Mathematically, $\alpha(t) \approx e^{-i\Omega t}\alpha(0)$. Nonetheless, its lifetime (which is tied to the still-vanishing real part of the pseudoeigenvalue), will still increase unbounded with system size. The approximate displacement symmetry, on the other hand, will now become an instantaneous, ``snapshot'' approximate symmetry: approximate invariance of $\mathcal{L}$ under $\mathscr{D}(z)$ in the rotating frame is equivalent to approximate invariance of $\mathcal{L}$ under the time-dependent displacement $\mathscr{D}(ze^{i\Omega_0t})$ in the lab frame. In turn, the rotating-frame quasi-steady states $\tilde{\rho}_\text{ss}(z)$ will coincide with Gaussian states $\rho_\text{ss}(z e^{i\Omega_0t})$, whose vector of first moments satisfies $\braket{\phi(t)} \approx e^{i\Omega_0t}\braket{\phi(0)}$. 

We note that our notion of frame-dependent topological metastability is closely related to the frequency-dependent winding numbers commonly employed in the study of topological amplifiers \cite{PorrasTopoAmp2019,WanjuraTopAmp2020,WanjuraTopAmp2021,PorrasIO2021}. In such works, it is shown that if the $\Omega$-dependent dynamical matrix has a nontrivial winding about the origin, then the system will allow for exponential end-to-end {\em steady-state} amplification of signals at frequency $\Omega$. While the topological condition for this feature is equivalent to the one we use to identify frame-dependent topological metastability, our focus, crucially, is instead on the {\em transient} dynamics of edge states arising in this regime. We direct the interested reader to Ref.\,\cite{FlynnBosoranasPRB2023}, where the relevant connections between this phenomena, dynamical metastability, and the emergence of topologically mandated, long-lived, edge modes are spelled out in more detail.

\section{Dynamical metastability in an interacting spin Lindbladian}
\label{sec: Lind}

\subsection{A spin model and its equations of motion}

\begin{figure}[t!]
    \centering    \includegraphics[width=\linewidth]{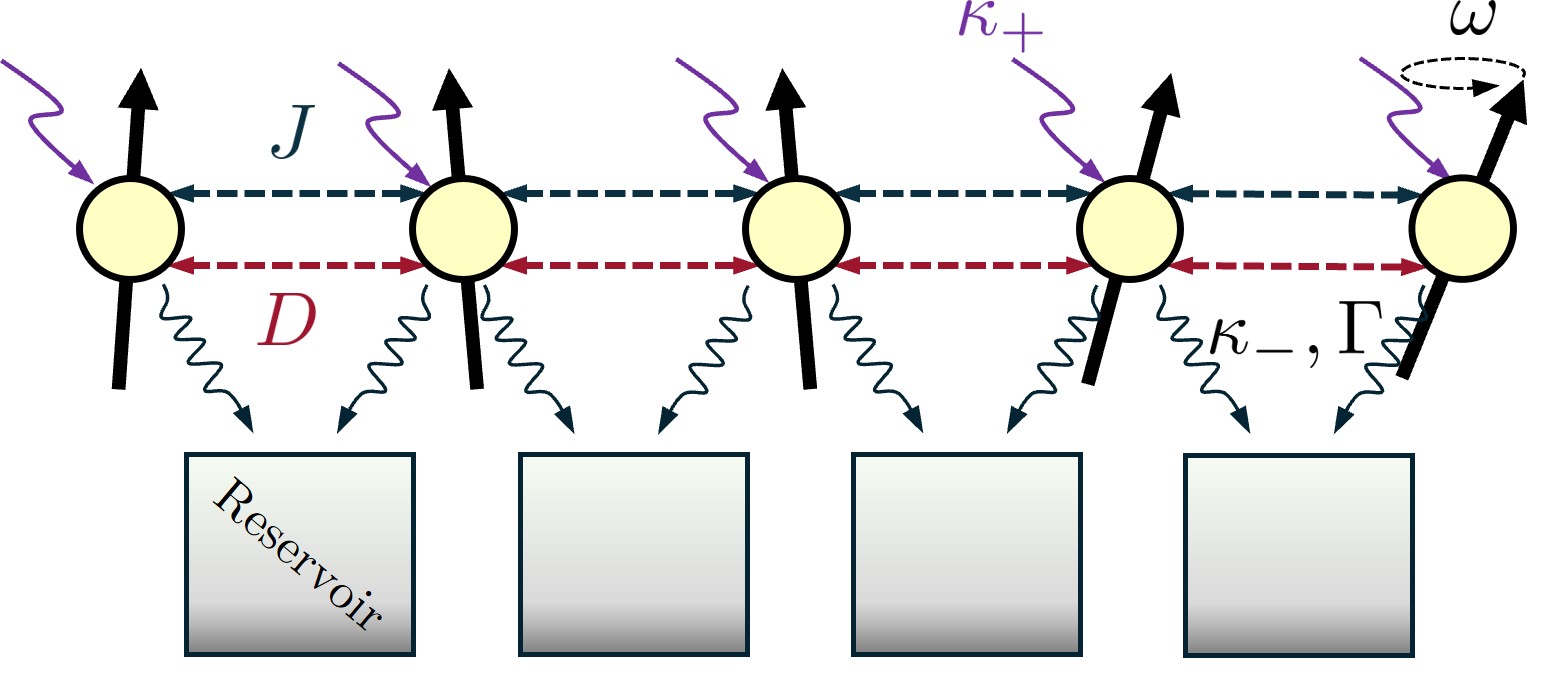}
    \caption{A schematic representation of the spin Lindbladian described by Eqs.\,\eqref{eq: spinLind}. Each spin precesses with a characteristic frequency $\omega$ about the $z$-axis. Nearest neighbors are coupled via real and imaginary exchange interactions of strength $J$ and $D$, respectively. Each spin dissipates into nearest-neighbor mutual reservoirs independently at a rate $\kappa_-$, with an additional correlated decay rate $\Gamma$. Additionally, each spin is pumped away from equilibrium at a rate $\kappa_+$.}
    \label{fig: Lindcartoon}
\end{figure}

We now introduce the first of the two concrete models that will enable us to explore how the framework of dynamical metastability introduced in Sec.\,II interplays with the effects of interactions -- which necessarily entails moving beyond quadratic Markovian dynamics. We consider $N$ spin-$s$ degrees of freedom, with Hamiltonian and dissipator respectively given by 
\begin{subequations}
\label{eq: spinLind}
\begin{align}
    \!\!H &= -\!\sum_j \omega S^z_j - \frac{1}{2}\sum_{j} \big[ (J+iD)S^-_j S^+_{j+1}+ \text{H.c.}\big], \label{eq: spinHam}
    \\
    \!\!\mathcal{D} &= \!\sum_{j} \lp\kappa_- \mathcal{D}[S^+_j] + \Gamma \mathcal{D}[S^+_j + S^+_{j+1}]+ \kappa_+ \mathcal{D}[S^-_j]\rp .
\label{eq: spinD}
\end{align}
\end{subequations}
Here, $\omega> 0$ is the spatially uniform Zeeman frequency, $J>0$ is the ferromagnetic exchange coupling, and $D$ is an antisymmetric exchange coupling. Meanwhile, $\kappa_-$, $\Gamma$, and $\kappa_+$ are the (nonnegative) spin damping, correlated decay, and spin pumping rates, respectively. We depict the model in Fig.\,\ref{fig: Lindcartoon}.

By construction, the spin Lindbladian resulting from the above $H$ and $\mathcal {D}$ is invariant under rotations about the $z$ axis, leading to invariance under the global U(1) transformation $S_j^z\mapsto S_j^z$ and $S_j^\pm \mapsto e^{\pm i\theta}S_j^\pm$. As we will see, this U(1) symmetry maps onto the same bosonic U(1) symmetry seen in the HN QBL of the preceding section in the magnon regime. Further, one may show that $S_j^2$ is conserved for all sites $j$. Notably, this model with $\kappa_+=0$ was considered in Ref.\,\cite{XinPaper_2025} as a case study for the NHSE in spin systems. For $\omega > 0$ and $D< J/\sqrt{3}$, the classical ground state of $H$ in Eq.\,\eqref{eq: spinHam} is the fully polarized state in the +$z$-direction \cite{KimHKMSpin2016}. The latter is also a steady state of the full spin Lindbladian $\mathcal{L}$ when $\kappa_+=0$. We expect that a nonzero spin pumping rate will turn the steady state into a mixed thermal state, provided that $\kappa_+$ is sufficiently small. For very large $\kappa_+$, however, spin pumping can drive the system toward a fully polarized state along  the $-z$-direction. 

The Heisenberg EOM provide insight into the competition between the various mechanisms the Lindbladian dynamics dictated by Eqs.\,\eqref{eq: spinLind} comprise:
\begin{widetext}
\begin{equation}
\left \{ \begin{array}{ll}
   \dot{S}_j^z &\!=(\kappa_-+2\Gamma) S_j^- S_j^+-\kappa_+S_j^+ S_j^-- \tfrac{1}{2}\lp J_R S_{j}^-S_{j-1}^+ + J_L S_j^- S_{j+1}^+  + \text{H.c.}\rp , \\
   \\
\dot{S}_j^+ &\!= -i\omega S_j^+ -\lp \kappa_- + 2\Gamma\rp S_j^z S_j^+ +\kappa_+ S_j^+ S_j^z +J_R S_j^z S_{j-1}^++J_LS_j^z S_{j+1}^+,
\end{array} \right. 
\label{eq: LindS} 
\end{equation}
\end{widetext}
where $J_R$ and $J_L$ are formally defined as below Eq.\,\eqref{eq: HNEOM}. We note that Ref.\,\cite{BeggNRCrit2024} identified similar effective nonreciprocal spin interactions in an XXZ chain with symmetric exchange, correlated Markovian dissipation, and dissipative gauge fluxes. In contrast, nonreciprocity in our model arises from the interplay between \textit{antisymmetric} exchange and nonlocal damping. Moreover, our framework independently incorporates both spin decay ($L \propto S^+$) and pumping ($L\propto S^-$), allowing us to sustain explicitly nonequilibrium steady states.

Let us now invoke a semiclassical  approximation $\langle S^\mu_j S^\nu_\ell\rangle \approx s_j^\mu s_\ell^\nu$, where $s_j^\mu = \langle S_j^\mu \rangle$. This approximation is valid in the large-$s$ limit, in which case Eqs.\,\eqref{eq: LindS} reduce to a simplified set of EOM for the spin expectation values:
\begin{equation}
\label{eq: LindSC}
\!\!\!\!\left\{ \begin{array}{ll}
   \dot{s}_j^z &\!\!=\kappa_\text{eff} |s_j^+|^2- \text{Re}\lb J_R s_{j}^-s_{j-1}^+ +J_L s_j^- s_{j+1}^+  \rb ,  \\
   \\
\dot{s}_j^+ &\!\!= -(\kappa_\text{eff} s_j^z+i\omega) s_j^+  + \lp J_R  s_{j-1}^++J_L s_{j+1}^+\rp  s_j^z,  
\end{array}\right. 
\end{equation}
where again $\kappa_\text{eff}$ is defined as below \,Eq.\,\eqref{eq: HNEOM}. We can thus immediately identify a class of semiclassical equilibria, characterized by $s_j^+ = 0$ and $s_j^z = \sigma_j s$, with $\sigma_j \in \{-1,1\}$. It is important to remark that here we use the term ``equilibria" to exclusively refer to dynamical equilibria, i.e., stationary solutions of Eqs.\,\eqref{eq: LindSC} that do not necessarily minimize the classical energy of Eq.\,\eqref{eq: spinHam}. More specifically, our main focus will be {\em fully translationally invariant equilibria}, namely, spin configurations where $\sigma_j$ is independent of $j$. In this case, all spins either all point up or down. With reference to the external magnetic field, in the following we call the equilibrium with all $\sigma_j=1$ $(-1)$  ``aligned" (``antialigned"), respectively.

\subsection{Magnon Lindblad dynamics}
\label{sec: MagLind}

The semiclassical EOM in Eq.\,\,\eqref{eq: LindSC} bear a striking resemblance to the Heisenberg equations obtained in Eq.\,\eqref{eq: HNEOM} for the HN QBL. This correspondence can be made precise by performing a large-spin Holstein-Primakoff  transformation \cite{HP}  at the level of the Lindbladian. Namely, we expand about $\braket{S_j^z} = +s$ and take $S_j^z = s-a_j^\dag a_j$ and $S_j^+\approx \sqrt{2s} \,a_j$. The magnon QBL describing the fluctuations about the aligned equilibrium resulting from Eqs.\,\eqref{eq: spinLind} is then $\mathcal{L}_+(\rho) = -i[H_+,\rho] + \mathcal{D}_+(\rho)$, where  the $+$ subscript refers to the fact we are expanding about the aligned state $\sigma_j=+1$, and the quadratic Hamiltonian $H_+$ and dissipator $\mathcal{D}_{+}$ are, respectively, given by 
\begin{align}
    H_+ & = \sum_j \omega (a_j^\dag a_j-s) -s\sum_{j}\lb (J+iD) a_j^\dag a_{j+1} + \text{H.c.}\rb,  \notag \\
\mathcal{D}_+& = 2s\sum_{j}\lp\kappa_-\mathcal{D}[a_j] + \Gamma\mathcal{D}[a_j + a_{j+1}]+\kappa_+\mathcal{D}[a_j^\dag]\rp.
\label{eq: SC+}
\end{align}
Note that, in the dilute magnon limit where the magnon density is low, $\langle a^\dag_j a_j\rangle \ll s$,  the QBL ${\cal L}_+$ is identical to the one in Eqs.\,\eqref{eq: HNHam}-\eqref{eq: HNdiss}, up to multiplicative factors and a constant energy shift, neither of which play a relevant role in our analysis of dynamics. Thus, the Heisenberg EOM that result from $\mathcal{L}_+^\star$ in Eq.\,\eqref{eq: SC+}, and govern the magnon fluctuations around the aligned equilibrium, are {\em precisely} the ones in Eq.\,\eqref{eq: HNEOM}. By virtue of this correspondence, we conclude that anomalous relaxation, dynamical metastability, and topological metastability are {\em all} possible at the level of magnon dynamics in this system. In particular, when parameters are chosen appropriately, the anomalous trajectories plotted in Fig.\,\ref{fig: HNSpecARDM}(c-d) and Fig.\,\ref{fig: HNSpecTDM}(b) correspond to the population dynamics of magnons at sites $j=N/2$ and  $j=1$, respectively. 

\begin{figure*}[t!]
    \centering
    \includegraphics[width=.95\linewidth]{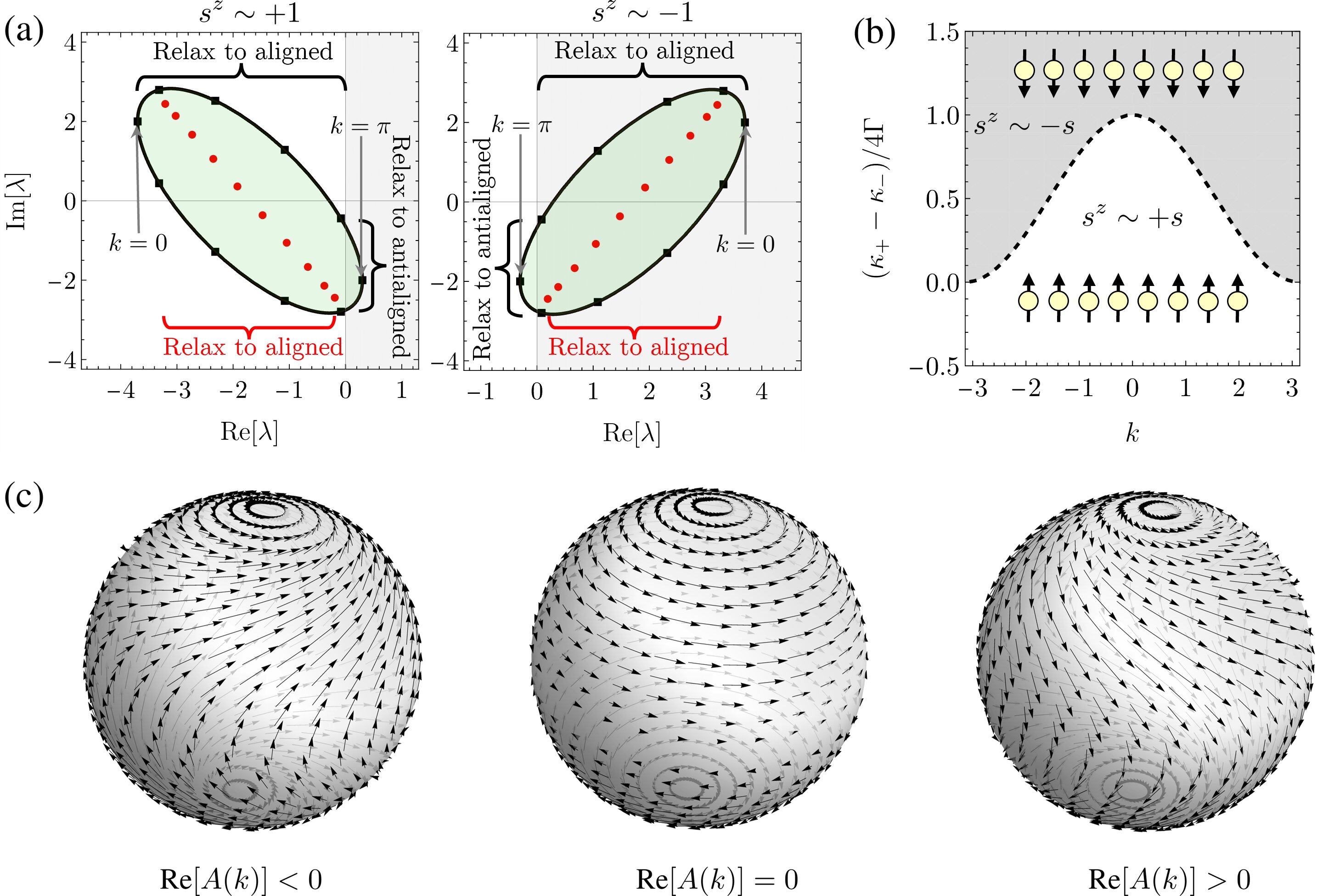}
    \vspace*{-1mm}
    \caption{
\textbf{(a)} 
The magnon rapidities for the spin Lindbladian models in Eqs.\,\eqref{eq: SC+} and \eqref{eq: SC-}, resulting from a Holstein-Primakoff expansion about the aligned spin state (${\cal L}_+$, left) and the antialigned spin state (${\cal L}_-$, right). For the aligned case, the OBC rapidities (red) are all strictly negative in real part, hence arbitrary nearby initial conditions converge to the aligned state. Meanwhile, under PBCs, convergence of bulk spin waves depends on whether $\text{Re}[A(k)]$ is positive or negative. In the antialigned case, the OBC rapidities are all strictly negative in real part, hence all nearby initial conditions diverge away. Meanwhile, under PBCs, those modes that previously were unstable around the aligned state, are now stable around the antialigned state, and vice-versa. The PBC rapidities of the modes $k=0$ and $k=\pi$ are indicated in each case.  \textbf{(b)} The steady-state phase diagram for spin waves under PBCs, resulting from Eq.\,\eqref{eq: ReAk}. As soon as the spin pumping $\kappa_+$ exceeds the spin damping $\kappa_-$, modes with wavevectors $k\sim \pm\pi$ begin to relax into the antialigned state. Once $\kappa_+$ exceeds $\kappa_-+4\Gamma$ (i.e., the decay rate of the $k=0$ mode), all spin waves relax to the antialigned state.
\textbf{(c)} The vector fields specified by Eq.\,\eqref{eq: bulkeom} defined on the sphere of radius $s$ for the three cases $\text{Re}[A(k)]<0$ , $\text{Re}[A(k)]=0$, and $\text{Re}[A(k)]>0$. In the first case, the north pole $s^z = +1$ is attractive, while in the last case, the south pole $s^z=-1$ is attractive. When $\text{Re}[A(k)]=0$, each circle of latitude at fixed $s^z$ is a closed trajectory with frequency dictated by Eq.\,\eqref{eq: closedlat}. The chirality of each trajectory depends on the hemisphere in which it is located, with the equator $s^z = 0$ being a circle of fixed points.
}
\label{fig: BulkSS}
\end{figure*}

\begin{figure*}[t!]
    \centering
    \includegraphics[width=0.8\linewidth]{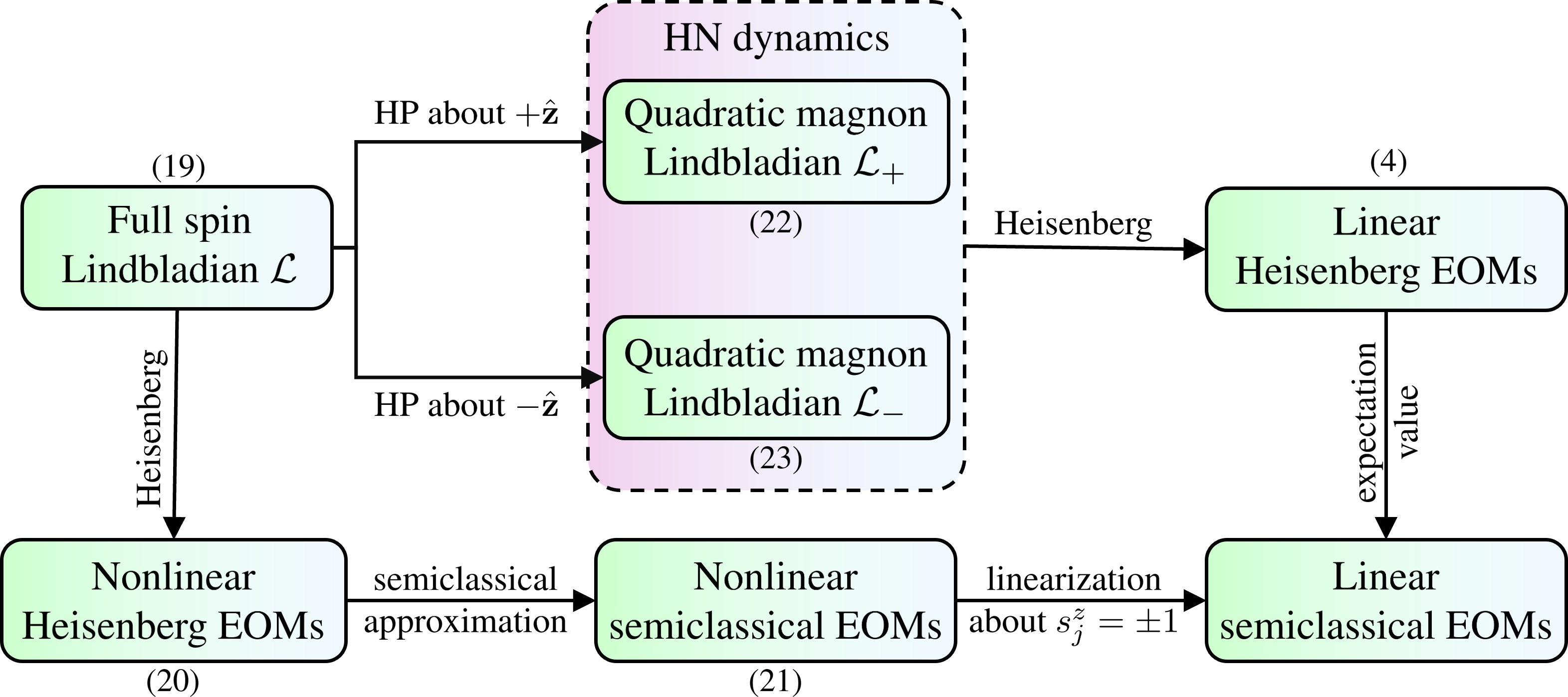}
    \caption{The hierarchy of EOM descendent from the full spin Lindbladian specified by Eqs.\,\eqref{eq: spinLind}. When possible, an equation number for the relevant equation is appended.}
    \label{fig: EOMHierarchy}
\end{figure*}

Whenever the magnonic QBL is dynamically stable (i.e., when the stability gap of $\mathcal{L}_+$ satisfies $\Delta<0$), the aligned state is  (at least locally) attractive, in the large-$s$ semiclassical limit. However, when the stability threshold is crossed, small perturbations to the initial aligned condition may cause the trajectory to be repelled away from this equilibrium and asymptotically converge to another equilibrium instead \footnote{Nonstationary attractors are also possible.}. For instance, if we expand Eqs.\,\eqref{eq: spinLind} about the $-z$ polarized state, $\braket{S_j^z} = -s$, by taking $S_j^z = -s+a_j^\dag a_j$ and $S_j^+\approx \sqrt{2s}\,a_j^\dag$, we find a new magnonic QBL, $\mathcal{L}_-(\rho) = -i[H_-,\rho] + \mathcal{D}_-(\rho)$, with Hamiltonian and dissipator being respectively given by
\begin{align}
    H_- & = \sum_j\omega(s-  a_j^\dag a_j) -s\sum_{j}\lb (J+iD) a_j a_{j+1}^\dag + \text{H.c.}\rb, \notag \\
\mathcal{D}_- & = 2s\sum_{j}\lp\kappa_-\mathcal{D}[a_j^\dag] + \Gamma\mathcal{D}[a_j^\dag + a_{j+1}^\dag]+\kappa_+\mathcal{D}[a_j]\rp.
\label{eq: SC-}
\end{align}
The magnon fluctuations $S^+_j \sim a^\dag_j$ about the antialigned equilibrium state now follow Heisenberg EOM that are determined by $\mathcal{L}_-^\star$ in Eq.\,\eqref{eq: SC-} and are still identical to those in Eq.\,\eqref{eq: HNEOM}, except the signs of $\kappa_\text{eff}$, $J_L$, and $J_R$ reversed. This results in an inversion of the dynamical stability phase diagram, while leaving the directionality of the nonreciprocal dynamics unaffected. We summarize these results in Fig.\,\ref{fig: BulkSS}(a), which contrasts the rapidity spectra about each of the $\pm z$ equilibria. We conjecture that local stability of the aligned (antialigned) state -- or, equivalently, stability of $\mathcal{L}_+$ $(\mathcal{L}_-)$ -- implies global attractivity under the semiclassical spin dynamics of Eq.\,\eqref{eq: LindSC}; that is, all initial configurations asymptotically relax to this state whenever nearby ones do. In Fig.\,\ref{fig: EOMHierarchy} we summarize the hierarchy of EOM descendant from the spin Lindbladian in Eqs.\,\eqref{eq: spinLind}, as discussed up until this point.

\subsection{Spin-wave steady-state phase diagram}
\label{sec: SWSSPDLind}

As a first step in analyzing the full nonlinear semiclassical dynamics,  we impose PBCs and consider the following translation-invariant spin-wave Ansatz for Eq.\,\eqref{eq: LindSC}:
\begin{align}
\label{eq: swansatz}
  s_j^+(t) = e^{ijk}s_k^+(t),\quad   s_j^z(t) = s^z(t),
\end{align}
where $k$ is a wavenumber in the discrete Brillouin zone. Such states transform naturally under the discrete translational symmetry of the system. The $j$-independent amplitudes $s^z$ and $s_k^+$ then obey the coupled nonlinear differential equations:
\begin{equation}
\left\{ 
\begin{array}{ll}
    \dot{s}^z &\!= -\frac{1}{s}\text{Re}[A(k)]|s_k^+|^2, \\
    \\
    \dot{s}^+_k &\!= -i\omega s_k^+ + \frac{1}{s}\tilde{A}(k)s^z s^+_k , 
    \end{array} \right.
    \label{eq: bulkeom}
\end{equation}
where $A(k)$ is the same rapidity band familiar from the HN QBL, Eq.\,\eqref{eq: Ak}, and we have denoted $\tilde{A}(k) \equiv A(k)\big|_{\omega=0}$ (thus, by construction, $\text{Re}[A(k)] = \text{Re}[\tilde{A}(k)]$). The fact that $k$-modes do not couple to modes of other momentum $q\neq k$, despite the nonlinearities, is a general consequence of U(1) symmetry, see Appendix \ref{app: nonlinsym}. The system in Eq.\,\eqref{eq: bulkeom}, can be solved analytically, which we do in Appendix \ref{app: ansolLind}. We identify two equilibria, corresponding to $s^+_k=0$ with $s^z = \pm s$. Stability analysis reveals that $s^z=s$ (respectively, $-s$) is globally stable if $\text{Re}[A(k)]<0$ (respectively, $\text{Re}[A(k)]>0$), where
\begin{align}
\label{eq: ReAk}
    \text{Re}[A(k)] = \kappa_+ -\kappa_- -4\Gamma \cos^2(k/2).
\end{align} 
The $k$-dependence in Eq.\,\eqref{eq: ReAk} arises from the $k$-dependence of the magnon decay rates, as determined by Eq.\,\eqref{eq: rk}. The resulting stability phase diagram is depicted in Fig.\,\ref{fig: BulkSS}(b). Fixing a wavevector $k$ and increasing $\kappa_+$ drives the system through a bifurcation, whereby the two equilibria swap stability. We see that the $k=\pi$ mode bifurcates as soon as the local spin pumping exceeds the spin damping, while the fastest decaying mode $(k=0)$ relaxes to the aligned state until $\kappa_+$ exceeds $\kappa_-+4\Gamma$. 

It is instructive to rewrite the equation for $s^+_k(t)$ in Eqs.\,\eqref{eq: bulkeom} in the form $\dot{s}_k^+ = - [ R_k(t)+i \Omega_k(t)]\, s_k^+,$ in terms of a state-dependent instantaneous decay rate $R_k (t)\equiv -\text{Re}[{A}(k)]s^z(t)/s$ and a state-dependent instantaneous  frequency $\Omega_k(t) \equiv \omega -\text{Im}[\tilde{A}(k)] s^z(t)$. In particular, if $\text{Re}[A(k)]=0$, one has $R_k(t) \equiv 0$, whereby $s^z(t) = \text{const} \equiv s^z_0$.  Thus, in this special case, 
\begin{align}
\label{eq: closedlat}
s^+_k(t) = e^{-i(\omega-\text{Im}[\tilde{A}(k)]s^z_0)t} s_k^+(0) \equiv e^{-i \Omega_k t} s_k^+ (0), 
\end{align}
meaning that the spin wave Eq.\,\eqref{eq: swansatz} with wavevector $k$ precesses at a constant frequency with a fixed $z$-component. These oscillatory, cyclic solutions facilitate the swapping of the stability of the two equilibria on the top and bottom of the unit sphere, which is illustrated in Fig.\,\ref{fig: BulkSS}(c). Two remarks are in order. (i) The oscillatory solutions emerging at the bifurcation do \textit{not} constitute limit cycle solutions. By definition, limit cycles locally attract or repel at least one trajectory. Arbitrary perturbations on top of these solutions, in contrast, neither grow nor decay. (ii) \textit{A priori}, there is no physical reason why the two equilibria could not be simultaneously stable or unstable for a given spin-wave initial condition. The exclusion of these two possibilities is entirely due to the structure of Eqs.\,\eqref{eq: bulkeom}, which forces the sign of $R_k(t)$ to be intrinsically tied to the sign of $s^z$. Thus, the EOM cannot support two attractive or repulsive antipodal points, as their decay (or amplification) rates would always have opposite sign.  

\subsection{Anomalous transient dynamics}
\label{sec: anomLind}

Let us now restrict ourselves to the parameter regime where the aligned spin configuration is stable and (at least) locally attractive. The NHSE present at the magnon level guarantees that the inequality in Eq.\,\eqref{eq: stabbound} is satisfied by the magnon stability gaps. Thus, at least at the (quadratic) magnon level, anomalous transient dynamics are expected for the QBLs ${\cal L}_+$ and ${\cal L}_-$ in Eqs.\,\eqref{eq: SC+} and \eqref{eq: SC-}. It is of interest then to see how, if at all, these anomalous magnon dynamics imprint onto the full (semiclassical) nonlinear dynamics of the system, given in Eq.\,\eqref{eq: LindSC}. Throughout this section, for convenience we will normalize the spin variables $s^\mu_j \mapsto s^\mu_j/s$ so that each real component of spin varies from $-1$ to $1$. We will further move to a frame rotating at frequency $\omega$, i.e., $s^+_j(t) \mapsto e^{i\omega t}s^+_j(t)$, which allows us to effectively take $\omega=0$ in all subsequent analysis.

\subsubsection{Anomalous relaxation in the presence of nonlinearities}
\label{sec: LindAR}

We first proceed to explore the parameter regime corresponding to the magnon rapidities shown in Fig.\,\ref{fig: HNSpecARDM}(a), in which the linearized magnon dynamics exhibits anomalous relaxation. Specifically, we set $\kappa_- >\kappa_+$, ensuring that the aligned state remains stable regardless of BCs. We assume an even system size $N$ and initialize the spins in the slowest-relaxing bulk spin-wave mode, namely, $k=\pi$:
\begin{align}
\label{eq: pimode}
   s_j^+(0) = (-1)^j s^+(0), \quad s^z_j(0) = s^z(0),
\end{align}
subject to $(s^z(0))^2+ |s^+(0)|^2 = 1$. As the initial value \( s^z(0) \) may be freely varied within \([-1, 1]\), we can control the proximity to either the aligned or antialigned equilibrium configurations. The analysis in Sec.\,\ref{sub: ARDM} shows that, under PBCs, this mode would relax to the aligned state with a characteristic relaxation rate $|\Delta^\text{Bulk}| = |\text{Re}[A(\pi)]| = \kappa_- - \kappa_+$. On the other hand, near equilibrium, the finite-size OBC system is predicted to asymptotically relax to the aligned state no slower than $|\Delta_N^\text{OBC}|$, which is, by Eq.\,\eqref{eq: stabbound}, strictly faster than $\kappa_--\kappa_+$.

\begin{figure}[t!]
    \centering
    \includegraphics[width=\linewidth]{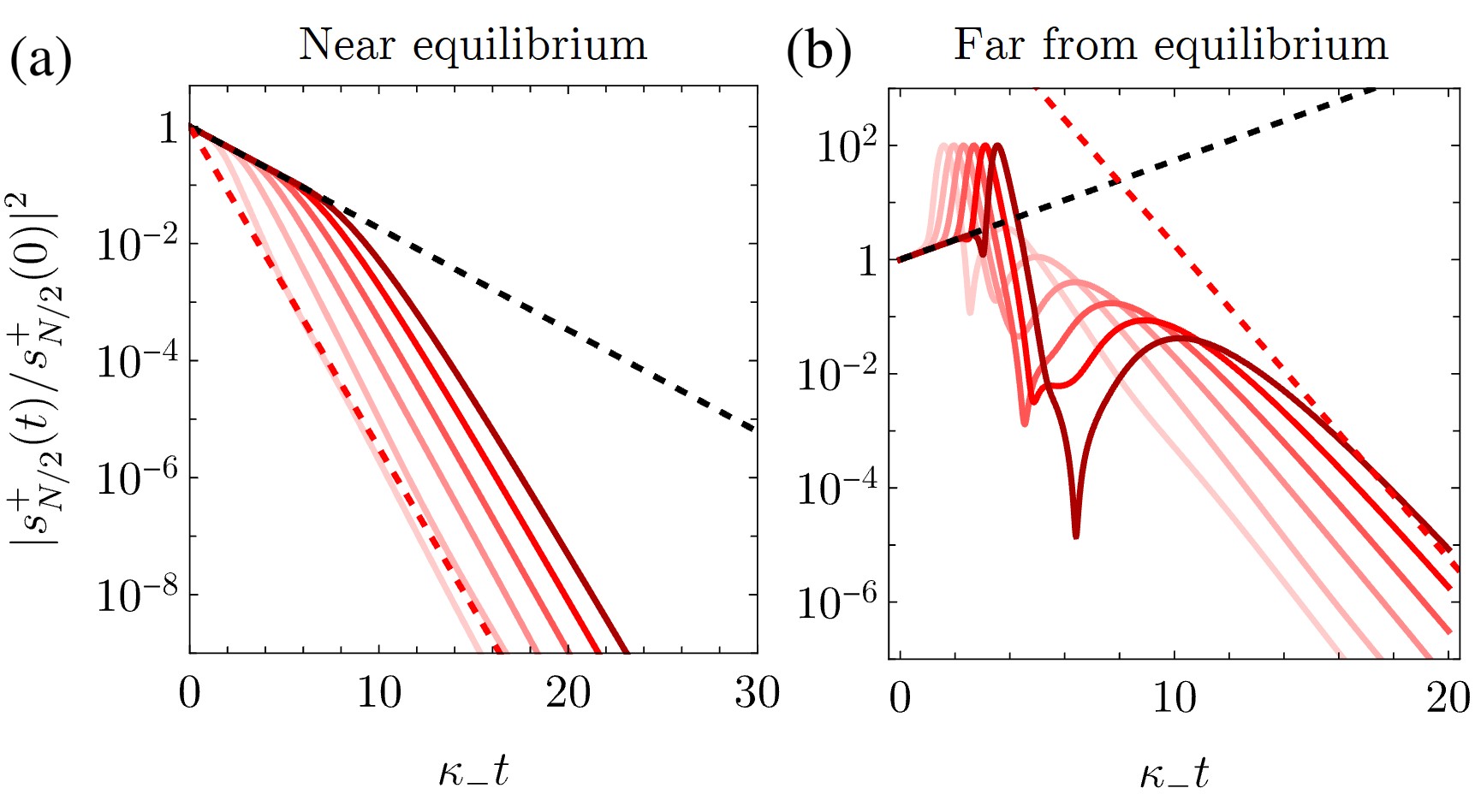}
    \vspace*{-2mm}
    \caption{Semiclassical nonlinear spin dynamics from Eq.\,\eqref{eq: LindSC} within the magnon anomalously relaxing regime. The dynamics of a bulk spin at site $j=N/2$ for an initial condition {\bf (a)} near equilibrium $(s^z(0) = \sqrt{1-0.1^2}$) and {\bf (b)} far from equilibrium $(s^z(0) = -\sqrt{1-0.1^2}$) are depicted. In (a), the  black (red) dashed line indicates the bulk (OBC) characteristic relaxation rate, corresponding to the linearized QBL dynamics in Eqs.\,\eqref{eq: SC+}. In (b), the  black (red) dashed line indicates the bulk (OBC) characteristic amplification (relaxation) rate, corresponding to the linearized about the aligned equilibrium QBL dynamics in Eqs.\,\eqref{eq: SC-} (Eqs.\,\eqref{eq: SC+}). In all cases, the parameters are chose to match those chosen in Fig.\,\ref{fig: HNSpecARDM}(a) and (c) and the system size increases from $N=10$ to $40$ in increments of $6$, with the darker curves corresponding to larger $N$.} \label{fig: LindAR}
\end{figure}

To investigate how the nonlinearities intrinsic to the spin dynamics modify the magnon anomalous relaxation regime, in Fig.\,\ref{fig: LindAR}(a) we plot the dynamics of the bulk spin at site $j=N/2$ following an initial condition near equilibrium $s^z(0)\sim 1$, for various system sizes. Unsurprisingly, given the proximity to the stable equilibrium, the trajectories closely match the linearized magnon population dynamics at site $1$ for the  HN QBL in Fig.\,\ref{fig: HNSpecARDM}(c). 

In Fig.\,\ref{fig: LindAR}(b), we instead prepare the system in a state far from the stable equilibrium, namely, near the unstable equilibrium $s^z(0) \sim -1$. If the system were under PBCs, this mode would amplify at a characteristic rate set by $|\text{Re}[A(\pi)]|$, until asymptotically relaxing over the same characteristic timescale as the spins become aligned. The OBC dynamics, however, are far more interesting. During the first stage of evolution (lasting until about $\kappa_-t=4$ for $N=40$), the bulk spin amplifies at the PBC amplification rate. We expect that, for larger and larger system sizes, a bulk spin prepared in a bulk eigenstates should behave as though the system were under PBCs, until the boundaries are ``detected". Surprisingly, however, following this stage we see a sharp amplification, at a rate {\em far faster} than either BCs characteristic amplification rate. This is followed up by a steep drop, again at a rate far faster than either BCs characteristic decay rates near the aligned equilibrium. A strictly linear analysis utilizing the framework of Sec.\,\ref{sec: DM} would suggest that the amplification rate would transition from the PBC rate to the OBC rate at some time $t$ roughly proportional to $N$. We thus conclude that these {\em uncharacteristically fast amplification and decay processes} must arise solely from the interplay with the underlying nonlinearities. Following these distinctive nonlinear behaviors, as the system approaches the aligned equilibrium, it relaxes at the true OBC relaxation rate, consistent with the established ``awareness" of boundaries.

\subsubsection{Dynamical metastability in the presence of nonlinearities}

\begin{figure}
    \centering
    \includegraphics[width=.75\linewidth]{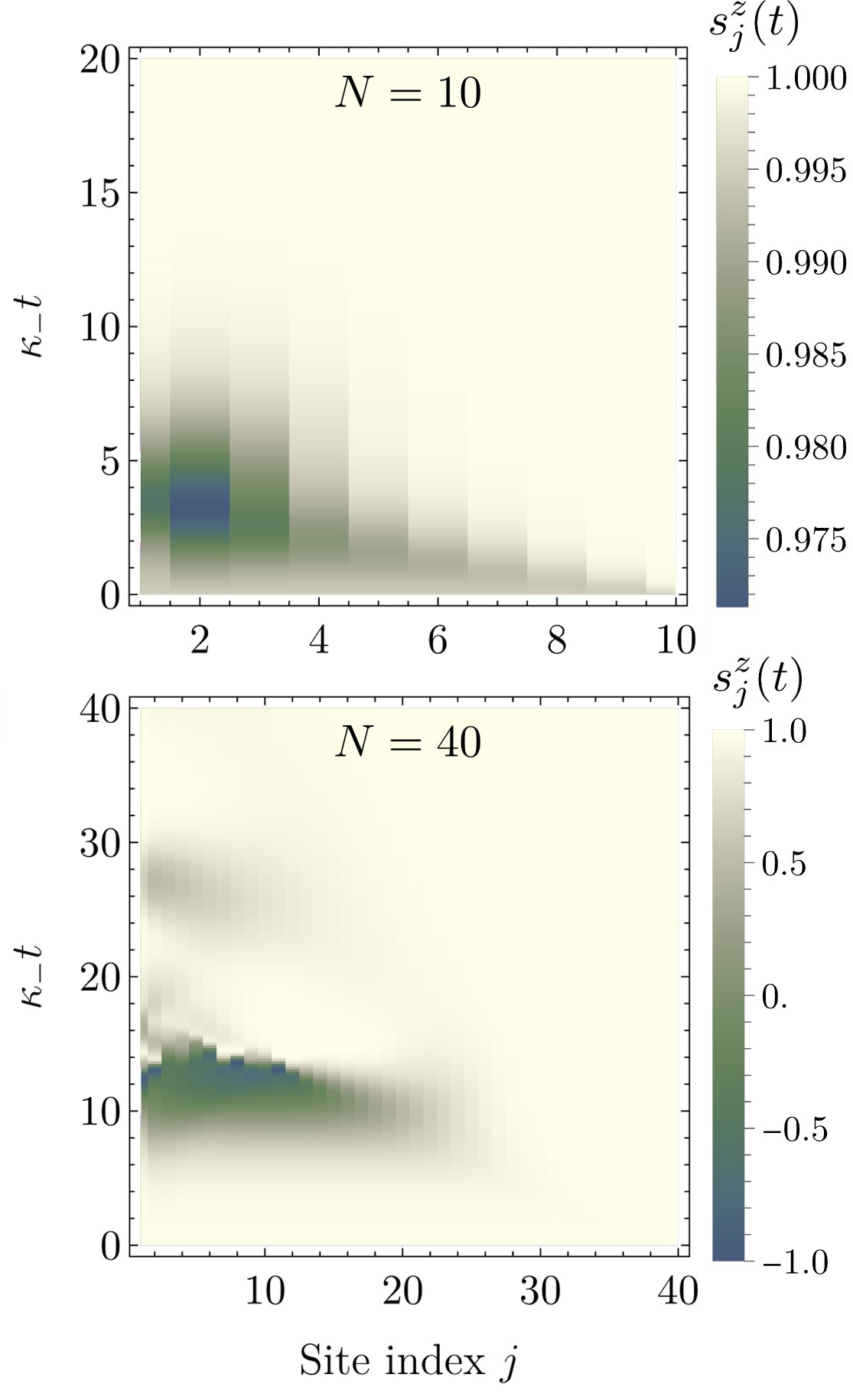}
    \vspace*{-3mm}
    \caption{Semiclassical nonlinear spin dynamics from Eq.\,\eqref{eq: LindSC} within the dynamically metastable magnon regime. The evolution of the $z$ spin component for each site $j$ is computed as a function of time, starting from an initial condition prepared in the $k=\pi$ mode as in Eq.\,\eqref{eq: pimode}, near the stable aligned equilibrium, $s_j^z(0) = \sqrt{1-0.1^2} \sim 1$. Despite the initial proximity to equilibrium, the spins are transiently pulled to the unstable antialigned state due to the transiently amplifying nature of the $k=\pi$ mode. The effect is significantly more pronounced for larger $N$, with the top figure depicting $N=10$ and the bottom depicting $N=40$. In both cases, parameters are chosen to match those of Fig.\,\ref{fig: HNSpecARDM}(b) and (d). }
    \label{fig: SpinDip} 
\end{figure}

We now turn to exploring the parameter regime in which the magnon fluctuations about equilibrium, described by the QBLs ${\cal L}_\pm$, are dynamically metastable, e.g., as depicted in Fig.\,\ref{fig: HNSpecARDM}(b). In this case, the stable equilibrium depends strongly on BCs and initial conditions. For instance, let us consider initializing the system in the most unstable bulk mode as described by Eq.\,\eqref{eq: pimode}. Under PBCs, this mode will relax to the antialigned state, independent of the choice of $s^z(0)$ (and hence, the nearness to the equilibrium). On the other hand, under OBCs, this mode will always relax to the aligned state. This sharp discrepancy in steady-state behavior, combined with the fact that increasingly large OBC chains should exhibit transient dynamics consistent with a dynamically unstable bulk, suggests the possibility of many intriguing dynamical scenarios.

To build insight, let us first consider the case where the $\pi$ mode is prepared close to the true OBC equilibrium. While intuition suggests that such a state should rapidly converge to the nearby steady state, a markedly different behavior is found, as illustrated in Fig.\,\ref{fig: SpinDip}. Consistent with the dynamics under PBCs, the mode initially amplifies as if the aligned state were an unstable equilibrium, causing the spins to transiently deviate -- or ``dip'' -- toward the antialigned configuration. Consistent with the distinctive size dependence expected from dynamical metastability, this effect becomes increasingly pronounced with larger system size $N$. As shown in the bottom panel of Fig.\,\ref{fig: SpinDip}, for $N=40$, the spins at sites $1\leq j \lessapprox 15$ nearly undergo a full reversal before eventually relaxing to the stable aligned state. On the contrary, for $N=10$, the maximum drop in $s_j^z$ is only about 2.5\%. The fact that the most pronounced dipping occurs on the left side of the chain reflects the directionality of magnon amplification: left-moving magnon modes grow preferentially, and their amplification is intrinsically accompanied by a reduction in the $z$-component of spin. Taken together, these results indicate that {\em dynamical metastability serves to make the unstable antialigned equilibrium transiently attractive}, at least for transiently unstable initial conditions like the $k=\pi$ mode we have examined. 

\begin{figure}[t]
    \centering
    \vspace*{-3mm}
    \includegraphics[width=\linewidth]{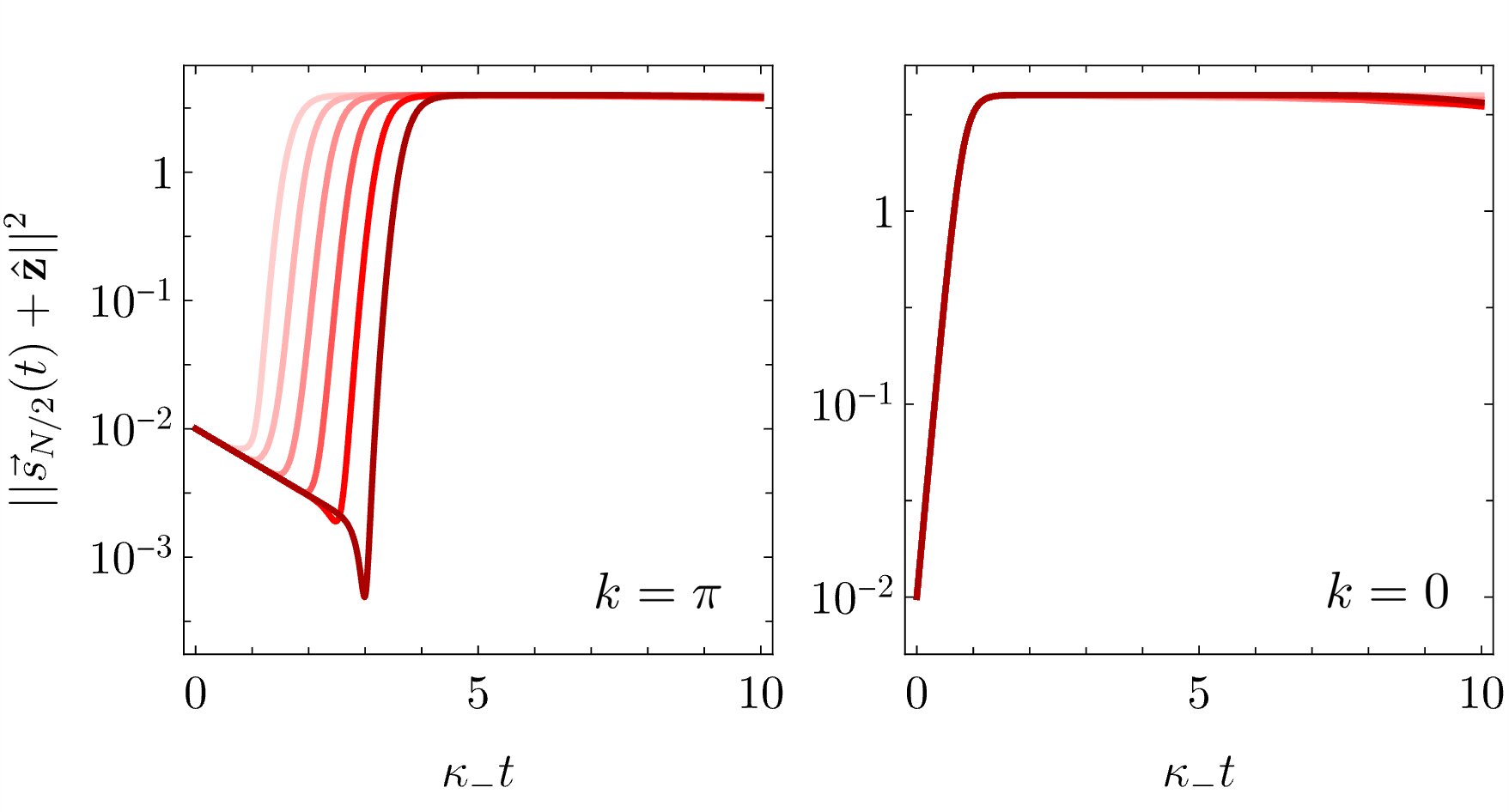}
    \caption{Deviation of the bulk spin vector $\vec{s}_{N/2}$ from the antialigned configuration $-\hat{\mathbf{z}}$ for the $k=\pi$ mode (left) and the $k=0$ mode (right), resulting from the nonlinear spin dynamics in Eq.\,\eqref{eq: LindSC} when in both cases the initial 
    state is prepared close to the antialigned state, $s^z(0) = -\sqrt{1-0.1^2}$. System parameters are the same as in Fig.\,\ref{fig: SpinDip}. }
    \label{fig: FalseEq}
\end{figure}

We can further demonstrate the transient attractivity of the antialigned state by initializing the spin configuration nearby it. In Fig.\,\ref{fig: FalseEq}, we consider both the $k=\pi$ mode and the $k=0$ mode to be initialized near the antialigned state, and plot the separation between the bulk spin vector $\vec{s}_{N/2}$ from the antialigned configuration $-\hat{\mathbf{z}}$. In the $k=\pi$ case (left panel), we see a clear exponential approach to the antialigned state during a transient whose duration increases with system size. Following this, there is a brief dip followed by an exponential approach to the true equilibrium configuration $+\hat{\mathbf{z}}$. This behavior is to be expected, since this mode is known to be stable about the antialigned state under PBCs, but not for OBCs. Hence, before the boundaries are ``detected", the mode begins to converge to the antialigned state. In contrast,  an exponential departure away from the antialigned state towards the true equilibrium state is seen for the $k=0$ mode (right panel). This is consistent with the fact that this mode is known to be unstable about the antialigned state, regardless of BCs. 

\subsection{The fate of Dirac bosons}
\label{sec: LindDB}

\subsubsection{Transient topological magnon modes in the presence of nonlinearities}

\begin{figure*} [th!]
    \centering
 \includegraphics[width=\linewidth]{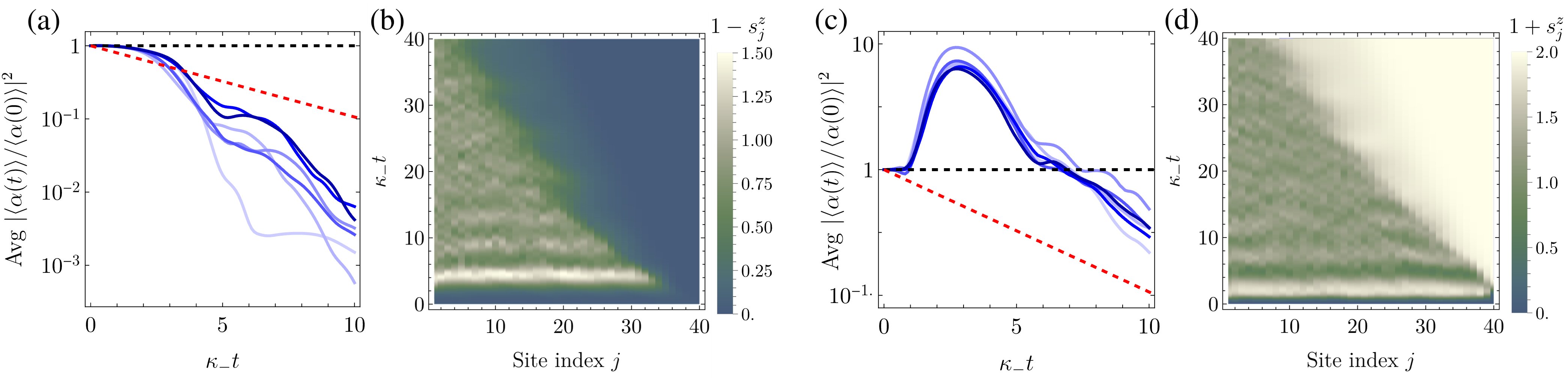}
 \vspace*{-3mm}
   \caption{\textbf{(a)} The nonlinear dynamics of the DB magnon zero mode $\alpha(t)$ in Eq.\,\eqref{eq: nldb}, averaged over 50 randomly sampled initial conditions prepared near the aligned state. The system size increases from $N=10$ (lightest curve) to $N=40$ (darkest curve), in increments of $6$. The black dashed line is fixed at 1, while the red dashed line corresponds to the characteristic decay rate of the system under OBCs. \textbf{(b)} The deviation of the local $z$ direction of spin $s_j^z(t)$ from the aligned equilibrium value $s_j^z = 1$ averaged over the same 50 initial conditions chosen in (a). \textbf{(c)} The same plot as in (a), but now with the sampled initial conditions prepared near the antialigned state. \textbf{(d)} The deviation of the local $z$ direction of spin $s_j^z(t)$ from the antialigned equilibrium value $s_j^z = -1$ averaged over the same 50 initial conditions chosen in (c). In all four panels, the parameters are chosen to match those used in Fig.\,\ref{fig: HNSpecTDM}.}
    \label{fig: NLDBZM} 
\end{figure*}

As a third relevant scenario, we now explore the parameter regime in which the linearized QBL dynamics are topologically metastable. In particular, we fix our system parameters so that the magnon rapidity spectrum matches the one in Fig.\,\ref{fig: HNSpecTDM}(a). Recall that, since we carry out our analysis in a frame rotating at $\omega$, any reported stationary dynamics will be rotating at $\omega$ in the physical lab frame. To avoid confusion, as in the previous sections we will only present frame-independent quantities, such as the modulus of complex spin variables.

We first study the approximate conservation of the DB mode $\alpha$, introduced in Eq.\,\eqref{eq: DBalpha}, under the full nonlinear dynamics. From the magnon analysis, we may expect that, for any state prepared sufficiently close to {\em either} equilibrium, the quantity
\begin{align}
    \alpha(t) \equiv \sum_{j=1}^N v_j^* s_j^+(t)
    \label{eq: nldb}
\end{align}
should be approximately conserved \footnote{Note that the coefficients $v_j^*$ are independent of which equilibrium we prepare our states close to. Due to the nature of the model, the dynamical matrices (and their Hermitian conjugates) describing the evolution of $s_j^+$ about each equilibrium share approximate zero eigenvectors.}. In Fig.\,\ref{fig: NLDBZM}, we put this to the test. In (a), we consider 50 random initial conditions prepared close to the aligned state and plot the mean value of $\alpha$ over time. This is the nonlinear analogue of Fig.\,\ref{fig: HNSpecTDM}(c) and uses the same random sample initial conditions for $s_j^+(0) \sim \braket{a_j}$. Immediately we see that, while the quantity does remain conserved for some appreciable transient time, the duration of this transient does {\em not} increase meaningfully with system size. The same behavior is observed in the case where we instead consider a random sample of initial conditions near the antialigned state, as shown in (c). 

While this result may seem surprising, it can be explained as a consequence of the interplay between transient magnon amplification and {\em total spin conservation}. For concreteness, consider an arbitrary initial condition prepared near the stable aligned equilibrium. The initial evolution of this state is described by the magnon QBL ${\cal L}_+$ of Eq.\,\eqref{eq: SC+} in the topologically metastable regime. Given some transient time $t\equiv t^*$, there must then exist a sufficiently large $N\equiv N^*$ such that, for all $N> N^*$, the state evolves as if the system were under PBCs for a time at least as long as $t^*$. During this time, the state undergoes transient amplification on the order of $e^{|\text{Re} A(\pi)| t^*}$, where again, $|\text{Re} A(\pi)|$ sets the $N$-independent maximal bulk amplification rate. However, the conservation of total spin dictates that amplification of $s_j^+(0)$ requires a correspondingly large deviation of $s_j^z$ from its initial unit value. We plot the average value of this deviation over all considered trajectories in Fig.\,\ref{fig: NLDBZM}(b). Since the deviation of $s_j^z$ from 1 reflects the strength of nonlinear effects, the linearized magnon approximation must eventually break down. As a result, $\alpha$ need not remain approximately conserved indefinitely in time, unlike for QBL dynamics. 

The above argument allows us to make a rough quantitative estimate of the lifetime of $\alpha$ under the full nonlinear (semiclassical) dynamics. If we assume the magnon approximation breaks down when $s_j^z \sim 1/2$, we expect $\alpha$ to deviate significantly from its initial value when $|s_j^+(0)| e^{|\text{Re} A(\pi)| t} \sim \sqrt{1-0.5^2}$. For $|s_j^+(0)| \sim 0.1$, this yields an estimated lifetime of $\kappa_- t\sim 2.16$ for the parameters chosen in Fig.\,\ref{fig: NLDBZM}, which is remarkably accurate, as confirmed in Fig.\,\ref{fig: NLDBZM}(a-b).  An analogous argument can be made for states prepared near the antialigned state, leading to nearly identical conclusions. The only difference is that, in this case, the maximum bulk amplification rate near the antialigned state is $|\text{Re} A(0)|$, which is greater than in the aligned case, see Fig.\,\ref{fig: BulkSS}(a). This results in a slightly lower lifetime. The dynamics of these states are plotted in Figs.\,\ref{fig: NLDBZM}(c-d).  

\begin{figure}[t]
    \centering
    \includegraphics[width=0.85\linewidth]{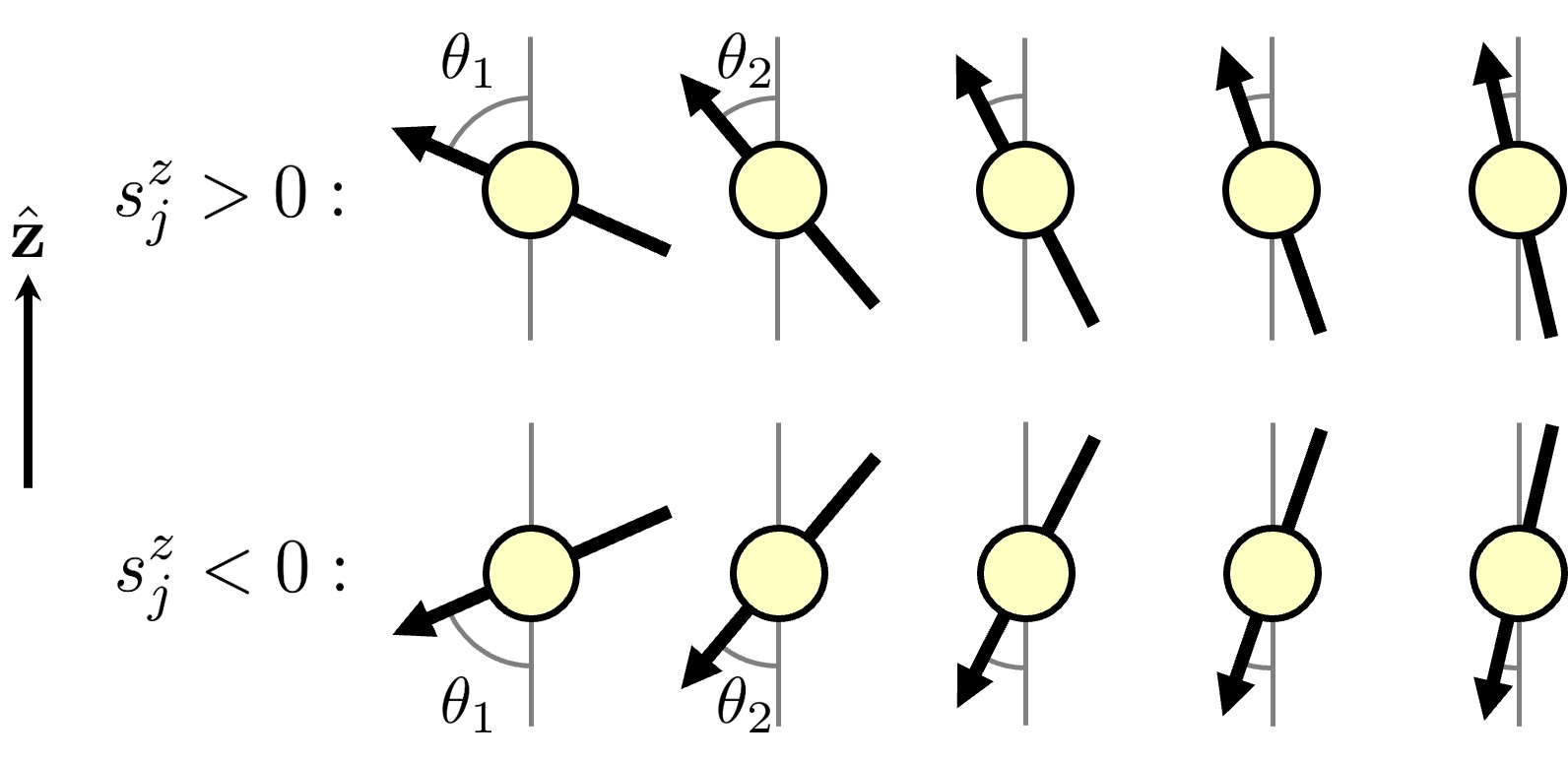}
    \vspace*{-2mm}
    \caption{Pictorial representation of the aligned (top) and antialigned (bottom) DB initial conditions defined in Eq.\,\eqref{eq: DBICs}. Each spin is projected into the plane spanned by the vectors $\hat{z}$ and $\vec{s}_j$.   }
    \label{fig: DBvis}
\end{figure}

Based on the above considerations, nonlinearities inhibit the lifetime scaling properties of the DB zero mode $\alpha$. What, then, is the fate of the approximate displacement symmetry generator $\beta$  in Eq.\,\eqref{eq: DBbeta}? To answer this, we prepare our system in a DB initial condition. More specifically, let $\vec{w}$ be the coefficient vector of $\beta$. We then define the DB initial conditions for our spin system as 
\begin{align}
\label{eq: DBICs}
    s_j^+(0) \equiv  \rho w_j,\quad s_j^z(0) \equiv \pm \sqrt{1-|\rho w_j|^2},
\end{align}
where the parameter $\rho\in[0,1]$ allows us to tune the relative distance from either equilibria. We call the state where $s_j^z(0)>0$ $(s_j^z(0)<0)$ the {\em aligned (antialigned) DB initial condition} (see Fig.\,\ref{fig: DBvis} for a visual representation). Because $|w_j|$ is exponentially bounded, the cosine of the deflection angle $\theta_j$ from the $z$ axis decreases exponentially from left to right in both cases. 

\begin{figure*}
    \centering
 \includegraphics[width=\linewidth]{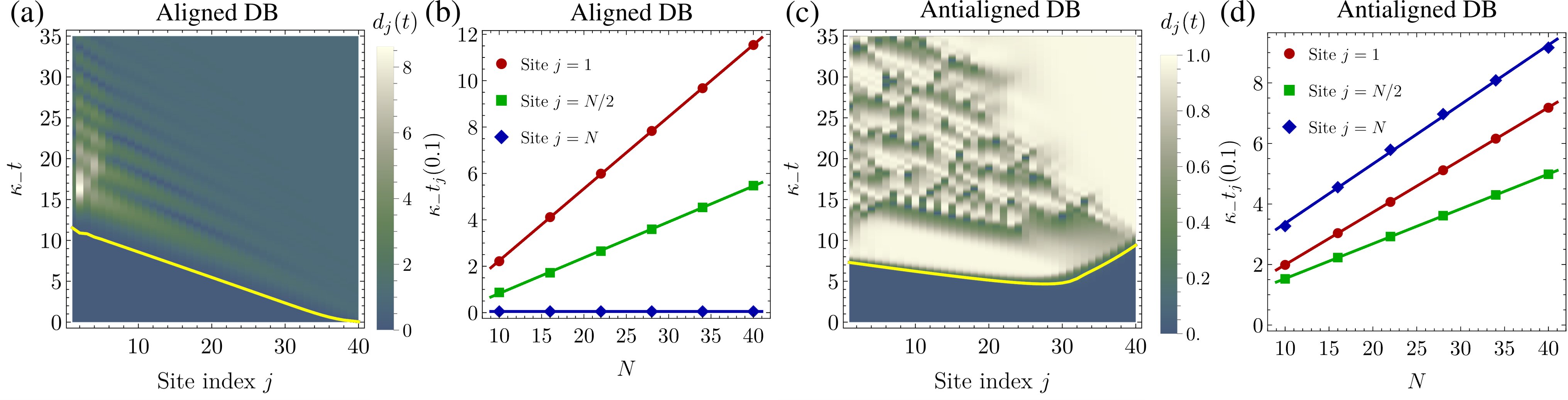}
 \vspace*{-3mm}
   \caption{\textbf{(a)} Spatial and time dependence of the local spin deviation $d_j(t)$ in Eq.\,\eqref{eq: dj} for an $N=40$ site chain prepared in the aligned DB initial condition with $s_j^z(0) >0$. The yellow line corresponds to the site-$j$ lifetime $t_j(\delta) =t_j(0.1)$.
 \textbf{(b)} The lifetimes of the spins at sites $1$, $N/2$, and $N$ for the evolution considered in (a), versus system size. The solid lines correspond to lines of best fit. \textbf{(c)} The same as in (a), but for the antialigned DB initial condition with $s_j^z(0)<0$.  \textbf{(d)} The same as in (b), but for the antialigned DB initial condition. In all cases, $\rho=1/2$ in Eq.\,\eqref{eq: DBICs} and system parameters are taken to match those in Fig.\,\ref{fig: HNSpecTDM}.}
    \label{fig: DBIC_Life} 
\end{figure*}

The linearized magnon analysis suggests that these spin configurations should remain long-lived, as long as nonlinear effects remain small. Since the strength of the nonlinearities is governed by the state variables themselves (i.e., the deviation of $s_j^z(t)$ from either $1$ or $-1$), we expect the nonlinear dynamics will preserve the above DB initial conditions provided that the system is initially in the linear regime -- and thus the dynamics is well-approximated by the QBL ${\cal L}_+$ or ${\cal L}_-$ -- thanks to the long-livedness of the state. To quantify the lifetimes of these states, we consider the following deviation measure:
\begin{align}
\label{eq: dj}
    d_j(t) \equiv \frac{\norm{\vec{s}_j(t) - \vec{s}_j(0)}}{\norm{\hat{\mathbf{z}}-\vec{s}_j(0)}}.
\end{align}
The numerator measures the separation of the spin at site $j$ from its initial value at time $t$, while the denominator serves to penalize initial conditions prepared very close to the aligned equilibrium, since states close to this point should not evolve much to begin with. In general, $d_j(0) = 0$ and $d_j(t) \to 1$ as $t\to\infty$, representing relaxation to equilibrium. 
A measure of the lifetime of site $j$ can then be defined by choosing a prescribed accuracy, say, $\delta>0$, and by letting $t_j(\delta)$ to be the first time at which $d_j(t) > \delta$. Specifically, with $\delta=0.1$ for our analysis, $t_j(\delta)$ is the first time the spin at site $j$ separates from its initial value by $10\%$ of its initial distance from equilibrium. 

In Fig.\,\ref{fig: DBIC_Life}, we plot both $d_j(t)$ and $t_j(\delta)$. In the aligned case, depicted in panels (a-b), a notable dependence of the lifetime on the site index is seen. In (a), we superimpose (in yellow) the lifetime $t_j(\delta)$ at $\delta=0.1$. Sites on the left side of the chain have a considerably larger lifetime and, in fact, experience a linear increase in lifetime as a function of $N$, as shown in (b). However, it is important to observe that, due to the exponential profile of the DB, the spin at site $N$ is already exponentially close to its equilibrium position. Non-reciprocity in the system serves to amplify magnons from right to left, and so, site $N$ simply relaxes exponentially fast. It is interesting to note that we observe the spin-dipping phenomena from the preceding section also for the aligned DB initial conditions, as evidenced by the sharp increase of $d_j(t)$ for sites on the left side of the chain. 

For the antialigned case, shown in panels (c-d), the lifetime of each spin is found increase linearly with system size, regardless of the spin's location. In contrast to the aligned case, however, the spin at site $N$ has the longest lifetime and, in fact, from a linear fit, we find that the slope of $t_N(\delta)$ versus $N$ is actually slightly larger than that of $t_1(\delta)$ versus $N$. Remarkably, the lifetimes of the spins (specifically, those close to the left side of the chain) are on the {\em same order of magnitude regardless of whether the DB is aligned or antialigned}. This is in spite of the fact that the antialigned state is manifestly unstable. In panel (c), following a period of ``frozen dynamics'' where the deviation $d_j$ remains nearly uniformly small across the chain, we observe a nearly uniform spin flip to the aligned state -- before a less structured regime of highly irregular dynamics sets in. This regime displays clear signatures of directional amplification in the form of stratified patterns of spin density. We further find that the state eventually relaxes to the aligned equilibrium around $\kappa_-t \simeq 80$ (data not shown), with sites near the left side of the chain converging the most slowly.

\subsubsection{Robustness against disorder}

In the discussion of the HN QBL, and as an important feature of topological metastability in general \cite{FlynnBosoranasPRL2021,FlynnBosoranasPRB2023}, we identified a degree of robustness that DBs inherited from their pseudospectral and topological origins. Importantly, the same degree of robustness is expected for generalized DBs, as long as U(1)-preserving, static disorder is considered. For both conceptual and practical reasons, it is important to understand the extent to which this robustness may persist in the presence of nonlinearities. Even under these basic assumptions, deviations from the intended dynamics may in principle occur in a variety of ways. In particular, either or both the dynamical generator and the initial conditions may be modified from their nominal values, and in turn the relevant modifications may affect selected subsets of spins differently. An in-depth exploration of this question would thus require significant additional analysis, which is beyond our present scope.

Here, in order to provide initial evidence on the persistence of meaningful DB signatures away from the intended dynamics, we present numerical results for an idealized disordered setting. Specifically, we study the system-size dependence of the antialigned DB lifetimes in a ``global disorder'' scenario where we allow each of the parameters entering the QBL magnon dynamics in Eqs.\,\eqref{eq: SC+}-\eqref{eq: SC-} (hence the nonlinear dynamics in Eq.\,\eqref{eq: LindSC}) to acquire a random spatial dependence. That is, we let $X\in\{ \omega,J,D,\kappa_-,\Gamma,\kappa_+\}$ and map $X\mapsto X_j$, with each $X_j$ being drawn from a normal distribution whose mean equals the value used in the topologically metastable configuration, and whose standard deviation equals 10$\%$ of the mean. We consider two initializations: 
\begin{itemize}
\item[(i1)] the initial state is the antialigned DB computed from the {\em clean linearized} system, Eq.\,\eqref{eq: DBICs};  
\item[(i2)] the initial state is the approximate DB of the {\em disordered linear} system, which we compute for each realization drawn from the ensemble under consideration. We call these latter states ``disordered DBs". 
\end{itemize}

Note that the existence of disordered DBs is ensured by the known response properties of the pseudospectrum to small perturbations, discussed in Sec.\,\ref{sec: TDMbg}. In case (i2), the initial state will depend in general on the specific disorder realization being considered, thus an average over both dynamical parameters and initial spin configurations is effectively implemented. 

\begin{figure}
    \centering
    \includegraphics[width=\linewidth]{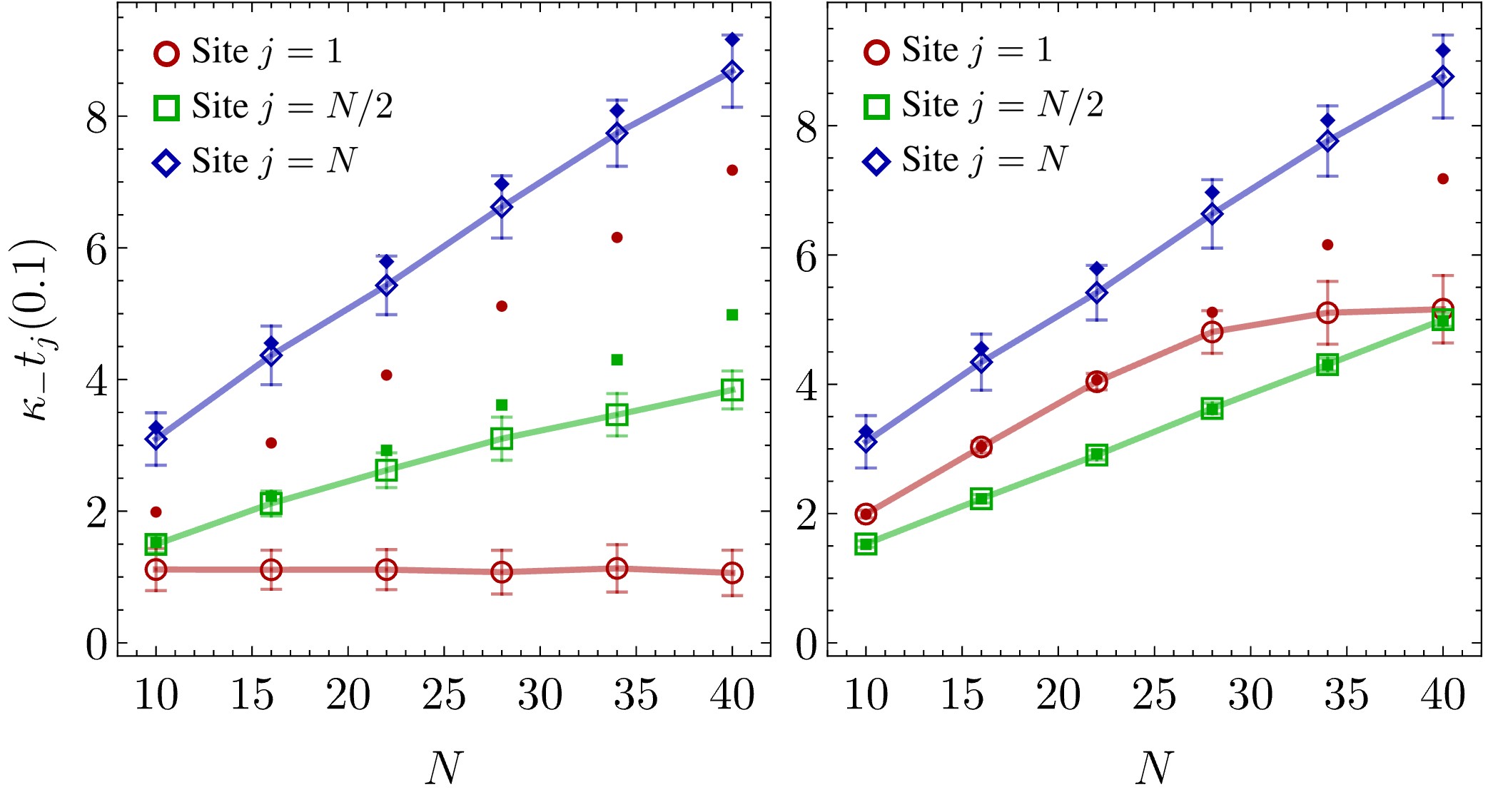}
    \vspace*{-3mm}
    \caption{
    The lifetimes of the spins at sites $1$, $N/2$, and $N$ for two different disorder implementations, versus system size. \textbf{(a)} Average lifetime of a fixed antialigned clean DB initial condition associated to the disorder-free Lindbladian under the dynamics generated by an ensemble of disordered Lindbladians - scheme (i1) in the main text. \textbf{(b)} Average lifetime of a disorder-dependent antialigned DB initial condition associated to the disordered Lindbladian under the corresponding disordered dynamics - scheme (i2) in the main text. In each case, 100 disorder realizations are considered where each parameter is normally distributed with mean given by the values used in Fig.\,\ref{fig: DBIC_Life} and standard deviations equal to 10\% of their means. Error bars indicate one standard deviation.}
    \label{fig: DisLife}
\end{figure}

In Fig.\,\ref{fig: DisLife}, we plot the average lifetimes and corresponding standard deviations for the spins at sites $1$, $N/2$, and $N$ for the two above-mentioned initialization schemes. For initialization (i1) (shown in (a)), the deviation of the lifetimes from the disorder-free case is most significant closer to the left side (site $j=1$) of the chain, with site $j=N$ the least affected by disorder. This is consistent with the right-to-left directional motion of magnons in the system. The spins at a site $1<j_0<N$ are most largely effected by the $N-j_0$ sites to the right, and thus, sites with $j_0\ll N$ experience an extensive degree of disorder compared to sites with $j_0\lessapprox N$. In other words, small fluctuations amplify as they propagate to the left, significantly reducing the lifetimes of sites on the left. 
For initialization (i2) (shown in (b)), the average lifetimes of the disordered DBs are within one standard deviation of the idealized clean case, save for the spin at site $1$ for $N \gtrsim 25$. Indeed, the lifetimes seem to become slightly more reduced as $N$ grows for all 3 sites considered, with the effect maximized for site $j=1$. We propose the same explanation (directional amplification of fluctuations) for the maximization of this effect at $j=1$. However, it is unclear why the bulk site $j=N/2$ has minimal overall variance among the three. Additionally, the variance seems to increase for larger size (especially for site $j=1$), likely again due to the extensive nature of disorder experienced by the left half of the chain.

\section{Signatures of dynamical metastability in classical magnetization dynamics}
\label{sec: LLG}

\subsection{The model and LLGS dynamics}

\begin{figure}[t]
    \centering
    \includegraphics[width=\linewidth]{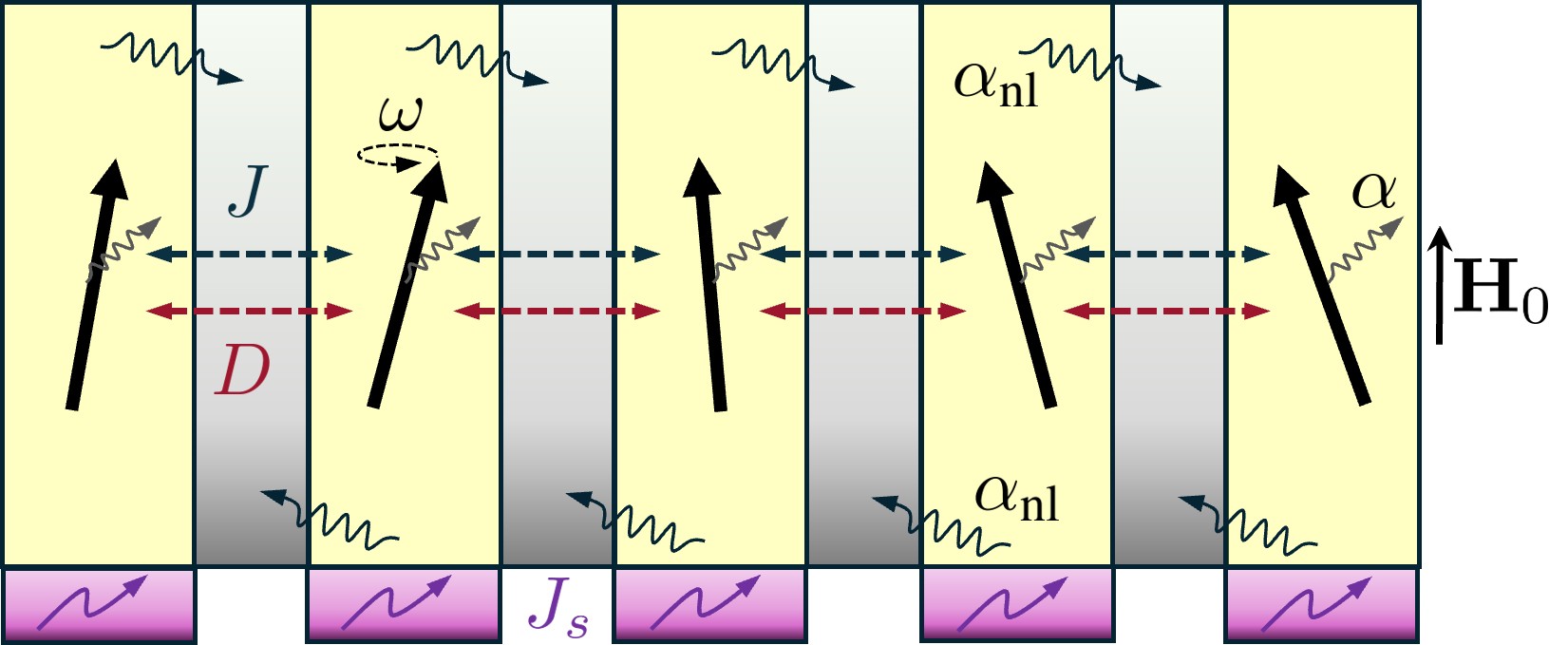}
    \caption{Pictorial depiction of the ferromagnetic heterostructure modeled by Eqs.\,\eqref{eq: LLGHam} and \eqref{eq: fullLLG}. The macrospin within each magnet (yellow) is coupled to its nearest neighbor via exchange interaction and DMI with strength $J$ and $D$, both of which are mediated by metallic spacers (gray). Additionally, each macrospin undergoes Gilbert damping of strength $\alpha$ and non-local (nearest-neighbor) damping of strength $\alpha_\text{nl}$, Eq.\,\eqref{eq: LLGdis}. Finally, each magnet is in contact with a metal layer (purple) subject to an electrical current, which in turn transfers spin current $J_s$  into the magnet via the spin Hall effect, applying a spin-transfer torque, Eq.\,\eqref{eq: LLGSTT}.}
    \label{fig: LLGcartoon}
\end{figure}

In this section, we introduce and analyze our second main illustrative model, namely, a classical spin model describing an experimentally accessible magnetic platform where the key ingredients entering the Lindbladian framework presented in Sec.\,\ref{sec: Lind} are present by design: dissipation and nonlinearities are built into the magnetization dynamics, while nonlocal interactions and external drive can be engineered and tuned within existing device architectures. Specifically, we  consider a driven-dissipative magnetic heterostructure as depicted in Fig.\,\ref{fig: LLGcartoon}, where $j=1, \ldots, N$ labels magnetic layers interacting via nonmagnetic metallic spacers. The coherent dynamics of the orientation of the magnetization vector $\mathbf{m}_j$ of the $j$'th macrospin, where $\norm{\mathbf{m}_j}=1$ for all $j$, are described by the Hamiltonian 
\begin{align}
    \mathcal{H} = -\sum_{j}[&\mu_0 M_s \mathbf{H}_0\cdot \mathbf{m}_j + J \mathbf{m}_j\cdot \mathbf{m}_{j+1}\nonumber
    \\
    &+D\hat{\mathbf{z}}\cdot\lp \mathbf{m}_j\times \mathbf{m}_{j+1}\rp],
    \label{eq: LLGHam}
\end{align} 
where $\mu_0$ is the vacuum permeability, $M_s$ is the saturation magnetization, and $\mathbf{H}_0= H_0 \hat{\mathbf{z}}$,
$H_0>0$, is the applied magnetic field.  Here, $J>0$ is a ferromagnetic Ruderman-Kittel-Kasuya-Yosida (RKKY)-like exchange interaction, and $D\in {\mathbb R}$ is the interfacial Dzyaloshinskii-Moriya interaction (DMI) strength \cite{DiDMI2015,FernDMI2019,HanExch2019}. The Hamiltonian in Eq.\,\eqref{eq: LLGHam} defines, for each layer $j$, an effective Landau-Lifshitz field $\mathbf{H}_j = -(1/M_s)\,\partial \mathcal{H}/\partial \mathbf{m}_j$. This leads to a coherent precession dynamics of the form
\begin{align}
\label{eq: LLGcoh}
    \dot{\mathbf{m}}_j\big|_{\text{coh}} = -\gamma \,\mathbf{m}_j\times \mathbf{H}_j,
\end{align}
where $\gamma$ is the gyromagnetic ratio and $\mathbf{H}_j$ is explicitly given by
\begin{align}
\label{eq: LLGhj}
    \mathbf{H}_j = \mu_0 H_0\hat{\mathbf{z}} &+ \frac{J}{M_s}\lp\mathbf{m}_{j+1} + \mathbf{m}_{j-1}\rp \nonumber \\
    &+\frac{D}{M_s}\lp \mathbf{m}_{j+1}-\mathbf{m}_{j-1}\rp\times \hat{\mathbf{z}} ,
\end{align}
where the fact that only neighboring macrospins at layers $j\pm 1$ contribute follows directly from the nearest-neighbor structure of $\mathcal{H}$.

In addition to the coherent evolution generated by $\mathcal{H}$, the magnetization dynamics is subject to both local and nonlocal dissipation which, within the LLG framework, can be phenomenologically accounted for with a local Gilbert damping term (with strength $\alpha> 0$) and a nonlocal damping term (with strength $\alpha_{nl}>0$) -- the latter contribution arising from angular-momentum transfer between adjacent magnetic elements \cite{Gilbert2004,YaroGilb2002,YaroPump2002}, i.e.,
\begin{align}
\label{eq: LLGdis}
\dot{\mathbf{m}}_j\big|_\text{dis} = \alpha\mathbf{m}_j\times \dot{\mathbf{m}}_j +\alpha_\text{nl}\mathbf{m}_j\times\lp \dot{\mathbf{m}}_{j-1}+\dot{\mathbf{m}}_{j+1}\rp.
\end{align}
Notably, if $D \neq 0$, nonlocal damping can induce {\em nonreciprocal} spin transport \cite{DuineUniNL2023,XinPaper_2025}. Microscopically, DMI breaks time-reversal symmetry in the magnon Hamiltonian and imprints a complex, gauge-unremovable phase on nearest-neighbor hopping, which skews the balance between coherent and dissipative couplings \cite{li2024cooperativenonreciprocalemissionquantum,PhysRevX.5.021025}. Upon linearization, as we show in detail below (Sec.\,\ref{sub: LLLG}), the interplay between DMI and nonlocal damping generates a nonreciprocal magnon dynamics that maps directly to the asymmetric hopping dynamics discussed for a HN chain.
 
Finally, access to the far-from-equilibrium, nonlinear regime of interest is enabled by the introduction of a drive term, namely, a Slonczewski-Berger spin-transfer torque (STT) that describes the interaction of the magnetic order parameter with the spin-polarized current \cite{SlonSTT1996,BergerGilb2001}. Mathematically, the resulting contribution is given by  
\begin{align}
\label{eq: LLGSTT}
    \dot{\mathbf{m}_j}\big|_\text{STT} = J_s \mathbf{m}_j\times\lp\mathbf{m}_j\times 
 \hat{\mathbf{z}} \rp,  
\end{align}
where $J_s$ is the strength of the spin current polarized along $\hat{\mathbf{z}}$. The complete LLGS EOM thus take the form
\begin{align}
\label{eq: fullLLG}
    \dot{\mathbf{m}}_j = \dot{\mathbf{m}}_j\big|_\text{coh}+\dot{\mathbf{m}}_j\big|_\text{dis}+\dot{\mathbf{m}}_j\big|_\text{STT}.
\end{align}

It is natural to try establishing some correspondence with the Lindbladian model in Eqs.\,\eqref{eq: spinLind}. In terms of their Hamiltonian components, the frequency, exchange, and DMI terms of the Lindbladian model in Eq.\,\eqref{eq: spinHam} correspond to $\gamma\mu_0 H_0$, $\gamma J/M_s$, and $\gamma D/M_s$, respectively, in the LLGS model of Eq.\,\eqref{eq: LLGHam}. Likewise, one might be tempted to associate the dissipative parameters $\kappa_-$, $\Gamma$, and $\kappa_+$ with the local Gilbert damping $\alpha$, the nonlocal damping $\alpha_\text{nl}$, and the spin injection rate $J_s$, respectively. While, conceptually, this mapping is reasonable, establishing a precise correspondence is nontrivial due to structural differences between the EOM the two models yield. Notably, the dissipative contributions in Eq.~\eqref{eq: LLGdis} involve time derivatives on the right-hand side as well. This causes the dissipative dynamics to depend explicitly on the coherent dynamics, while in the Lindblad case, these are, \textit{a priori}, independent \footnote{Within microscopic derivations of Lindblad master equations, the Hamiltonian and dissipative contribution cannot be treated as independent under standard weak-coupling assumptions. Independence can be recovered, in principle, by working under a singular-coupling limit assumption, as often assumed in dissipation-engineering approaches. We stress that our use of Lindblad master equations here is {\em phenomenological}, as providing effective (Markovian) models whose validity should be operationally assess for the class of systems at hand}. For example, if the Hamiltonian and, for simplicity, the drive are set to zero in both models, nontrivial evolution still ensues in the Lindbladian setting (relaxing the system to $s_j^z=+1$), while the LLGS dynamics completely freeze, despite a nonzero $\alpha$. We will uncover and analyze important ramifications of these structural differences at the linearized level in the next section.

To simplify our analysis, in what follows we will assume a hierarchy between the strengths of the various interactions that is consistent with current material properties and experimental capabilities. Specifically, we will work under the following assumptions:

\begin{itemize}
\item[(a1)] The exchange and DMI fields $J/M_s$ and $D/M_s$ are both significantly weaker than the applied field $\mu_0 H_0$.

\item[(a2)] The strengths of the damping terms are taken to obey $\alpha_\text{nl}/2< \alpha\ll 1$, so that terms of order of $\alpha^2$, $\alpha_\text{nl}^2$, $\alpha\alpha_\text{nl}$ (or higher) can be safely ignored (note that the factor of $1/2$ in the above reflects the stability requirement that the total nonlocal damping due to the two nearest neighbors in Eq.\,\eqref{eq: LLGdis} cannot exceed the local Gilbert damping \cite{YuanDissTorque2019,DuineUniNL2023}).

\item[(a3)] The STT strength $J_s$ is left variable, but will typically be set around the characteristic damping rate $\alpha\omega$, leading to terms like $\alpha J_s$ and $\alpha_\text{nl}J_s$ being also negligible. 
\end{itemize}

Altogether, these assumptions allow for a considerable simplification of the EOM in Eq.\,\eqref{eq: fullLLG}. By recursively replacing each time derivative on the right hand-side with the full expression for $\dot{\mathbf{m}}_j$, and neglecting the aforementioned terms, we obtain the simplified form
\begin{align}
\label{eq: approxLLG}
\dot{\mathbf{m}}_j  \approx-\gamma \mathbf{m}_j\times \mathbf{H}_{\text{eff},j} + J_s \mathbf{m}_j\times\lp\mathbf{m}_j\times \hat{\mathbf{z}}\rp ,
\end{align}
where the influence of local and non-local damping is now included through the effective magnetic field,  
\begin{align*}
\mathbf{H}_{\text{eff},j} & \equiv \mathbf{H}_j + \alpha \mathbf{m}_j\times \mathbf{H}_j 
\\
&+ \alpha_\text{nl}\lp \mathbf{m}_{j-1}\times \mathbf{H}_{j-1}+\mathbf{m}_{j+1}\times \mathbf{H}_{j+1}\rp .
\end{align*}
By introducting the Larmor frequency $\omega \equiv \gamma\mu_0 H_0$, we further define the dimensionless reduced exchange and DMI strengths as 
$$\bar{J} \equiv \gamma J/(M_s\omega), \qquad \bar{D} \equiv \gamma D/(M_s\omega),$$ 
respectively, which quantify the relative strength of the respective interactions to the applied field. Rescaling time $t\mapsto \tau = \omega t$ and the drive strength $J_s\mapsto \bar{J}_s = J_s/\omega$, Eq.\,\eqref{eq: approxLLG} can be rewritten in dimensionless form as
\begin{align}
\label{eq: approxLLGdiml}
\dot{\mathbf{m}}_j  \approx- \mathbf{m}_j\times \mathbf{h}_{\text{eff},j} + \bar{J}_s \mathbf{m}_j\times\lp\mathbf{m}_j\times \hat{\mathbf{z}}\rp ,
\end{align}
where now $\dot{\mathbf{m}}_j = d\mathbf{m}_j/d\tau$ and $\mathbf{h}_{\text{eff},j} \equiv \mathbf{H}_{\text{eff},j}/\mu_0 H_0$.

\subsection{Linearized LLGS dynamics}
\label{sub: LLLG}

The LLGS EOM in Eq.\,\eqref{eq: fullLLG}, and its approximate form in Eq.~\eqref{eq: approxLLGdiml}, retain a global rotational U(1) symmetry about $\hat{\mathbf z}$, inherited from the axial symmetry of the Hamiltonian in Eq.~\eqref{eq: LLGHam} and the form of the STT in Eq.~\eqref{eq: LLGSTT}. While this symmetry no longer implies conservation of $m^z_j$ in a driven-dissipative setting, it suggests the existence of rotationally invariant equilibria, of the form $\mathbf{m}_j = \sigma_j \hat{\mathbf{z}}$, with $\sigma_j\in\{1,-1\}$. Indeed, for such states, $\mathbf{H}_j$  becomes parallel to the $z$-axis and thus they constitute equilibria of Eq.~\eqref{eq: approxLLGdiml} (and, in fact, the exact dynamics Eq.~\eqref{eq: fullLLG}). 

As in the Lindbladian model of Sec.~\ref{sec: Lind}, we will focus on the aligned ($\sigma_j = +1$) and antialigned ($\sigma_j = -1$) equilibria, and expand Eq.\,\eqref{eq: approxLLGdiml} to first order in the fluctuations about them. Informed by the underlying U(1) symmetry, we move to complex state variables $m_j^\pm \equiv m^x_j\pm i m^y_j$, and consider the dynamics of fluctuations $m_j^+ = \delta m^+_j$, $m^z_j = \sigma (1-\delta m_j^z)$, with $\sigma=+1 (-1)$ in the (anti)aligned case. The EOM for the fluctuations, linearized around the equilibrium $m^z_j = \sigma$, then read
\begin{align}
    \delta\dot{m}_j^+ = &-\lp \kappa^\sigma -i\eta^\sigma\rp \delta m^+_j+J^\sigma_{R,1} \delta m^+_{j-1} + J^\sigma_{L,1} \delta m^+_{j+1}\nonumber
    \\
    &+J^\sigma_{R,2} \delta m^+_{j-2} + J^\sigma_{L,2} \delta m^+_{j+2},
    \label{eq: LLGlinEOM}  
\end{align}
where $\kappa^\sigma \equiv \sigma(\alpha\eta^\sigma - \bar{J}_s) - 2 \sigma\alpha_\text{nl} \bar{J}$ is the effective onsite loss rate, $\eta^\sigma  \equiv 1+2\sigma \bar{J}$ is the effective onsite (dimensionless) frequency, and $J^\sigma_{R,\ell}$ ($J^\sigma_{L,\ell}$) represents the hopping amplitude from site $j$ to $j-\ell$ $(j+\ell)$. By further introducing the complex (dimensionless) strength $\tilde{J} \equiv \bar{J}+i\bar{D}$, we may write 
\begin{equation}
\left\{ 
\begin{array}{ll}
    J_{R,1}^\sigma &= (\alpha-i\sigma)\tilde{J} - \sigma\alpha_\text{nl}\,\eta^\sigma, 
    \\
    J_{L,1}^\sigma &= (\alpha-i\sigma)\tilde{J}^* - \sigma\alpha_\text{nl}\,\eta^\sigma, 
    \\
    J^\sigma_{R,2} &= \alpha_\text{nl} \tilde{J} = J^{\sigma \ast}_{L,2} . 
\end{array}\right. 
\label{eq: LLGhop}
\end{equation}
Collecting the fluctuations into a vector array $\phi \equiv [\delta m^+_1,\ldots,\delta m_N^+]^T$, and defining the dynamical matrix implicitly via $\dot{\phi} \equiv  \mathbf{A}^\sigma \phi$, as before, we can diagnose stability of the state $m^z_j = \sigma$ by diagonalizing $\mathbf{A}^\sigma$ in different parameter regimes and for different BCs. In analogy with the bosonic Lindbladian formalism, we will call the eigenvalues of $\mathbf{A}^\sigma$ the \emph{(classical)} rapidities (associated with the equilibrium state $\sigma$). Once again, stability of the equilibrium state labeled by $\sigma$ requires that all the rapidities have strictly negative real parts (equivalently, the stability gap must be negative).  

Before undergoing a stability analysis of the two equilibria, we note that the anticipated mapping between the fluctuations of the linearized LLGS dynamics and the HN chain is made explicit by the following identity: 
\begin{align}
\label{eq: LLGNR}
|J_{L,1}^\sigma|^2 - |J_{R,1}^\sigma|^2 = 4(\alpha_\text{nl} \eta^\sigma)\bar{D}.
\end{align}
Thus, whenever {\em both} $\alpha_\text{nl}$ and $D$ are nonzero, the effective left-hopping strength exceeds that of the right. This is directly analogous to the result obtained for the HN chain in Eqs.\,\eqref{eq: HNEOM} and \eqref{eq: HNnonrec}, with $\alpha_\text{nl}\eta^\sigma$ and $\bar{D}$ playing the roles of $\Gamma$ and $D$ in the HN chain, respectively. There are two key differences between the two models, however. First, since $\eta^\sigma$ depends on the exchange interaction $J$ and the equilibrium state $\sigma$, the degree of nonreciprocity is influenced by these two factors {\em only} in the LLGS case.  As an extreme example, if the the applied field is arranged so that $\bar{J}=1/2$, we have $\eta^+ = 2$ but $\eta^-=0$, meaning that the magnon dynamics about $\sigma=+1$ will be nonreciprocal, while those around $\sigma=-1$ will not be. This stands in contrast to the HN Lindbladian case, where the degree of nonreciprocity is controlled by the state-independent parameter $\Gamma D$.
Secondly, as seen from Eq.\,\eqref{eq: LLGhop}, the non-local damping is always accompanied by \textit{next}-nearest neighbor hopping as well, which happens to be reciprocal,  $|J_{L,2}^\sigma| = |J_{R,2}^\sigma|$. 

As a further important difference, note that the local loss rate $\kappa^\sigma$ and all four hopping amplitudes in Eq.\,\eqref{eq: LLGhop}, contain contributions that are {\em independent} of $\sigma$ (e.g., we may rewrite $\kappa^\sigma= 2\alpha \bar{J}+\sigma(\alpha - \bar{J}_s-2 \alpha_\text{nl}\bar{J}$)). Comparing to the semiclassical EOM for the magnon Lindblad dynamics in Eq.\,\eqref{eq: LindSC}, with $s_j^z\approx \sigma$, we see that all such terms change sign upon taking $\sigma\mapsto-\sigma$. This simple relation enabled us, in Sec.\,\ref{sec: MagLind}, to perform simultaneous stability analyses about each equilibrium for the Lindbladian model and, in Sec.\,\ref{sec: SWSSPDLind}, to conclude that only one of $\sigma=+1$ and $\sigma=-1$ can be attractive for a given generic set of parameters. For the linearized LLGS dynamics, the $\sigma$-independent contribution to Eq.\,\eqref{eq: LLGlinEOM} renders these arguments invalid as, in general, the rapidity bands (and hence stability gaps) about $\sigma=+1$ and $\sigma=-1$ are not simply related. In what follows, we will denote by $\Delta_\text{X}^{(\sigma)}$, with $\text{X}\in\{\text{OBC},\text{ PBC}\}$, the stability gap of the system linearized about $m^z_j\sim \sigma$ under $\text{X}$ BCs. Adapting this notation to the Lindbladian case, we always have $\Delta_\text{X}^{(+1)} = -\Delta_\text{X}^{(-1)}$ therein. While we forgo a detailed explanation of the origin of this difference here, we conjecture that it is a consequence of the intrinsic dependence of the dissipative dynamics on the coherent dynamics in the LLG formalism and, more generally, the aforementioned structural distinction between the classical LLG and the quantum Lindblad master equation. As we show next, the differences in the emerging dynamical behaviors are enhanced in the presence of a strong external drive, leading to a significantly different structure of the nonequilibrium steady-state phase diagram. 

\begin{figure*}[t!]
    \centering
    \includegraphics[width=.95\linewidth]{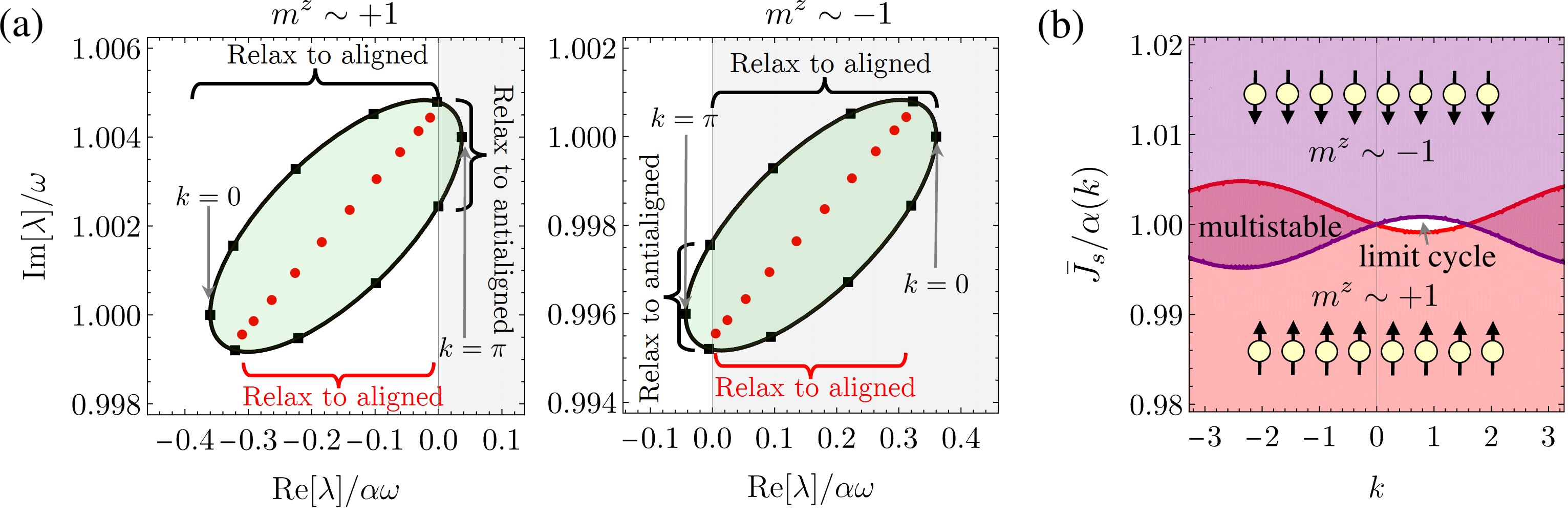}
    \vspace*{-3mm}
    \caption{\textbf{(a)} The rapidities for the linearized LLGS model in Eq.\,\eqref{eq: LLGlinEOM} for $\sigma=+1$ (left) and $\sigma=-1$ (right) under OBCs (red disks) and PBCs (black squares). The dynamical analysis follows exactly as in the Lindbladian model (see the caption of Fig.\,\ref{fig: BulkSS}. The yellow star in the right figure marks a point $i\Omega$ with $\Omega=0.996$, about which the rapidity band winds. The relevant parameters are $\bar{J}=\bar{D}=10^{-3}$, $\alpha=10^{-2}$, $\alpha_\text{nl}=10^{-3}$, $\bar{J}_s = 0.84\alpha$, and $N=10$ layers.
    \textbf{(b)} The steady-state phase diagram for spin waves under PBCs for the LLGS model Eq.\,\eqref{eq: LLGbulkeom}, resulting from Eq.\,\eqref{eq: Jssigmabound}. The red (purple) curve corresponds to $\bar{J}_s^+(k)$ ($\bar{J}_s^-(k)$) in Eq.\,\eqref{eq: Jspm}. The red (purple) region is where the aligned (antialigned) state $\sigma=1$ $(-1)$ is locally stable. The overlapping region corresponds to the multistable regime where both equilibria are attractive, while the excluded white region corresponds to the regime where both are repulsive and an attractive limit cycle is supported. The parameters chosen are the same as in (a), but with $\bar{J}_s$ varied.}
\label{fig: LLGBulkSS}
\end{figure*}

In Fig.\,\ref{fig: LLGBulkSS}(a)-(b) we plot the rapidities for the equilibrium $\sigma$ determined from the classical linearized EOM in Eq.\,\eqref{eq: LLGlinEOM}, for a set of parameters consistent with experimental capabilities. In particular, these example rapidities demonstrate that the LLGS model can support phases whereby OBCs and PBCs have different stability properties for the same set of system parameters, thus opening the door for anomalous transient dynamics.

\subsection{LLG spin-wave steady-state phase diagram}

Following the approach of Sec.\,\ref{sec: SWSSPDLind}, we investigate the nonlinear bulk dynamics of Eq.\,\eqref{eq: approxLLG} by introducing a spin-wave Ansatz of the form:
\begin{align}
\label{eq: LLGswansatz}
 m_j^+(t) = e^{ijk}m_k^+(t),\quad   m_j^z(t) = m^z(t), \quad \forall j. 
\end{align}
The $j$-independent amplitudes $m^z$ and $m_k^+$ then obey the following coupled nonlinear differential equations:
\begin{equation}
\left\{ 
\begin{array}{ll}
    \dot{m}^z &\!= \big(\alpha(k)-\bar{J}_s+\alpha(k)\eta(k)m^z\big)\, |m_k^+|^2, \\
    \\
    \dot{m}^+_k &\!= \big(i+(\bar{J}_s-\lambda(k))m^z -\alpha(k)\eta(k)m^z{}^2\big) \,m_k^+ , 
    \end{array} \right.
    \label{eq: LLGbulkeom}
\end{equation}
in terms of $k$-dependent parameters
\begin{subequations}
\begin{align}
    \alpha(k) &= \alpha + 2\alpha_\text{nl}\cos(k), 
    \label{eq: alphak} 
    \\
    \eta(k) &= 2\lb \bar{J}(1-\cos(k)) - \bar{D}\sin(k)\rb,
    \label{eq: etak}
    \\
    \lambda(k) &= \alpha(k)-i\eta(k).
\end{align}
\end{subequations}
While one may formally interpret $\alpha(k)$ in Eq.\,\eqref{eq: alphak} as a $k$-local decay rate and $\eta(k)$ in Eq.\,\eqref{eq: etak} as a dispersion, the nonlinear dependence on $m^z$ in Eq.\,\eqref{eq: LLGbulkeom} renders this interpretation delicate: comparing directly to Eq.\,\eqref{eq: bulkeom}, the appearance of $m^z$ and $m^z{}^2$ in the equations for $\dot{m}^z$ and $\dot{m}_k^+$, respectively, are key structural differences which alter the stability analysis.

Specifically, upon linearization of Eq.\,\eqref{eq: LLGbulkeom}, the stability condition for the state $m^z(t) = \sigma$ is
\begin{align}
\label{eq: Jssigmabound}
    \big(\bar{J}_s-\alpha(k)\big)\sigma -\alpha(k)\eta(k)<0 .
\end{align}
It follows that stability of the aligned $\sigma=+1$ and antialigned states $\sigma=-1$ is, respectively, guaranteed if 
\begin{subequations}
\label{eq: Jspm}
\begin{align}
    \bar{J}_s &< \bar{J}_s^+(k) = \alpha(k)(1+\eta(k)),
    \\
    \bar{J}_s &> \bar{J}_s^-(k) = \alpha(k)(1-\eta(k)). 
\end{align}
\end{subequations}
In words, $\bar{J}_s^+(k)$ $(\bar{J}_s^-(k)$) sets the upper (lower) bound on the STT below (above) which the spin wave of momentum $k$ relaxes to the aligned (antialigned) state. The dependence of Eqs.\,\eqref{eq: Jspm} upon $\eta(k)$ opens up the possibility for two scenarios not encountered in the Lindblad model, namely, both $\sigma=+1$ and $\sigma=-1$ can be either (i) stable or (ii)  unstable. In Fig.\,\ref{fig: LLGBulkSS}(b), the full steady-state diagram is shown for representative parameters. In case (i), which equilibrium a given spin wave relaxes to depends on its initial amplitude $m^z(t=0)$, with values closer to $m^z=+1$ relaxing to $1$, and vice-versa. That is, case (i) supports {\em dynamical multistability}. In case (ii), one may show that the system develops an {\em attractive limit cycle} with a finite canting angle $|m^z_\text{LC}|\neq 1$. As it turns out, this regime hosts \textit{spin-superfluid} behavior which is explored in detail in Ref.\,\cite{LCpaper}. Henceforth, we will work in parameter regimes where {\em only one} of the two equilibria are stable, i.e., outside of the cases (i) and (ii).

\subsection{Anomalous transient LLGS dynamics}

In this section, we show that the anomalous transient dynamics and topologically mandated edge modes uncovered within the Lindbladian framework (Secs.\,\ref{sec: anomLind}-\ref{sec: LindDB}) persist within the LLGS formalism under experimentally realistic parameter regimes. In particular, our analysis predicts the emergence of:

\begin{itemize}
\item[(i)] Anomalous relaxation -- a two-step decay of the magnetization, as in Fig.\,\ref{fig: LindAR}(a).
\item[(ii)] Spin-dipping -- a sudden drop in the magnetization of one or more layers, despite near the aligned equilibrium state, as shown in Fig.\,\ref{fig: SpinDip}.
\item[(iii)] Transient attractivity of an unstable equilibrium, as in Fig.\,\ref{fig: FalseEq}.
\item[(iv)] A long-lived edge state, analogous to the DB showcased in Fig.\,\ref{fig: DBIC_Life}.
\end{itemize}

\noindent 
In each case, the effect becomes more pronounced with increasing layer number -- providing a clear ``smoking gun'' signature of dynamical metastability and the associated anomalous relaxation discussed in Sec.\,\ref{sec: DM} and \ref{sec: Lind} in the Lindbladian setting.

Observation of anomalous dynamics is ultimately a question of experimental feasibility rather than theoretical possibility. The LLGS model already exhibits the nonreciprocity (Eq.\,\eqref{eq: LLGNR}) needed to generate a mismatch between the OBC and PBC rapidity spectra (Fig.\,\ref{fig: LLGBulkSS}(a)). In the Lindbladian case, these dynamics arise by initializing the spins in a pseudoeigenvector state whose pseudoeigenvalue lies well outside the exact linear spectrum. The accuracy $\epsilon$, which is inversely proportional to the anomalous transient timescale,  is controlled by a number of factors. In particular, $\epsilon = \epsilon_N$ decreases with system size (recall, e.g., Eq.\,\eqref{eq: bkEOM} for the special case of a spin-wave initial condition with $\epsilon_N\sim 1/\sqrt{N}$). In principle, the transient window can be made parametrically long by increasing the system size. In multilayer stacks, however, the accessible layer number is limited by fabrication and materials constraints; thus, realistically, the observable transient remains finite (i.e., $N_\text{max} \lesssim 50$ \cite{MultilayerBook2013}).

For conservatively small systems (say, $N\leq 10$), the finite-size accuracy $\epsilon_N$ may be too coarse to cleanly resolve the anomalous regime. To test this potential limitation, we evolve a spin-wave initial condition
\begin{align}
\label{eq: LLGswIC}
   m_j^+(0) = e^{ikj} m_k^+(0), \quad m^z_j(0) = m^z(0),
\end{align}
under the LLGS equation in Eq.\,\eqref{eq: approxLLGdiml} and OBCs. For $m^+(0)$ sufficiently small, with $m^z(0)\sim \sigma$, the dynamics are approximately linear. Since OBCs are imposed, $\phi_k(0) = [e^{ik},\ldots,e^{iNk}]m_k^+(0)$ is only an approximate eigenvalue of $\mathbf{A}^\sigma$, with eigenvalue $$A(k) = i+(\bar{J}_s-\lambda(k))\sigma - \alpha(k)\eta(k),$$ as obtained from Eq.\,\eqref{eq: LLGbulkeom} by setting $m^z=\sigma$. Concretely, we estimate
\begin{align}
     \norm{ \big(\mathbf{A}^\sigma-A(k)\big)\phi_k(0)} \approx \sqrt{\frac{2(\bar{J}^2+\bar{D}^2)}{N}} + \mathcal{O}(\alpha^2),
\end{align}
which provides an upper bound on $\epsilon_N$. This sets a rough estimate for the anomalous time scale of $\omega t_\text{anom} \sim 1/\epsilon_N$, which, for $\omega=1$ GHz, and $\bar{J}=\bar{D}=10^{-3}$, evaluates to $t_\text{anom} \sim \sqrt{N}\cdot 0.5$ $\mu$s. Crucially, dynamics on this timescale are readily resolvable with established probes \cite{AMRExp_Schumacher2007,Moke_Kirilyuk2010,BarmanMagReview2020}; this is essential to ensure that finite-size inaccuracy does not obscure the anomalous size-dependent phenomena of interest.

In what follows, we will use the same parameters as in Fig.\,\ref{fig: LLGBulkSS}(a), while varying $\bar{J}_s$ accordingly to adjust the dynamical phase. We will plot all quantities versus the rescaled dimensionless time $\tau' = \alpha\tau = \alpha\omega t \equiv t/ \tau_\text{LLG}$, where $\tau_\text{LLG} \equiv 1/(\alpha\omega)$ sets the characteristic decay time for the LLG dynamics; with the parameters we consider (with $\omega = 1$ GHz), we have $\tau_\text{LLG} =0.1\,\mu$s. While other criteria may be envisioned, with these choices any behavior persisting for a time $t > \tau_\text{LLG}$  ($\tau' > 1$) remains well within the resolution of standard all-electrical or time-resolved pump-probe measurement schemes \cite{AMRExp_Schumacher2007, Moke_Kirilyuk2010}.

\subsubsection{Anomalous relaxation under LLGS dynamics}

Following a similar progression to the one of the interacting spin Lindbladian, we first explore under which conditions anomalous relaxation and dynamical metastability behaviors may emerge within the LLGS framework. By virtue of Eq.\,\eqref{eq: LLGlinEOM}, note that the STT contribution enters the dynamical matrix as an identity shift, namely, $\mathbf{A}^\sigma = \sigma\bar{J}_s \mathds{1}_N + \mathbf{A}^\sigma|_{\bar{J}_s=0}$. Since changing $\bar{J}_s$ simply translates the rapidities along the real axis and shown in Fig.\,\ref{fig: LLGBulkSS}(a), we may decrease $\bar{J}_s$ so that both equilibria are stable under both OBCs and PBCs. The nonreciprocity is not affected by this shift and the stability-gap disagreement persists. By tuning $\bar{J}_s$, we can arrange for $\Delta_\text{OBC}^{(+1)}<\Delta_\text{PBC}^{(+1)}\lesssim 0$, so that the decay timescale  of the $k=\pi$ mode near the aligned state under PBCs, i.e., $1/|\Delta_\text{PBC}^{(+1)}|$, is longer than the anomalous transient timescale. At the same time, we can ensure $\Delta_\text{OBC}^{(-1)}>\Delta_\text{PBC}^{(-1)}\gtrsim 0$, so that the amplification timescale  of the $k=\pi$ mode near the antialigned state under PBCs, i.e., $1/|\Delta_\text{PBC}^{(-1)}|$, is {\em also} longer than the anomalous transient timescale. 

\begin{figure}[t!]
    \centering
    \includegraphics[width=\linewidth]{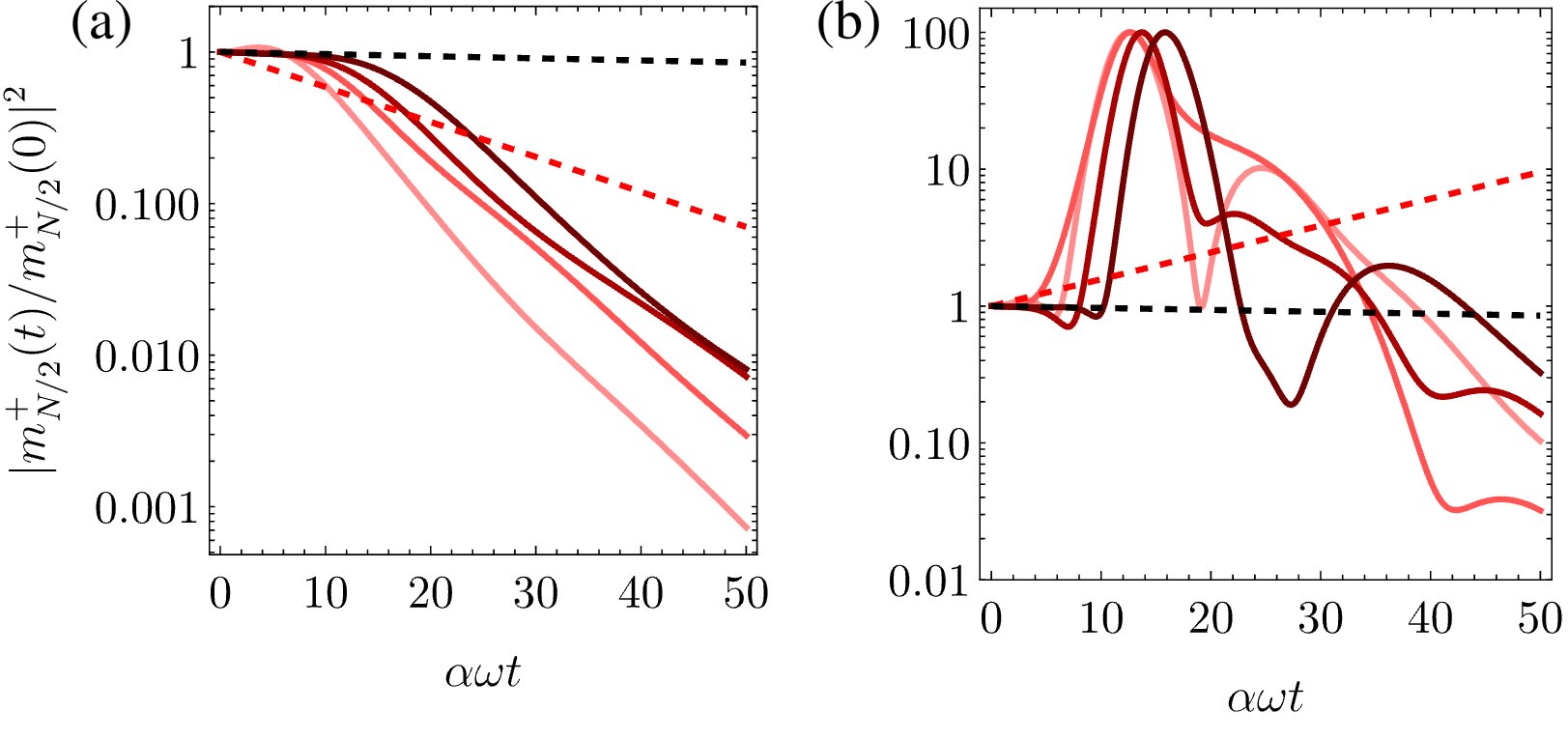}
    \vspace*{-2mm}
    \caption{LLGS nonlinear spin dynamics from Eq.\,\eqref{eq: approxLLGdiml} within the linearized anomalously relaxing regime. The dynamics of a bulk spin at site $j=N/2$ for an initial condition {\bf (a)} near equilibrium $(m^z(0) = \sqrt{1-0.1^2}$) and {\bf (b)} far from equilibrium $(m^z(0) = -\sqrt{1-0.1^2}$) are depicted. In (a), the slope of the black (red) dashed line is the PBC (OBC) characteristic decay rate (i.e., the stability gap) $\Delta_{\text{PBC}}^{(+1)}$ ($\Delta_{\text{OBC}}^{(+1)}$). The same is true in (b); however, the stability gaps are now computed with respect to the unstable antialigned state. In both (a) and (b), the parameters are the same as in Fig.\,\ref{fig: LLGBulkSS}(a) but with $\bar{J}_s = 0.8\alpha$. The different curves correspond to $N=4,6,8$, and $10$ with darker color corresponding to more layers.
}
    \label{fig: LLGAR}
\end{figure}

Concretely, we choose $\bar{J}_s = 0.8\alpha$, initialize the OBC system in the  spin-wave mode corresponding to $k=\pi$ in Eq.\,\eqref{eq: LLGswIC}, and plot the resulting dynamics of the bulk spin at site $j=N/2$ in Fig.\,\ref{fig: LLGAR}. In (a), we situate the spins near the aligned state $m^z_j\sim +1$ and increase the system size from $N=4$ to $10$, with the darker curves corresponding to more layers. Evidently, the spins follow the bulk decay rate $\Delta_\text{PBC}^{(+1)}<0$ (black dashed line) for a time increasing with $N$. After this system-size-dependent transient period, they relax at the rate set by the true OBC stability gap $\Delta_\text{OBC}^{(+1)}<0$ (red dashed line). In (b), the spins are instead initialized near the antialigned state $m_j^z\sim -1$. As in the Lindbladian case (recall Fig.\,\ref{fig: LindAR}(b)), the spins initially amplify at the rate set by the bulk stability gap $\Delta_\text{PBC}^{(-1)}>0$ (black dashed line), again over a timescale increasing with $N$. They next undergo a rapid amplification followed by a rapid decay, ultimately relaxing to the aligned equilibrium. The amplification and decay are {\em significantly} faster than those set by the OBC stability gaps $\Delta_\text{OBC}^{(\sigma)}$: as we argued in Sec.\,\ref{sec: LindAR}, we attribute this to a intrinsically nonlinear effect.

\subsubsection{Dynamical metastability under LLGS dynamics}

Continuing along the lines of the previous discussion, we can now set the STT strength to the same value used in Fig.\ref{fig: LLGBulkSS}(a) (namely, $\bar{J}_s = 0.84\alpha$), so that the OBC system unconditionally relaxes to the aligned state, whereas the PBC system can relax to either the aligned or antialigned state, depending on the initial condition. That is, the linearized dynamics are dynamically metastable.

\begin{figure}[t]
    \centering
    \includegraphics[width=\linewidth]{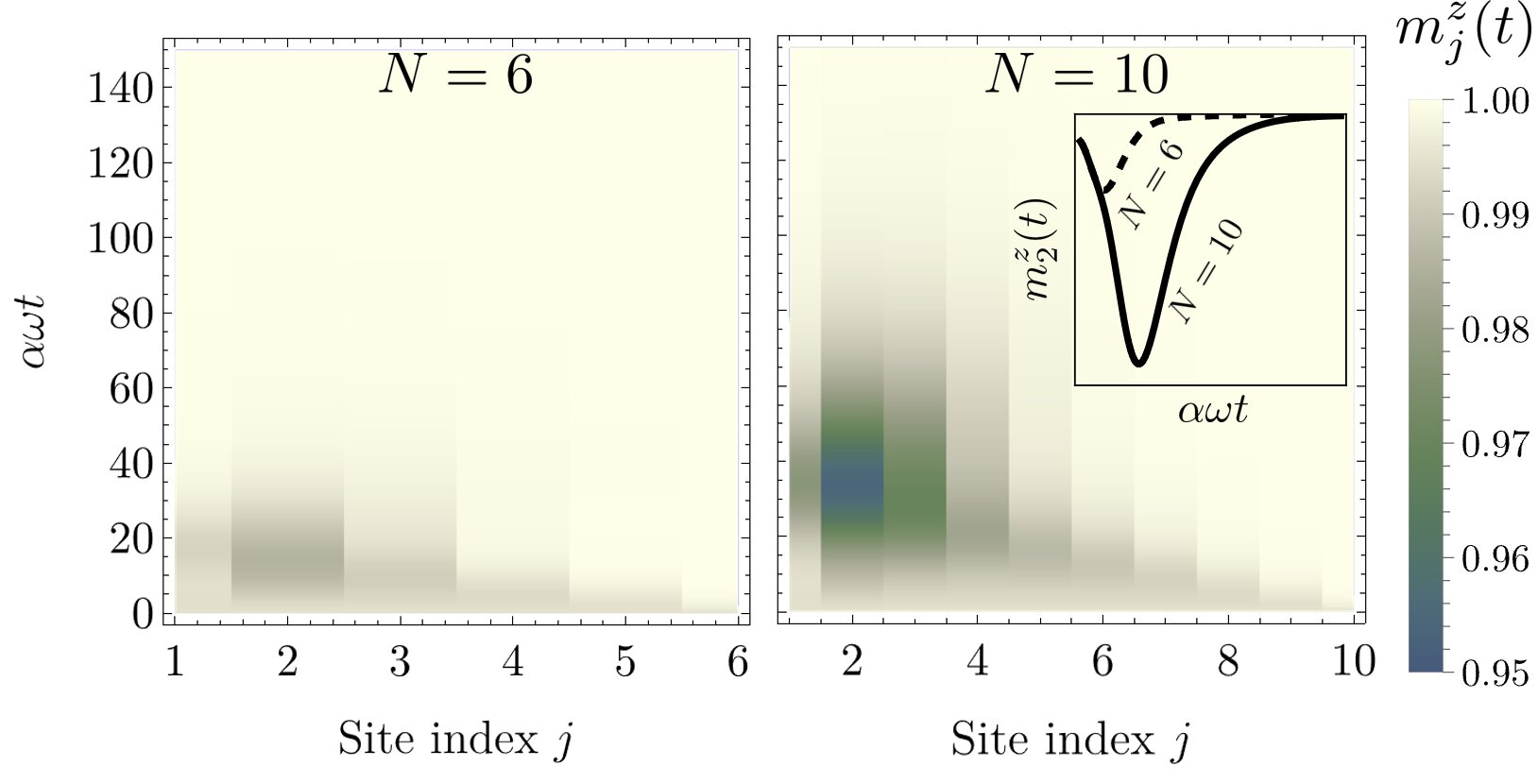}
    \vspace*{-3mm}
    \caption{LLGS nonlinear spin dynamics from Eq.\,\eqref{eq: approxLLGdiml} within the dynamically metastable regime. The evolution of $m_j^z$ is computed as a function of time, starting from an initial condition prepared in the $k=\pi$ mode as in Eq.\,\eqref{eq: LLGswIC} with $k=\pi$, near the stable aligned equilibrium, $m_j^z(0) = \sqrt{1-0.1^2} \sim 1$. Despite the initial proximity to equilibrium, the spins are transiently attracted to the unstable antialigned state, due to the transiently amplifying nature of the $k=\pi$ mode. The effect is significantly more pronounced for larger $N$, with the left figure depicting $N=6$ and the right figure depicting $N=10$. The inset for $N=10$ depicts the dynamics of $m_2^z$ over the same time range as the main plot's vertical axis, with the vertical scale of the inset restricted to $0.95 \leq m_2 \leq 1$. In both cases, parameters are chosen to match those of Fig.\,\ref{fig: LLGBulkSS}(a).}
    \label{fig: LLGSpinDip} 
\end{figure}

First, we again initialize the OBC system in the $k=\pi$ spin-wave state (Eq.\,\eqref{eq: LLGswIC}) with $m_j^z\sim 1$. If the system were under PBCs, this mode would relax to the antialigned state. Thus, we expect to see a transient reduction of the $z$ spin component, whose duration and degree increases with system size. Indeed, this is observed in Fig.\,\ref{fig: LLGSpinDip}, wherein $m_j^z(t)$ is density-plotted as a function of $j$ and $t$ for $N=6$ and $N=10$. There is a clear spin-dipping localized around site $j=2$, with the effect roughly doubling when the number of layers is increased from $N=6$ to $N=10$.

\begin{figure}[t]
    \centering
    \vspace*{-3mm}
    \includegraphics[width=\linewidth]{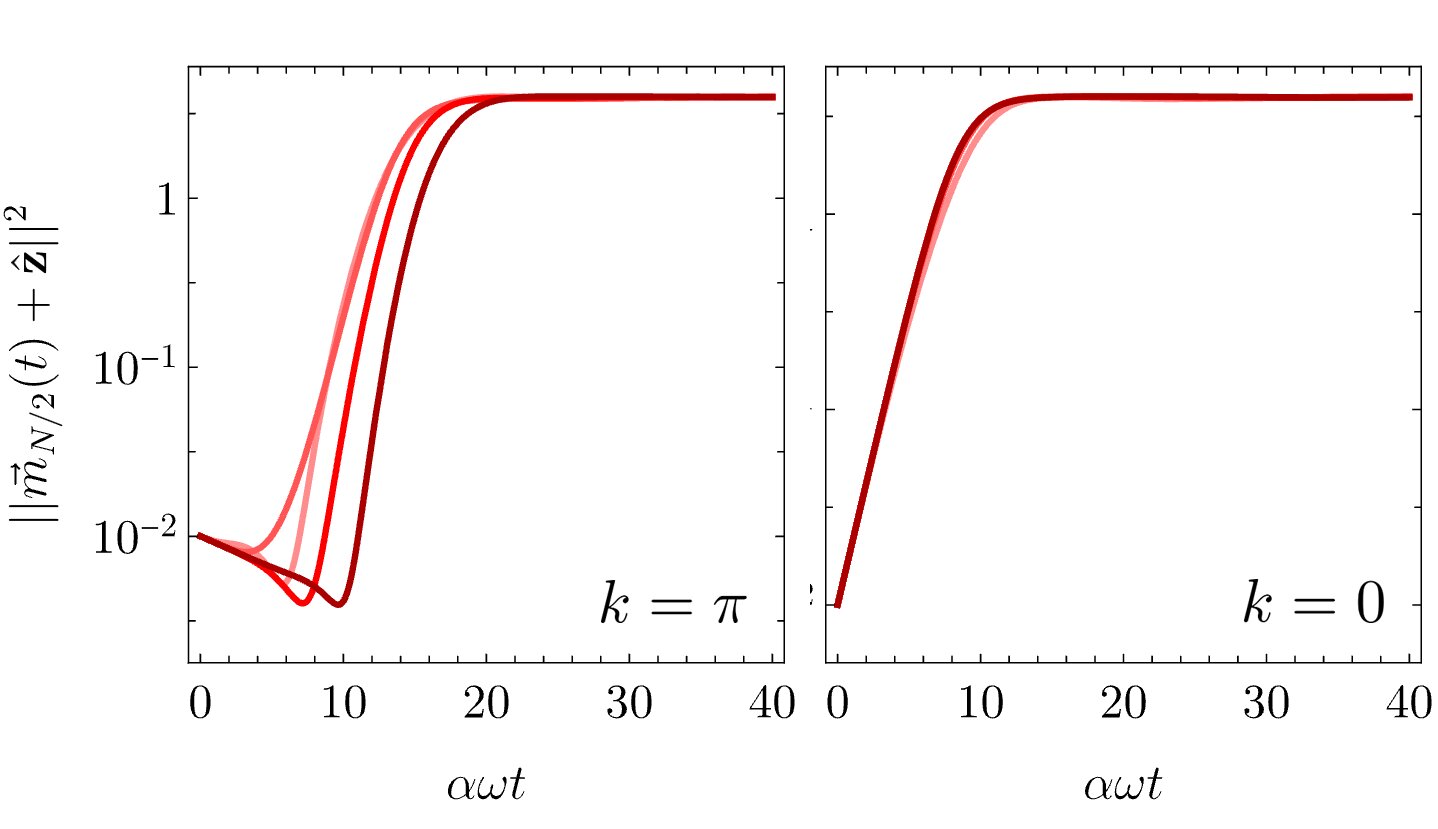}
    \caption{Deviation of the bulk spin vector $\vec{m}_{N/2}$ from the antialigned configuration $-\hat{\mathbf{z}}$ for the $k=\pi$ mode (left) and the $k=0$ mode (right), resulting from the LLGS nonlinear dynamics in Eq.\,\eqref{eq: approxLLGdiml} when in both cases the initial state is prepared close to the antialigned state, $m^z(0) = -\sqrt{1-0.1^2}$. System parameters are the same as in Fig.\,\ref{fig: LLGBulkSS}(a). }
    \label{fig: LLGFalseEq}
\end{figure}

Suppose we now instead take $m_j^z\sim -1$, and consider the $k=0$ spin-wave. Under PBCs, the $k=\pi$ mode would rapidly decay to the antialigned state, while the $k=0$ mode would amplify to the aligned state (see the right plot of Fig.\,\ref{fig: LLGBulkSS}(a)). Imposing OBCs and evolving these modes, we find the behavior shown in Fig.\,\ref{fig: LLGFalseEq}. The $k=\pi$ mode exhibits a {\em transient decay} to the unstable (under OBCs) antialigned equilibrium, with the duration and degree of the decay increasingly monotonically with system size. On the contrary, the $k=0$ mode displays a {\em system-size independent amplification} towards the stable (under OBCs) aligned equilibrium. As it is clear by comparing Fig.\,\ref{fig: LLGSpinDip} and Fig.\,\ref{fig: LLGFalseEq} to their Lindbladian counterparts, Fig.\,\ref{fig: SpinDip} and Fig.\,\ref{fig: FalseEq}, respectively, the qualitative behaviors that emerge are remarkably similar.

\subsubsection{Dirac bosons under LLGS dynamics}

As a final point of comparison, we exhibit
a long-lived edge state which constitutes the classical magnetization analogue of the DBs discussed in Sec.\,\ref{sec: frameTDM} and discussed for the nonlinear semiclassical dynamics of the Lindbladian model in Sec.\,\ref{sec: LindDB}. With reference to the yellow star in the right plot of Fig.\,\ref{fig: LLGBulkSS}(a), the rapidity band for the linearization about $m^z\sim -1$ winds around the point $i\Omega$ with $\Omega = 0.996$, implying that there exists a pseudoeigenvector $\vec{w}$ of $\mathbf{A}^{(-1)}$ corresponding to pseudoeigenvalue $i\Omega$. Thus, in a frame rotating at $\Omega$, the initial condition
\begin{align}
\label{eq: LLGDBICs}
    m_j^+(0) \equiv  \rho w_j,\quad m_j^z(0) \equiv - \sqrt{1-|\rho w_j|^2},
\end{align}
with $\rho$ sufficiently small, should remain frozen over a significant timescale. In the physical frame, this corresponds to an \textit{antialigned generalized DB rotating at $\Omega$}. As we noted in discussing the linearized LLGS dynamics (Sec.\,\ref{sub: LLLG}), the rapidity bands for the aligned and antialigned states are not related in a simple way, unlike in the Lindbladian scenario, but they rather enclose distinct intervals of the imaginary axis. For a topologically metastable system, this implies that each equilibrium supports families of generalized DBs with distinct characteristic frequencies. For simplicity, we continue to focus on the antialigned DB rotating at $\Omega$.

\begin{figure}[t]
    \centering
    \includegraphics[width=\linewidth]{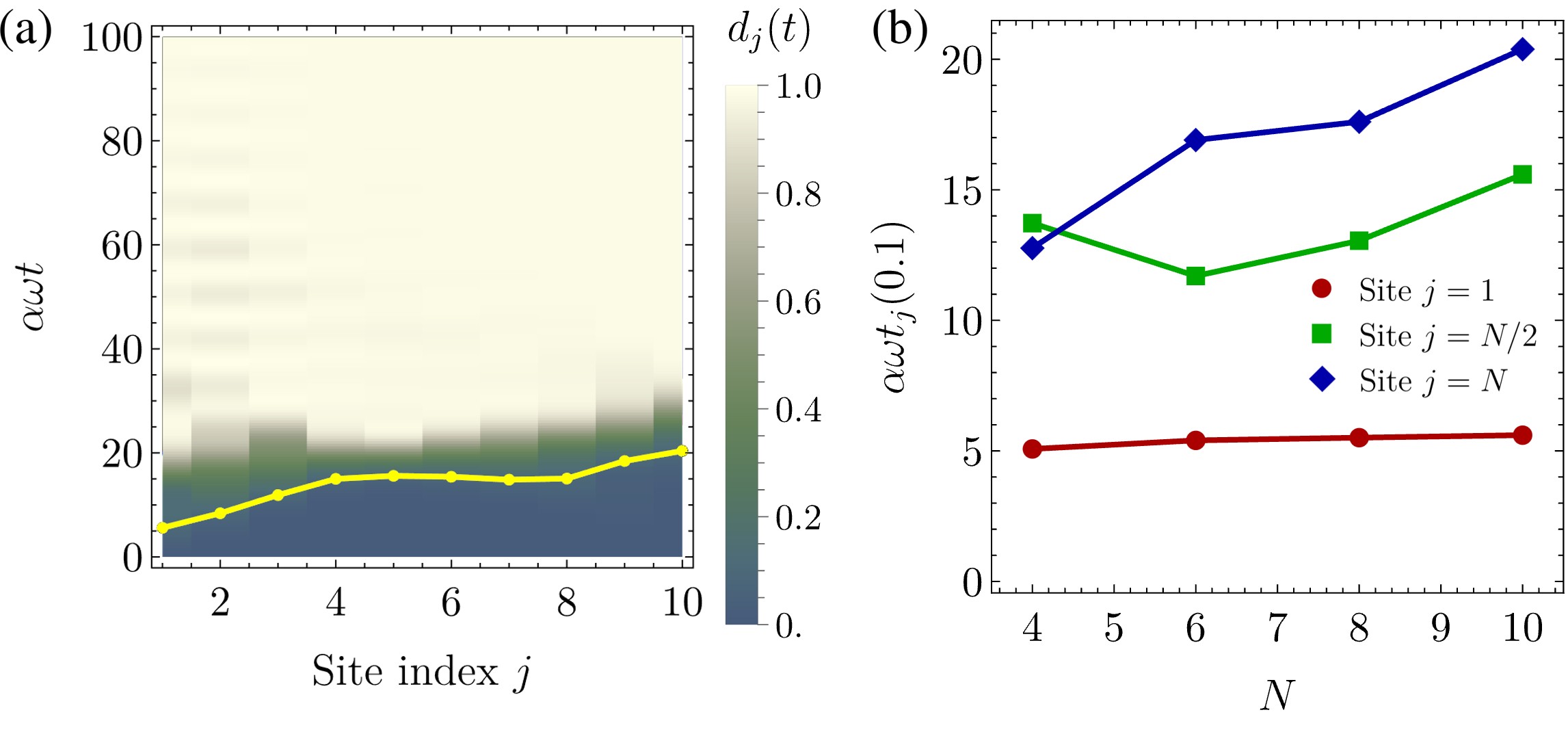}
    \vspace*{-3mm}
    \caption{ \textbf{(a)} Dynamics of the deviation measure in Eq.\,\eqref{eq: dj}, adapted for classical magnetizations $\vec{s}_j\mapsto \vec{m}_j$, for the initial condition Eq.\,\eqref{eq: LLGDBICs} evolving under the simplified LLGS dynamics in Eq.\,\eqref{eq: approxLLGdiml} for $N=10$ ferromagnetic layers in a frame rotating at $\Omega=0.996$ (within which the system is topologically metastable - see Fig.\,\ref{fig: LLGBulkSS}(a)) As in Fig.\,\ref{fig: DBIC_Life}(a) and (c), the yellow line indicates the point at which the deviation measure exceeds $0.1$. The yellow dots indicate the exact value for each layer $j=1,\ldots,10$. \textbf{(b)} The lifetimes for the magnetizations at sites $j=1$, $N/2$, and $j=N$ for the evolution considered in (a), versus system size.  }
    \label{fig: DBLLG} 
\end{figure}

In Fig.\,\ref{fig: DBLLG}, we plot a deviation measure similar to the one in Eq.\,\eqref{eq: dj} for the DB initial condition in Eq.\,\eqref{eq: LLGDBICs} evolving under LLGS dynamics, along with the corresponding lifetimes of layers $j=1$, $N/2$, and $N$ for $N=4,6,8,$ and 10. Again, the overall trend remains qualitatively the same as in the Lindbladian model (compare to Figs.\,\ref{fig: DBIC_Life}(c)-(d)). In particular, in Fig.\,\ref{fig: DBLLG}(a), we observe, for an $N=10$ layer system, an initial transient of meaningful length during which the magnetizations remain {\em effectively frozen}, with maximal lifetime at site $j=10$. Following this transient, the spins relax to the expected equilibrium $m_j^z = +1$. In (b), we note a general increase in lifetime with system size, with the magnetization at site $j=10$ nearly doubling from $N=4$ to $N=10$. While the idealized linear increase seen in Fig.\,\ref{fig: DBIC_Life}(d) is not as cleanly present here, we have verified that it becomes manifest for larger system sizes exceeding our conservative upper bound $N\sim 10$ [data not shown]. This can be explained simply as a small-system size effect: topologically metastable (and dynamically metastable, more generally) phenomena are naturally most pronounced in large systems.

\section{Conclusion and outlook}
\label{sec: end}

In this work, we have developed a unified framework for dynamical metastability beyond the linear, noninteracting regime, with a focus on magnetic systems -- thereby bridging open-quantum spin dynamics and classical magnetization dynamics relevant to experimental spintronics setups. Building on the notion that long-lived transient behavior can originate from the spectral geometry and topology of a non-Hermitian evolution operator rather than from energetic barriers and timescales separations, we have shown that this mechanism persists -- and is {\em qualitatively enriched} -- once interactions and nonlinearities are fully accounted for in the underlying quantum or classical dynamical model, respectively. 

Starting from a fully quantum description of Markovian driven-dissipative dynamics, we have introduced an interacting spin Lindbladian, whose linearized magnon dynamics realize a dynamically metastable Hatano-Nelson chain -- thereby featuring anomalous relaxation, transient amplification, and topologically mandated Dirac boson edge modes. Crucially, we showed that dynamical metastability serves as a natural organizing principle that persists beyond the linear regime. While the linearized dynamics already encode boundary-sensitive amplification and topologically mandated edge structure, we demonstrate that these spectral features shape the full nonlinear evolution: in the semiclassical regime they seed distinctly nonlinear phenomena -- including system-size-dependent spin dipping and transient attraction to unstable equilibria. A central result of our analysis is that {\em topological metastability} also survives beyond quadratic dynamics, albeit in a modified form. While the lifetime of Dirac-boson zero modes is ultimately limited by nonlinear effects, we identified classes of edge-localized dynamical modes -- rooted in approximate displacement symmetries derived in the linear theory -- 
that retain parametrically (in system size) long lifetimes and even persist in the presence of disorder.

By extending our analysis to LLGS dynamics of magnetic multilayers, we established that all nonlinear incarnations of dynamical metastability found in the Lindbladian model reappear in realistic magnetic heterostructures. In this setting, DMI, nonlocal damping, and spin-transfer torque emerge as experimentally tunable knobs that control nonreciprocity, band topology, and bulk-boundary stability mismatch. At the same time, the LLGS framework admits additional nonlinear phenomena, such as multistability and limit cycles, that are absent in the Lindbladian description -- highlighting a clear separation between quantum-dissipative and classical nonlinear effects. Taken together, these results constitute the first systematic demonstration that dynamical metastability is a robust and experimentally relevant organizing principle for nonlinear magnetization dynamics.

Looking ahead, several directions naturally emerge from this investigation. First, it would be valuable to extend the present analysis beyond one dimension, where non-Hermitian topology supports richer structures, including higher-order skin effects and corner-localized modes \cite{Kawabata2020,Zhang2022,FlebusNHSEMag2022,Hou2024} as well as other exotic size-dependent dynamical behavior \cite{Yikang}. Understanding how such features interplay with nonlinear spin dynamics could uncover new forms of transient localization and amplification. Second, magnetization dynamics in heterostructures is routinely probed using established optical, microwave, and electrical techniques -- including Brillouin light scattering, broadband ferromagnetic resonance, magnetization-noise and propagating spin-wave spectroscopy~\cite{FlebusMagRoad2024}, all-electrical detection schemes based on anisotropic magnetoresistance on microsecond timescales~\cite{AMRExp_Schumacher2007}, and time-resolved MOKE pump-probe measurements~\cite{Moke_Kirilyuk2010}. While these techniques can in principle access the timescales in which the novel phenomena we predict are expected to manifest, extending our analysis of disorder effects to account for device-specific disorder and non-idealities is likely a necessary step to quantitatively test our predictions in a real-world setting. Third, the coexistence of dynamical metastability with limit cycles in the LLGS setting suggests a deeper connection between non-Hermitian topology, nonlinear attractor structure, and even spin-superfluidity \cite{LCpaper} -- potentially linking transient topological modes to synchronization and pattern formation in driven-dissipative quantum networks \cite{BucaSynch,Tancredi,zhao2025}. 

From a fundamental open-system perspective, the key structural differences that our side-by-side analysis reveals  between the Lindbladian and the LLGS description of the magnetization dynamics -- both in the nature of their equilibria and the resulting nonlinear dynamics -- reinforce the need for further understanding in recasting the phenomenological LLGS approach within a consistent many-body quantum framework -- in line with ongoing efforts \cite{Wieser2013,BranislavLLGLind2024,BranislavKeldyshLLG2024,WatanabeLLGLind2025,UhrigLLGLind2025}. We thus expect our study to naturally spur new investigations along these lines and anticipate that dynamical metastability will play an important role not only in spintronics and magnonics, but also in a wider class of driven-dissipative systems where topology, interactions, and nonlinear dynamics intersect.

\acknowledgements 

It is a pleasure to thank Michiel Burgelmann and Xin Li for stimulating discussions. 
Work at Boston College was supported by the Office of Naval Research under Award No.\,N000142412427.
Work at Dartmouth was supported by the National Science Foundation through award No.\,PHY-2412555.


\appendix

\section{Localization length and lifetime \\ of Dirac bosons}
\label{app: DBlife}

In this appendix, we provide more details on the localization lengths and lifetimes of DBs in topologically metastable, U(1)-symmetric QBLs. For any such system, the array $\phi = [a_1,\ldots,a_N]^T$ satisfies Eq.\,\eqref{eq: phiEOM} for some $N\times N$ dynamical matrix $\mathbf{A}$. In general, $\mathbf{A}$ has no particular structure, unless other symmetries are imposed. Here, we consider systems that are translation-invariant (disorder-free), up to boundaries. As mentioned in the main text, his forces $\mathbf{A}$ to be a Toeplitz matrix, in the case of OBCs, and a circulant matrix, in the case of PBCs. We recount basic properties of these classes of matrices. All such properties can be found in Chapter II of Ref.\,\cite{TrefethenPS2005}.

\subsection{Basic properties of Toeplitz and \\ circulant matrices}

Toeplitz and circulant matrices are most easily defined by introducing appropriate {\em left-shift operators} $\mathbf{T}$ and $\mathbf{V}$, which acts on the one-dimensional lattice as
\begin{align*}
    \mathbf{T}\vec{e}_j \equiv \begin{cases}
\vec{e}_{j-1}, & j>1,
\\
0, & j=1,
    \end{cases} \qquad  \mathbf{V}\vec{e}_j \equiv \begin{cases}
\vec{e}_{j-1}, & j>1,
\\
\vec{e}_N, & j=1,
    \end{cases}
\end{align*}
where $\vec{e}_j$ is the $j$'th standard basis vector. The shift operator $\mathbf{T}$ intrinsically takes into account OBCs and represents a left-shift on a finite chain, whilst $\mathbf{V}$ encodes PBCs and left-shifts on a finite ring. We then have
\begin{align}
\label{eq: gentoe}
    \mathbf{A} = A_0\mathds{1}_N + \begin{cases}
        \sum_{r=1}^R\lp A_r \mathbf{T}^r + A_{-r} \mathbf{T}^\dag{}^r\rp, & \text{OBCs},
        \\
        \sum_{r=1}^R\lp A_r \mathbf{V}^r + A_{-r} \mathbf{V}^\dag{}^r\rp, & \text{PBCs},
    \end{cases}
\end{align}
where the coefficients $A_r\in\mathbb{C}$ describe the couplings between sites $j$ and $j+r$, up to BCs, and $R$ is the maximum range of interaction, which is assumed to be finite. For example, for a system with at most nearest-neighbor interactions, $R=1$. We will also always assume $R<N/2$ \cite{GBTJPA2017}, so that there is a meaningful notion of a ``bulk site". We note that this formulation allows us to easily identify, for instance, nonreciprocal systems as those in which $|A_r| \neq |A_{-r}|$, for some interaction distance $r$.

The {\em symbol} of a Toeplitz/circulant matrix $\mathbf{A}$ is defined by the complex-valued function
\begin{align*}
    A(z) \equiv  A_0 + \sum_{r=1}^R\lp A_r z^r + A_{-r} z^{-r}\rp,
\end{align*}
which is trivially obtained from Eq.\,\eqref{eq: gentoe} by replacing $\mathbf{T}$ (or $\mathbf{V}$) with $z$ and $\mathbf{T}^\dag$ (or $\mathbf{V}^\dag$) with $z^{-1}$. 
The utility of this function becomes evident by considering the action of $\mathbf{A}$ on the  Bloch states,
\begin{align*}
    \vec{v}_k = \sum_{j=1}^N e^{ikj} \vec{e}_j.
\end{align*}
One may verify that, if $k = 2m\pi/N$ for some integer $m$, then, under PBCs, $\mathbf{A}\vec{v}_{k} = A(e^{ik})\vec{v}_{k}$. That is, Bloch states are eigenstates of $\mathbf{A}$ under PBCs, with eigenvalues $A(e^{ik})$. We recognize $A(k) \equiv A(e^{ik})$ as the rapidity band introduced in Sec.\,\ref{sub: ARDM}. In general, $A(k)$ traces out a closed curve in the complex plane as $k$ varies by $2\pi$. This allows one to define, for any number $\lambda\in\mathbb{C}$ {\em not} in the rapidity band, the {\em winding number}
\begin{align*}
    \nu_\lambda \equiv \frac{1}{2\pi i}\oint_{|z|=1}\frac{A'(z)\,dz}{A(z) - \lambda}.
\end{align*}
For our purposes, the winding number is useful for identifying pseudoeigenvalues and pseudoeigenvectors. Theorem 7.2 in Ref.\,\cite{TrefethenPS2005} implies that, if $\nu_\lambda \neq 0$, then, for all sufficiently large $N$, there exists a $\delta<1$ such that 
\begin{align*}
    \norm{(\mathbf{A} - \lambda)\vec{v}} < \delta^N, 
\end{align*}
for some normalized vector $\vec{v}$ \footnote{Note that, due to our slightly different definition of the symbol compared to Ref.\,\cite{TrefethenPS2005}, our winding number sign conventions are the opposites of one another. We also take $\delta = 1/M$, where $M$ is defined in the actual theorem statement.}
and
\begin{align*}
    |v_j| \leq \begin{cases}
        \delta^{j}, & \nu_\lambda >0,
        \\
        \delta^{N-j}, & \nu_\lambda <0.
    \end{cases}
\end{align*}
That is to say, {\em any} $\lambda$ enclosed by the rapidity band is an exponentially (in system size) accurate pseudoeigenvalue, and the sign of the winding number determines which edge the corresponding pseudoeigenvectors are localized. More importantly, $\delta$ can be chosen to be any number such that $A(z) \neq \lambda$ inside the annulus $\delta \leq |z| \leq 1$ for $\nu_\lambda >0$, or $1 \leq |z| \leq 1/\delta$ for $\nu_\lambda <0$. We can then define a characteristic {\em localization length} of as follows:
\begin{equation}
\xi \equiv \frac{1}{\ln(1/\delta)}\;\Rightarrow\;\delta = e^{-1/\xi}.
\label{eq: xi}
\end{equation} 
Thus, $\xi$ is minimized by taking the minimal $\delta$ satisfying the above stated conditions.

\subsection{Localization length of the Dirac bosons \\ in the 
Hatano-Nelson Lindbladian}
\label{app: locl}

We can specialize the above results to estimate the localization length of the DBs supported by the dissipative HN chain described by the QBL in Eqs.\,\eqref{eq: HNHam}-\eqref{eq: HNdiss}. Recall that the zero mode DB comes from the zero pseudoeigenvector of $\mathbf{A}^\dag$, while the approximate symmetry generator comes from the zero pseudoeigenvector of $\mathbf{A}$. The corresponding symbols (in the appropriate rotating frame) have the simple form
\begin{align*}
    A^\dag(z) &= -\kappa_\text{eff} + J_R^* z + J_L^* z^{-1},
    \\
    A(z) &= -\kappa_\text{eff} + J_L z + J_Rz^{-1}.
\end{align*}
In particular, the winding number of $A^\dag(z)$ about any point has the opposite sign of that of $A(z)$. This leads to each DB being localized on opposite end of the chain.

Since we are interested in approximate zero modes, we consider $\lambda=0$, and thus $|\nu_{\lambda=0}|$ is the number of times $A(k)$ $(A^\dag(k))$ winds around the origin. Focusing on $A(z)$ (the case of $A^\dag(z)$ follows trivially), by Cauchy's Argument Principle, the winding number is the number of zeros of $A(z)$ inside the unit circle, minus the number of poles inside the unit circle. There is one pole at $z=0$. The number of zeros is the same number of zeros as $zA(z)$, which is a quadratic polynomial. Thus, the winding number is only nonzero if both roots (counting multiplicity) are inside or outside the unit circle. Using this, and the fact that $|J_L|^2 > |J_R|^2$ in this model, one may verify that the condition for nonzero winding is 
\begin{align*}
    |\kappa_\text{eff}| < \frac{|J_L|^2 - |J_R|^2}{|J_L - J_R^*|},
\end{align*}
or, in terms of the system parameters,
\begin{align}
\label{eq: windcond}
    |\kappa_ - + 2\Gamma - \kappa_+ |< \frac{2\Gamma D}{|J+iD|}.
\end{align}

\begin{figure}[t]
    \centering
    \includegraphics[width=0.65\linewidth]{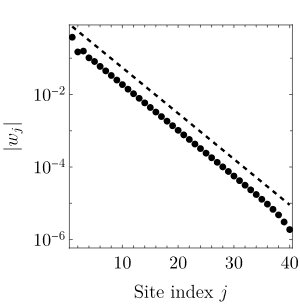}
    \vspace*{-3mm}
    \caption{Modulus of  the coefficients $w_j$ of the DB $\beta$ [Eq.\,\eqref{eq: DBbeta}] obtained from the topologically metastable HN QBL using the parameters in Fig.\,\ref{fig: HNSpecTDM}. The dashed line corresponds to the curve $e^{-1/\xi}$, with the localization length $\xi$ obtained from Eq.\,\eqref{eq: xi}.}
    \label{fig: DBloc}
\end{figure}

Assuming that the nonzero winding condition is met, let $z_+$ and $z_-$ denote the two poles of $A(z)$ inside the unit circle. We can then take $\delta = \max(|z_\pm|)$, since this defines the the maximal inner radius of the largest annulus $\delta < |z| < 1$, such that $A(z) \neq 0$. This is difficult to obtain analytically, in general, but can be accomplished easily for any given set of numerical parameters. For instance, for the topologically metastable parameters used throughout the text (given in the caption of Fig.\,\ref{fig: HNSpecTDM}), we find $\xi \sim 3.4$. Physically, this means that the (approximate) symmetry generator $\beta$ is localized primarily within the first 3-5 sites (see Fig.\,\ref{fig: DBloc}).

\subsection{Lifetimes of Dirac bosons}

We now estimate the lifetime of DBs. More specifically, we derive the scaling relation in Eq.\,\eqref{eq: tauDB}, detailing how the lifetime $\tau_\text{DB}$ of $\rho_\text{ss}(z)$ and $\braket{\alpha}$ scale with $N$ and $\xi$.

Consider an arbitrary initial state $\rho(0)$ and let the mean vector $\vec{m}(0) \equiv  \braket{\phi} = \tr[\phi \rho(0)]$. It follows that $\braket{\alpha(t)} = \vec{v}^\dag \vec{m}(t)$, where $\vec{m}(t) = e^{\mathbf{A}t}\vec{m}(0)$ is obtained from Eq.\,\eqref{eq: phiEOM} by taking expectation values on both sides and solving the differential equation. This allows us to write
\begin{align*}
    \braket{\alpha(t)} = \vec{v}^\dag e^{\mathbf{A} t}\vec{m}(0) = \big( e^{\mathbf{A}^\dag t}\vec{v} \big)^\dag \vec{m}(0),
\end{align*}
or, equivalently, $\vec{v}(t)^\dag \vec{m}(0)$, with $\vec{v}(t) = e^{\mathbf{A}^\dag t}\vec{v}$. As noted in the main text, $\vec{v}$ is a zero pseudoeigenvector of $\mathbf{A}^\dag$, with $\norm{\mathbf{A}^\dag \vec{v}} < \delta^N$ and $\delta <1$ derived according to the previous subsection. Now, we have the identity,
\begin{align*}
    \vec{v}(t)-\vec{v}(0) = \int_0^t \frac{d}{ds}\vec{v}(s)\,ds = \int_0^t \mathbf{A}^\dag \vec{v}(s)\,ds.
\end{align*}
The right hand-side is bounded above as
\begin{align*}
\norm{\int_0^t \mathbf{A}^\dag \vec{v}(s)\,ds} &= \norm{\int_0^t e^{\mathbf{A}^\dag s}\mathbf{A}^\dag \vec{v}(0)\,ds}   
\\
&\leq \int_0^t \left\Vert {e^{\mathbf{A}^\dag s}}\right \Vert \norm{\mathbf{A}^\dag \vec{v}(0)}\,ds
\\ &
< \delta^N \int_0^t \norm{e^{\mathbf{A}^\dag s}}\,ds . 
\end{align*}
The norm of the matrix exponential can be shown to be bounded by $e^{\Delta^\text{Bulk} s}$, where $\Delta^\text{Bulk}$ is, again, the maximal bulk amplification rate (see the supplement of Ref.\,\cite{FlynnBosoranasPRL2021} for more details). Thus,
\begin{align*}
 \norm{\vec{v}(t)-\vec{v}(0)} < \frac{\delta^N}{\Delta^\text{Bulk}}    \lp e^{\Delta^\text{Bulk} t}-1\rp.
\end{align*}
For instance, $\vec{v}(t)$ is guaranteed to stay within an accuracy $\eta>0$ of its initial value as long as 
\begin{align*}
    t &
    < \frac{1}{\Delta^\text{Bulk}}\ln\lp \delta^{-N} \Delta^\text{Bulk} \eta+1\rp 
    \\
    &
    = \frac{1}{\Delta^\text{Bulk}}\ln\lp e^{N/\xi}\Delta^\text{Bulk} \eta+1\rp .
\end{align*}
Moreover, since $\braket{\alpha(t)} - \braket{\alpha(0)} = \lp\vec{v}(t) -\vec{v}(0)\rp^\dag \vec{m}(0),$ the Cauchy-Schwarz inequality bounds the separation of the expectation value of $\alpha$ according to
\begin{align*}
   |\braket{\alpha(t)} - \braket{\alpha(0)}|^2 \leq \norm{\vec{v}(t)-\vec{v}(0)} \norm{\vec{m}(0)} .
\end{align*}
For $\norm{\vec{m}(0)}$ reasonably bounded, and taking $\eta \sim 1$ for reference, we thus obtain $\tau_\text{DB}$ as stated in Eq.\,\eqref{eq: tauDB} \footnote{Note that in physical systems, the dynamical matrix (and thus also the $\epsilon$'s associated with pseudospectra) carries units of frequency. It follows that $\norm{\mathbf{A}^\dag \vec{v}} < \omega_0 \delta^{-N}$ where $\omega_0$ is some system-dependent, but system size-\textit{independent} frequency scale.}. An analogous analysis may be applied to $\rho_\text{ss}(z)$ by noting that  $\vec{m}(0) = \vec{w}$ is a zero pseudoeigenvector of $\mathbf{A}$, and thus evolves slowly away from its initial value under the dynamics in Eq.\,\eqref{eq: HNEOM}.

\section{Invariant symmetric subspaces of U(1)-symmetric nonlinear systems}
\label{app: nonlinsym}

In this Appendix, we present an auxiliary result used throughout the main text, mainly to derive Eqs.\,\eqref{eq: bulkeom} and \eqref{eq: LLGbulkeom}. Consider a generic dynamical system on $\mathbb{C}^n$:
\begin{subequations}
\label{eq: gends}
\begin{align}
\dot{\vec{\alpha}}(t) &= F(\vec{\alpha}(t)),\,\, t\geq t_0 ,
\\
\vec{\alpha}(t_0) &= \vec{\alpha}_0\in\mathbb{C}^n,
\end{align}
\end{subequations}
where $F:\mathbb{C}^n\to \mathbb{C}^n$ is some sufficiently smooth function.  We will show that global U(1) symmetry implies that the invariant subspaces of {\em any} other unitary symmetry are invariant under the flow generated by $F$. This applies directly to our systems, with discrete translational symmetry playing the role of the auxiliary unitary symmetry, thus leading to invariance of the momentum $k$-subspaces in Eqs.\,\eqref{eq: bulkeom} and \eqref{eq: LLGbulkeom}.

\vspace{1em}
\noindent 
\textbf{Theorem.}  {\em Let $F$ defining the dynamical system in Eq.\,\eqref{eq: gends} satisfy $F(e^{i\theta}\vec{\alpha}) = e^{i\theta}F(\vec{\alpha})$, for all $\theta\in\mathbb{R}$ and $\vec{\alpha}\in\mathbb{C}^n$. Suppose there exists a unitary $\mathbf{U}\in$ U$(n)$ such that $F(\mathbf{U}\vec{\alpha}) = \mathbf{U}F(\vec{\alpha})$, for all $\vec{\alpha}\in\mathbb{C}^n$. Then the eigenspaces of $\mathbf{U}$ are invariant under the flow generated by $F$. That is, if $\vec{\alpha}(t)$ is a solution to Eq.\,\eqref{eq: gends} and at some time $t_s\geq t_0$ we have $\mathbf{U} \vec{\alpha}(t_s) = \lambda\vec{\alpha}(t_s)$, then  $\mathbf{U} \vec{\alpha}(t) = \lambda\vec{\alpha}(t)$ for all $t\geq t_s$. }

\vspace{1em}
\noindent\textit{Proof.} Unitarity of $\mathbf{U}$ means $\lambda = e^{i\theta}$ for some $\theta\in\mathbb{R}$. Note that, in view of the global U(1)-symmetry and $\mathbf{U}$-symmetry assumed on $F$, it follows that $F$ commutes with the linear composed map $\mathbf{S} = e^{-i\theta}\mathbf{U}$. Moreover, the set $I=\set{\vec{\alpha}\in\mathbb{C}^n : \mathbf{S}\vec{\alpha} = \vec{\alpha}}$ is precisely the eigenspace of $\mathbf{U}$ associated to eigenvalue $\lambda$. Now fix a solution $\vec{\alpha}(t)$ to Eq.\,\eqref{eq: gends}, and consider $\vec{\beta}(t) \equiv \mathbf{S}\vec{\alpha}(t) - \vec{\alpha}(t)$. We have
\begin{align*}
\dot{\vec{\beta}}(t) &= \mathbf{S}\dot{\vec{\alpha}}(t) - \dot{\vec{\alpha}}(t) \\
&
= \mathbf{S}F(\vec{\alpha}(t)) - F(\vec{\alpha}(t)) 
\\
&= F(\mathbf{S}\vec{\alpha}(t)) - F(\vec{\alpha}(t)). 
\end{align*}
In particular, $\dot{\vec{\beta}}(t) = 0$ whenever $\vec{\alpha}(t)\in I$. Thus, if $\vec{\alpha}(t_s)\in I$, then $\vec{\beta}(t_s) = 0$ and $\dot{\vec{\beta}}(t_s) = 0$. Thus, $\vec{\beta}(t)=0$ for all $t\geq t_s$, which is our desired result. \hfill$\Box$

\section{Analytical plane-wave solutions of \\ the semiclassical spin Lindbladian}
\label{app: ansolLind}

In this appendix, we analytically solve the system of equations in Eq.\,\eqref{eq: bulkeom}, describing the nonlinear semiclassical evolution of a bulk spin wave in the Lindbladian model Eq.\,\eqref{eq: LindSC} under PBCs. 

For simplicity, we take $s=1$ and move to a rotating frame $s^+_k\mapsto e^{i\omega t} s_k^+(t)$. We also write $\tilde{A}(k) \equiv -(r_k + i \omega_k)$, with $r_k$ and $\omega_k$ the decay rate and (rotating-frame) frequency of mode $k$, respectively. The relevant EOM then take the form
\begin{align*}
 \left \{ 
\begin{array}{lcl}
\dot{s}^z &=& r_k|s_k^+|^2 , \\
\dot{s}_k^+ &=& -(r_k+i\omega_k)s^z s_k^+.
\end{array}
\right. 
\end{align*}
The total spin $|s^+_k|^2+(s^z)^2 = (s_k^x)^2 + (s_k^y)^2 + (s^z)^2 = 1$ is conserved, and so we may consider the above flow to be defined on the unit sphere. There are two fixed points, independent of the values for $r_k$ and $\omega_k$. Namely, $(s_k^+,s^z) = (0,\pm 1)$. For $r_k>0$, $(0,1)$ is stable, while for $r_k<0$, $(0,-1)$ is stable. If $r_k=0$, there are periodic solutions of the form $s_k^+ = \sqrt{1-s^z(0)^2}\,e^{-i\omega_ks^z(0) t}$, $s^z(t)=s^z(0)$. These trajectories rotate in opposite directions on the upper and lower hemispheres. 

Conservation of total spin makes it convenient to move to spherical coordinates, i.e., $s_k^+ = \sin \theta_k e^{i\phi_k}$, $s^z(t) = \cos \theta_k$. In these coordinates, the fixed points then correspond to $\theta_k = 0$ or $\pi$, and the new state variables $\phi_k$ and $\theta_k$ evolve according to 
\[ \left \{ \begin{array}{lcl}
\dot{\phi}_k &= &-\omega_k\cos\theta_k ,\\
\dot{\theta}_k &= &-r_k\sin\theta_k. 
\end{array} \right. , \]
subject to initial conditions $\phi_{k,0} \equiv \phi_k(0)$ and $\theta_{k,0} \equiv \theta_k(0)$. The second equation can be solved for $\theta_k(t)$:
\begin{align*}
\cos\theta_k(t) = \tanh\lb r_k t + \text{arctanh}\lp\cos \theta_{k,0}\rp\rb ,
\end{align*}
which, in turn, provides a solution for $\phi_k(t)$:
\begin{align*}
\phi_k(t) &= \phi_{k,0} - \frac{\omega_k}{r_k}\ln \left\{\sin \theta_{k,0}\cosh\lb r_kt+\text{arctanh}\lp \cos \theta_{k,0}\rp\rb \right\} .
\end{align*}

\vfill

\end{document}